%% file: mathematical_estimate.tex
\documentclass[a4paper,10pt,twoside]{article}

\usepackage{psfrag}
\usepackage{chemarr}
\usepackage{appendix}
\usepackage{subfigure}
\usepackage{amssymb}
\usepackage{multirow}
\usepackage{amsmath}
\usepackage{psfrag}
\usepackage{array}
\usepackage[sort&compress,numbers]{natbib}
\usepackage{url}
\usepackage{pdflscape}

\newcommand{\ecoli}{{\em E.coli}}
\newcommand{\one}{($i$) }
\newcommand{\two}{($ii$) }
\newcommand{\three}{($iii$) }
\newcommand{\four}{($iv$) }
\newcommand{\five}{($v$) }

\newcommand{\onel}{($A$)}
\newcommand{\twol}{($B$)}
\newcommand{\threel}{($C$)}
\newcommand{\fourl}{($D$)}
\newcommand{\fivel}{($E$)}
\newcommand{\sixl}{($F$)}

\newcommand{\dmel}{\emph{D. melanogaster}}

\newcommand{\sm}{\emph{Appendix}}
\newcommand{\mm}{\emph{Materials and Methods}}

\usepackage[margin=2cm]{geometry}
\date{}

\usepackage{graphicx}

\begin{document} 
\title{Estimating binding properties of transcription factors from genome-wide binding profiles} 
\author{Nicolae Radu Zabet$^{1,2,\dagger}$ and Boris Adryan$^{1,2,\ast}$      \\
\mbox{}\\
\footnotesize $^1$Cambridge Systems Biology Centre, University of Cambridge, Tennis Court Road, Cambridge CB2 1QR, UK\\
\footnotesize $^2$Department of Genetics, University of Cambridge, Downing Street, Cambridge CB2 3EH, UK\\
\footnotesize $^\ast$ Corresponding author: ba255@cam.ac.uk
\footnotesize $^\dagger$ Correspondence can also be addressed to: n.r.zabet@gen.cam.ac.uk
}

\maketitle

\begin{abstract}
The binding of transcription factors (TFs) is essential for gene expression. One important characteristic is the actual occupancy of a putative binding site in the genome. In this study, we propose an analytical model to predict genomic occupancy that incorporates the preferred target sequence of a TF in the form of a position weight matrix (PWM), DNA accessibility data (in case of eukaryotes), the number of TF molecules expected to be bound specifically to the DNA and a parameter that modulates the specificity of the TF. Given actual occupancy data in form of ChIP-seq profiles, we backwards inferred copy number and specificity for five \emph{Drosophila} TFs during early embryonic development: Bicoid, Caudal, Giant, Hunchback and Kruppel. Our results suggest that these TFs display thousands of molecules that are specifically bound to the DNA and that, while Bicoid and Caudal display a higher specificity, the other three transcription factors (Giant, Hunchback and Kruppel) display lower specificity in their binding (despite having PWMs with higher information content). This study gives further weight to earlier investigations into TF copy numbers that suggest a significant proportion of molecules are not bound specifically to the DNA.
\end{abstract}

\section{Introduction}

Site-specific transcription factors (TFs) bind to the DNA and control the transcription rate of genes. Identifying the parameters influencing the interactions between TFs and DNA is essential in unveiling the gene regulatory program and better understanding the gene regulatory process. Significant insight has been gained by deriving the genome-wide binding profiles of TFs and, often, two complementary approaches have been combined to determine and analyse these genomic binding events, namely: \one experimental determination of regions of genomic occupancy through chromatin immunoprecipitation experiments (ChIP-chip or ChIP-seq) \cite{park_2009} and \two computational inference of the very binding sites using various bioinformatics and biophysics approaches. In most cases, these computational approaches are based on scanning the DNA with a preferred DNA word, the so-called motif (often represented in the form of position weight matrix - PWM) \cite{wasserman_2004}. However, this approach discards effects from steric hindrance and competition on the DNA \cite{hermsen_2006,hoffman_2010,sheinman_2012} or saturation of the binding sites due to high abundance of the TF \cite{djordjevic_2003,foat_2006,roider_2007,zhao_2009,simicevic_2013,zabet_2013_occupancy,ilsley_2013}. 

An alternative to the bioinformatics approach is the statistical thermodynamics framework, which models the binding of TF molecules to DNA segments using the principles of physical chemistry \cite{djordjevic_2003,foat_2006,roider_2007,zhao_2009,simicevic_2013,sadka_2009,he_2009,wasson_2009,hoffman_2010,kaplan_2011,cheng_2013}. This approach considers both steric hindrance and the number of molecules that are bound to the DNA. Briefly, this framework computes the statistical weight for each possible configuration of the system, where a configuration represents the specific combination of locations on the DNA segment that are occupied by TF molecules. However, given the number of possible configurations, the computations of all statistical weights becomes challenging with increasing DNA segment size. To address this problem, we used several approximations within the statistical thermodynamics framework \cite{ackers_1982,bintu_2005_model,chu_2009,simicevic_2013}, which lead us to develop an analytical solution. This analytical model now allows us to compute binding profiles with the benefits of thermodynamics methods on a genomic scale (e.g. we computed the ChIP-seq profile of five TFs over $92\ Mbp$ of DNA in less than one day using one CPU), instead of being restricted to a few loci compared to the classical approach as it was the case in some previous studies \cite{sadka_2009,he_2009,wasson_2009,hoffman_2010,kaplan_2011,cheng_2013}. This model takes as input four parameters: \one a PWM,  \two DNA accessibility data, \three the predicted or measured number of molecules that are specifically bound to the DNA and \four a factor that modulates the specificity of the TF \cite{berg_1987}. While the first two parameters are often known -- the PWM from \emph{in vitro} experiments such as DNAse I footprinting, EMSA, SELEX or PBM \cite{stormo_2010} and the DNA accessibility data from genome-wide DNase I-seq experiments -- the last two parameters are usually unknown and difficult to measure. Here, we show that the number of specifically bound molecules and the specificity of the TF can be computed by fitting the predictions of the model to experimentally determined binding profiles. 

We applied our model on binding data of five TFs (Bicoid, Caudal, Giant, Hunchback and Kruppel) in the \dmel\ stage 5 embryo \cite{bradley_2010,li_2011}. Using the aforementioned rationale, we identified the number of DNA bound molecules and the specificity for each of these TFs that fit the ChIP-seq signal with good accuracy. In particular, we estimate that the abundance of each of the TFs in the system is in the range of thousands of molecules that are specifically bound to the DNA per cell/nuclei. Finally, we also found that, while Bicoid and Caudal display high specificity (being able to better discriminate better between ``good'' and ``bad'' DNA words), Giant, Hunchback and Kruppel display lower specificity. 

\section{Materials and Methods}

\subsection{Analytical model}
In our previous work \cite{zabet_2013_occupancy}, we investigated the genomic occupancy of TFs using a comprehensive model that simulated the dynamics of TF molecules in the cell. We consider the \emph{facilitated diffusion} mechanism, which assumes that the molecules perform 3D diffusion in the cytoplasm/nucleoplasm and 1D random walk on the DNA \cite{riggs_1970a,berg_1981,kabata_1993,blainey_2006,elf_2007,mirny_2009,hager_2009,vukojevic_2010,hammar_2012,zabet_2012_model,zabet_2012_review}. Our results showed that the genomic occupancy of TFs is mainly influenced by: \one TF abundance and \two PWM information content \cite{zabet_2013_occupancy}. These results suggest that the statistical thermodynamics framework could accurately predict the genomic occupancy of TFs, although at a high computation cost. We addressed this issue, and in the \sm\ we derive an analytical model based on statistical thermodynamics framework to compute the probability that a binding site is occupied as (see section \ref{sec:appendixAnalyticalModel})
\begin{equation}
P^{bound}_j\left(\lambda, w, N, a\right)=\frac{N\cdot a_j\cdot e^{\left(\frac{1}{\lambda}w_j\right)}}{N\cdot a_j\cdot e^{\left(\frac{1}{\lambda}w_j\right)} + L\cdot n\cdot\left\langle a_i e^{\left(\frac{1}{\lambda}w_i\right)}\right\rangle_i}\label{eq:tauPWMabundanceAcc}
\end{equation}
where $N$ is the number of molecules bound to the DNA, $a_j$ represents the accessibility at site $j$, $\lambda$ represents a scaling factor of the PWM score \cite{berg_1987,roider_2007}, $w_j$ represents the PWM score at site $j$, $L$ the length of the DNA and $n$ is the ploidy level (the number of copies of the genome, e.g.  for diploid genomes $n=2$). When DNA accessibility data is discarded, then $a_j=1, \forall j$.

\subsection{Data sets}

\begin{figure*}[htp]
  \begin{center}
  		\includegraphics[width=\textwidth]{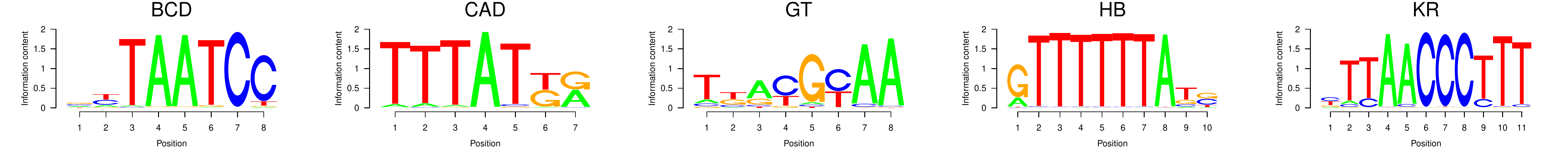}
  \end{center}
\caption{\emph{PWMs for the five TFs}. The graph show the sequence logos for the following TFs: \one Bicoid, \two Caudal, \three Giant, \four Hunchback and  \five Kruppel as also used in \cite{kaplan_2011}. When computing the PWMs we used a pseudo-count of $1$. The information content for the five motifs is: \one $I_{BCD}=11.3$, \two $I_{CAD}=10.7$, \three $I_{GT}=8.7$, \four $I_{HB}=15.6$ and  \five $I_{KR}=17.3$.}\label{fig:PWMmotifs}
\end{figure*}

In Figure \ref{fig:PWMmotifs}, we plot the sequence logos of the PWMs for the five TFs included in our analysis (Bicoid, Caudal, Giant, Hunchback and Kruppel); Berkeley \emph{Drosophila} Transcription Network Project (\url{bdtnp.lbl.gov}) \cite{kaplan_2011}. To generate occupancy profiles we used a method originally introduced by \cite{kaplan_2011}, for which we selected a mean segment length of $200\ bp$, a standard deviation of $200\ bp$ and the profile was smoothed over $250\ bp$; see also \cite{zabet_2013_occupancy}. First, we consider all the loci from \cite{kaplan_2011}, which are also listed in Table \ref{tab:kaplan2011loci} in the \sm.

In our analysis, we also consider DNA accessibility data derived from DNase-seq experiments in stage 5 \dmel\ embryos. The raw data are from \cite{thomas_2011} and was used to compute the probability of accessible DNA; see  \cite{kaplan_2011} and equation \eqref{eq:paccessible} in the \sm. In addition, we also used in our analysis the set of DNA accessible regions computed at a $5\%$ false discovery rate ($14.5\ Mbp$) \cite{li_2011} and represented accessible sites by $a=1$ and inaccessible regions by $a=0$; note that the dm3 release 5 coordinates of these regions were downloaded from \url{ftp://hgdownload.cse.ucsc.edu/goldenPath/dm3/database/bdtnpDnaseAccS5.txt.gz}.  The \dmel\ genome consists of the euchromatin (chromosomes chr2L, chr2R, chr3L, chr3R, chr4, chrX), heterochromatin (chr2LHet, chr2RHet, chr3LHet, chr3RHet, chrXHet, chrYHet), unmapped regions (ChrU and chrUextra) and mitochondrial genome (ChrM) \cite{adams_2000}. Only $5.6\%$ of the DNase-seq reads from \cite{thomas_2011} map to heterochromatin or unmapped regions and  only $3.1\%$ of the total accessible DNA is in heterochromatin or on ChrU. The contribution of the heterochromatin and ChrU to our analysis is negligible and, thus, we considered in our analysis only the euchromatic genome ($\approx 120\ Mbp$). 

\subsection{Quantifying the differences between the analytical model and experimental data}

To quantify the difference between our analytical model and the experimental data, we consider two measures, namely: \one the Pearson correlation coefficient ($\rho$) and \two the normalised mean squared error over $1\ Kbp$ ($MSE$).

\section{Results}

We applied our analytical model (derived in \mm\ and \sm) to investigate the ChIP-seq data set published in \cite{bradley_2010}, which lead to a direct comparison to the method proposed in \cite{kaplan_2011}. This data set consists of ChIP-seq profiles determined in stage 5 \dmel\ embryos for five TFs: \one Bicoid (BCD), \two Caudal (CAD), \three Giant (GT), \four Hunchback (HB) and \five Kruppel (KR).

Our analytical model requires four parameters: \one the PWM for the factor under investigation, \two the DNA accessibility data over the locus that is analysed, \three the number of bound molecules (as can be inferred from genomic binding data) and \four the specificity of the TF for the DNA (through the $\lambda$ factor); see equation \eqref{eq:tauPWMabundanceAcc}. We used the PWMs presented in Figure \ref{fig:PWMmotifs} and treated DNA either as `naked' or used DNA accessibility from previously published work \cite{kaplan_2011,thomas_2011,li_2011}. 

The number of bound molecules and TF specificity $\lambda$ are usually unknown. Here, we estimate these parameters by identifying the values that produce the best fit with the experimentally measured profile \cite{foat_2006,roider_2007,cheng_2013,simicevic_2013}. First, we converted the binding probability determined by equation \eqref{eq:tauPWMabundanceAcc} into an occupancy profile (an artificial ChIP-seq signal; using equation \eqref{eq:PboundToChIPseq} in the \sm) and then we generate the ChIP-like \emph{in silico} profiles using a method described by \cite{kaplan_2011}, (selecting a mean segment length of $200\ bp$ and a standard deviation of $200\ bp$ and then smoothing the profile over $250\ bp$); the R implementation of this method is described in \cite{zabet_2013_occupancy}.

\begin{figure*}
\centering
\includegraphics[width=\textwidth]{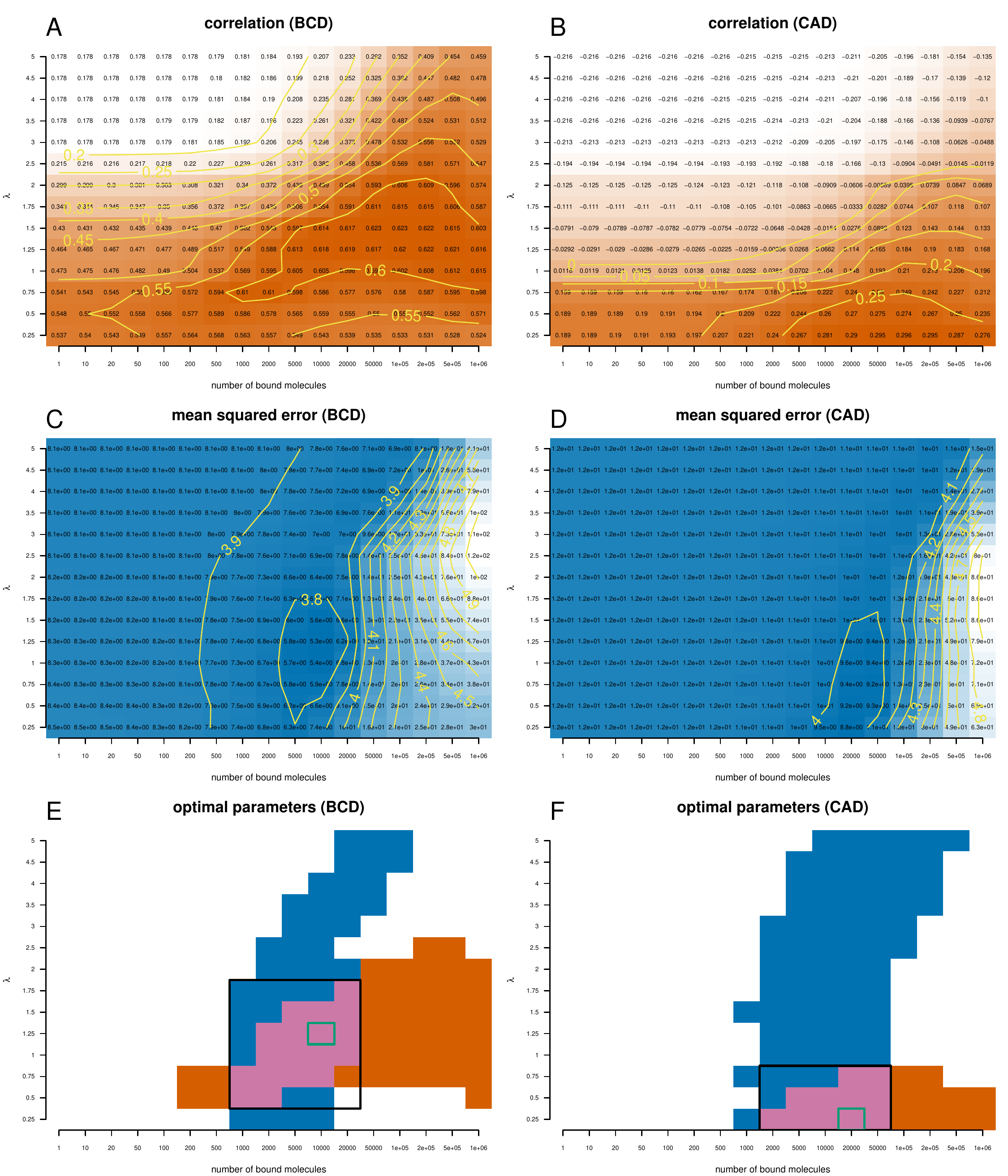}
\caption{\emph{Quantifying the distances between Bicoid and Caudal ChIP-seq profiles and the profiles derived with the analytical model}. We plotted heatmaps for the correlation \onel\ and \twol\ and mean squared error \threel\ and \fourl\ between the analytical model and the ChIP-seq profile of Bicoid ($A,C$) and Caudal ($B,D$). We computed these values for different sets of parameters: $N\in[1,10^6]$ and $\lambda\in[0.25,5]$. We considered only the sites that have a PWM score higher than $70\%$ of the difference between the lowest and the highest score. ($A,B$) Orange colour indicates high correlation between the analytical model and the ChIP-seq profile, while white colour low correlation.  ($C,D$) Blue colour indicates low mean squared error between the analytical model and the ChIP-seq profile, while white colour high mean squared error. ($E,F$) We plotted the regions where the mean square error is in the lower $12\%$ of the range of values (blue) and the correlation is the higher $12\%$ of the range of values (orange). With green rectangle we marked the optimal set of parameters in terms of mean squared error and with a black rectangle the intersection of the parameters for which the two regions intersect.  \label{fig:heatmapChIPseq_BCD_CAD_NoaccGroup070}}
\end{figure*}

Figure \ref{fig:heatmapChIPseq_BCD_CAD_NoaccGroup070} plots the heatmap of the correlation and mean squared error for Bicoid and Caudal when the number of bound molecules and $\lambda$ factor are varied. We performed a grid search to identify the sets of parameters (TF abundance and specificity) that maximises the correlation and that minimises the mean squared error. Our results show that the set of parameters that minimises the mean squared error lead to only a negligible reduction in the correlation compared to its maximum value. In contrast, the parameters that maximises the correlation lead to a strong increase in the mean squared error compared to its minimum. This means that changes in the $\lambda$ factor and the TF abundance have a stronger effect on the mean squared error than on correlation. In a similar way, we determined the set of parameters that minimise the mean squared error and maximise the correlation for the rest of TFs (Giant, Hunchback and Kruppel) by analysing the data on the heatmaps in Figure \ref{fig:heatmapChIPseq_GT_HB_KR_NoaccGroup070} in the \sm. Table \ref{tab:paramsModelNoAcc} lists the optimal set of parameters for all five TFs.

\input{./OptimalSetOfParametersTableNoAccessibility.tex}

\subsection{DNA accessibility improves the model predictions}

One of the main results of \cite{kaplan_2011,simicevic_2013,cheng_2013} is that DNA accessibility data improves the computational prediction when estimating ChIP-seq profiles with PWMs. To investigate this result, we also consider DNA accessibility regions from stage 5 \dmel\ embryos computed with a $5\%$ false discovery rate \cite{li_2011} and represented accessible sites by $a=1$ ($14.5\ Mbp$) and inaccessible sites by $a=0$. 

\begin{figure*}
\centering
\includegraphics[width=\textwidth]{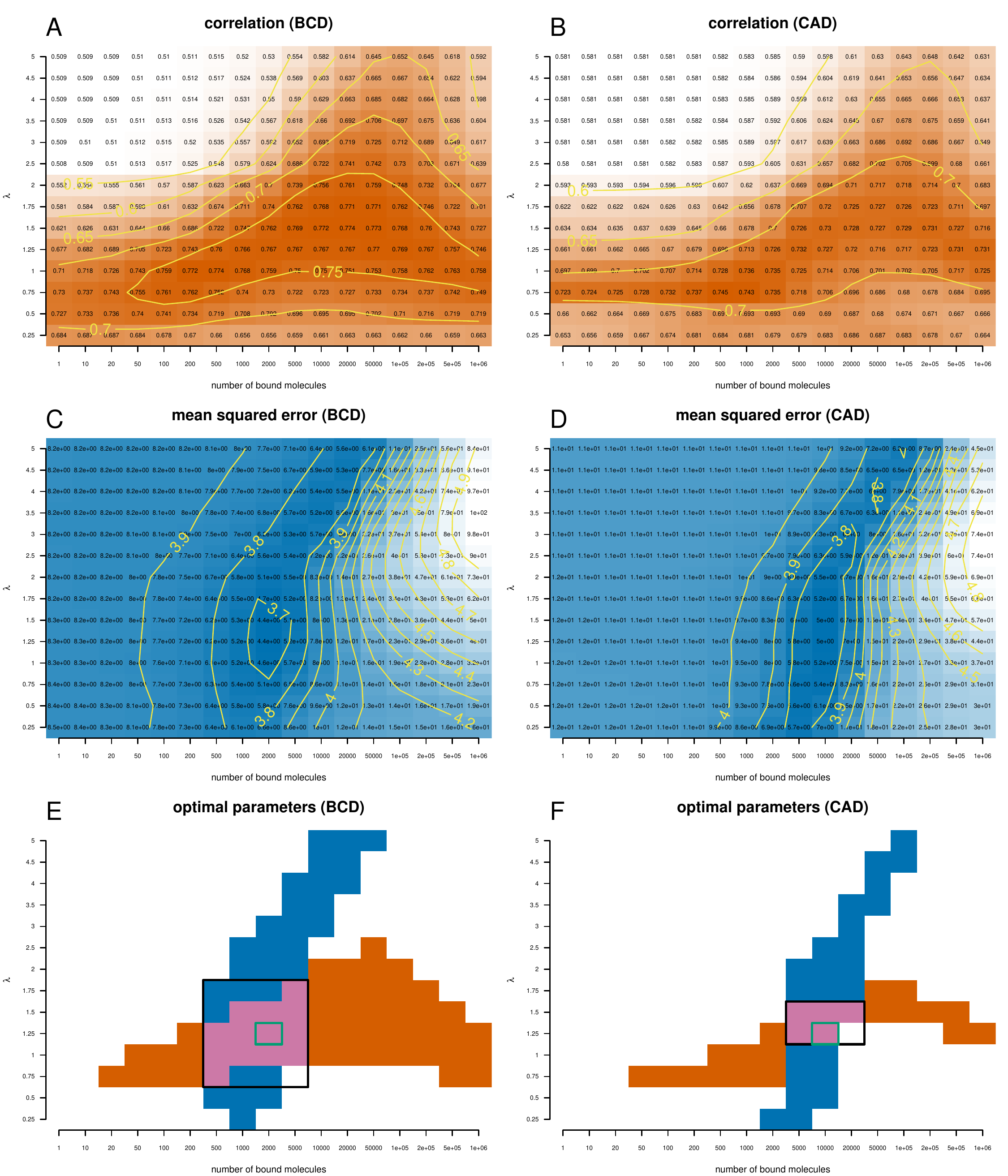}
\caption{\emph{Quantifying the distances between Bicoid and Caudal ChIP-seq profiles and the profiles derived with the analytical model which includes DNA accessibility data}. This is the same as Figure \ref{fig:heatmapChIPseq_BCD_CAD_NoaccGroup070}, except that we included binary DNA accessibility data in the analytical model. \label{fig:heatmapChIPseq_BCD_CAD_AccRegionsGroup070}}
\end{figure*}

Using the same approach as in the case of all DNA being accessible, we plot the heatmaps for correlation and mean squared error for Bicoid, Caudal, Giant, Hunchback and Kruppel and then we performed a grid search to identify the combination of parameters that minimises the differences between the ChIP-seq profiles and the profiles predicted by equations \eqref{eq:tauPWMabundanceAcc} and \eqref{eq:PboundToChIPseq} in the \sm; see Figure \ref{fig:heatmapChIPseq_BCD_CAD_AccRegionsGroup070} and Figure \ref{fig:heatmapChIPseq_GT_HB_KR_AccRegionsGroup070} in the \sm. 

\begin{figure*}
\centering
\includegraphics[width=\textwidth]{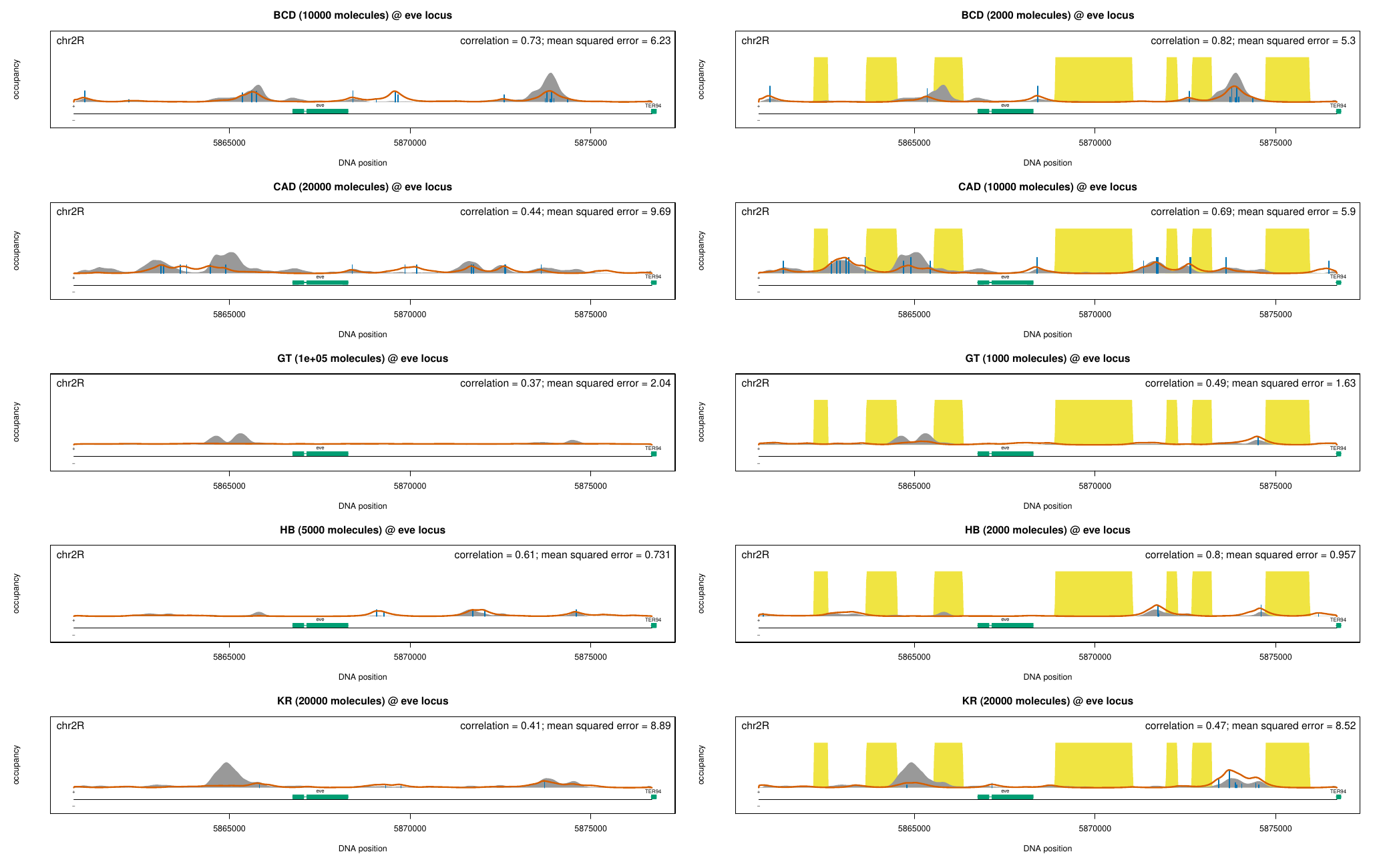}
\caption{\emph{Binding profiles at eve locus}. The grey shading represents a ChIP-seq profile, the red line represents the prediction of the analytical model, the yellow shading represents the inaccessible DNA and the vertical blue lines represent the percentage of occupancy of the site (we only displayed sites with an occupancy higher than $5\%$).  We plotted the profiles for the five TFs: \one Bicoid, \two Caudal, \three Giant, \four Hunchback and \five Kruppel. (\textbf{Left}) The analytical model assumed a naked DNA (the entire genome is accessible) and used the set of parameters listed in Table \ref{tab:paramsModelNoAcc}. (\textbf{Right})  The analytical model included DNA accessibility data from \cite{kaplan_2011,li_2011} and used the set of parameters listed in Table \ref{tab:paramsModelAccRegions}. \label{fig:bindingProfilesEve}}
\end{figure*}

Table \ref{tab:paramsModelAccRegions} lists the optimal set of parameters for the five TFs in the case of DNA accessibility. Again, one can see that selecting the set of parameters that minimises the mean squared error leads to only negligible reduction in the correlation. Our results confirm that DNA accessibility data improves the model prediction (increases the correlation and reduces the mean squared error) and, thus, supports the finding that DNA accessibility is a significant factor that drives the genomic occupancy of TFs; compare Table \ref{tab:paramsModelNoAcc} to Table \ref{tab:paramsModelAccRegions} and see Figure \ref{fig:bindingProfilesEve}. Overall the correlation between our model predictions and the ChIP-seq data sets is similar to the one found in \cite{kaplan_2011}.

\input{./OptimalSetOfParametersTableAccessibilityRegions.tex}

Interestingly, Figure \ref{fig:heatmapChIPseq_BCD_CAD_AccRegionsGroup070} \fivel\ and \sixl\ show that there is a high correlation between our model and the ChIP-seq data for a wide range of values for the number of DNA bound molecules (orange area). However, the range of values of $\lambda$ that result in high correlation is much smaller. In contrast, the mean squared error is reduced only for a small interval of TF abundances, but it is optimal for a wide range of values for $\lambda$ (blue area). This result suggests that the correlation between the model and the ChIP-seq profile cannot be used to estimate TF abundance (as previously done), but can accurately estimate the range of values for $\lambda$. In contrast, the mean squared error can be used to estimate the number of bound molecules, but cannot be used to estimate the $\lambda$ factor. Thus, to get a better estimate for both the number of DNA bound molecules and $\lambda$ one needs to consider both correlation and mean squared error and identify the range of parameters where both measures are optimal.

It is also worthwhile noting that, for all TFs, the number of bound molecules that best fit the data is in the range of thousands of molecules bound to the genome (and is on average five times lower than the case of naked DNA). In addition, we identified that, for Hunchback and Kruppel, the values of $\lambda$ that optimises the analytical model are significantly higher than $1$, which means that although the motifs of these two TFs have high information content (see Figure \ref{fig:PWMmotifs}), the two TFs have low specificity and cannot distinguish well between different DNA words \cite{stormo_2010}. In contrast, we found that Bicoid, Caudal and Giant display values of $\lambda$ around $1$, which indicates that the specificities of these TFs are equal to their information content, as defined by Stormo and Zhao \cite{stormo_2010}.

\subsection{Additional factors that influence the binding profiles}

Despite the generally accurate predictions of the model, there are locations on the genome where the model fails to predict the ChIP-seq profile. In the \sm, we plotted the estimates of the binding profiles at all twenty-one loci using the optimal set of parameters (from Table \ref{tab:paramsModelAccRegions}); see Figures \ref{fig:profileAllPositivesBCD},  \ref{fig:profileAllPositivesCAD},  \ref{fig:profileAllPositivesGT},  \ref{fig:profileAllPositivesHB} and  \ref{fig:profileAllPositivesKR}. Next, we systematically investigated several assumptions in our model that could account for the differences between the ChIP-seq data set and the model predicted profiles.  

\subsubsection{Range of PWM scores included in the model}

For our analysis we only considered predicted binding sites that display a PWM score which is higher than $70\%$ of the difference between the strongest and the weakest site ($\forall j \textrm{, where } w_j \geq 0.7\times \left[ \max_{i}(w_i) - \min_{i}(w_i)\right]$). To understand why we discard non-specific sites, it is important to remember that conventional ChIP experiments display a population average over millions of cells/nuclei. While specific sites will be occupied in the majority of the cells (nuclei), a particular non-specific site will be occupied in a few cells, because there are many similar low affinity sites in the genome \cite{mueller_2013}. This means that ChIP data reflects binding at the specific sites, and that low affinity sites can be discarded. To test whether our threshold selection affected the results, we also considered the case of a lower threshold of $30\%$.

We found that weaker binding sites do not affect our model estimate for the profile for TFs that have low values of $\lambda$ (Bicoid, Caudal and Giant); see Table \ref{tab:paramsModelAccRegions030} and Figure \ref{fig:heatmapChIPseq_BCD_CAD_GT_AccRegionsGroup030} in the \sm. However, for TFs with higher values of $\lambda$ (Hunchback and Kruppel), weaker binding sites affect the binding profiles, but leading only to a negligible reduction in the quality of the profile generated by our model; see Figure \ref{fig:heatmapChIPseq_HB_KR_AccRegionsGroup030} in the \sm. Including lower affinity binding into our model also leads to a similar set of parameters (TF abundance and $\lambda$) that optimises the fit for four TFs (Bicoid, Caudal, Giant and Hunchback), thus, indicating that our estimates are robust. For Kruppel, when including lower affinity binding sites, our method estimates a lower TF abundance and $\lambda$. However, $\lambda$ remains significantly higher than $1$ and the quality of the fit are slightly worse that in the case of including only sites  that have a PWM score higher than $70\%$ of the difference between the lowest and the highest score; see Table  \ref{tab:paramsModelAccRegions030} in the \sm.

\subsubsection{DNA accessibility data}

Some DNA loci  are marked as being `inaccessible' in the DNA accessibility data, but, at the same time,  display binding of TFs in ChIP experiments; see  Figures \ref{fig:profileAllPositivesBCD}-\ref{fig:profileAllPositivesKR} in the \sm. This suggests that regions with an intermediary level of DNA accessibility could have been marked as inaccessible despite allowing binding of TFs. To investigate this aspect, we also considered the case of different levels of DNA accessibility (the data is represented by continuous values between 0 and 1) and we converted the read density in probability of a site being accessible by using the approach described in \cite{kaplan_2011}; see equation \eqref{eq:paccessible} in the \sm. We found that, when using non-binary accessibility data, the difference between the predictions of our model and the ChIP-seq data are similar to the case of using binary DNA accessibility data; see Table \ref{tab:paramsModelAccProbability070} and Figures \ref{fig:heatmapChIPseq_BCD_CAD_AccProbabilityGroup070} and \ref{fig:heatmapChIPseq_GT_HB_KR_AccProbabilityGroup070} in the \sm.

\subsubsection{PWMs}
Finally, we investigated whether the choice of PWMs affected our results by performing the analysis using alternative PWMs  from the JASPAR database  \cite{portales-casamar_2010}; see Figure \ref{fig:PWMmotifsJaspar} in the \sm. One should note that, while the motifs of Bicoid and Caudal are similar in both BDTNP \cite{kaplan_2011} and JASPAR \cite{portales-casamar_2010}, the motif for Giant has a higher information content and the motifs for Hunchback and Kruppel have a lower information content in JASPAR \cite{portales-casamar_2010} compared to BDTNP \cite{kaplan_2011}. Our results show that, by using a different set of PWMs, we obtained slightly worse values for correlation and mean squared error compared to the case of using the PWMs from BDTNP \cite{kaplan_2011}; see Figures \ref{fig:heatmapChIPseq_BCD_CAD_AccRegionsJasparGroup070} and \ref{fig:heatmapChIPseq_GT_HB_KR_AccRegionsJasparGroup070} and Table \ref{tab:paramsModelAccRegionsJaspar070} in the \sm. It remains to be investigated if this could be a generalized approach to determine the quality of PWMs from different sources.

Interestingly, we found that the values of  $\lambda$ that optimise the model for Giant,Hunchback and Kruppel differ significantly between the case of the PWMs from BDTNP \cite{kaplan_2011} and the PWMs from JASPAR \cite{portales-casamar_2010}. This suggests that independent of the actual PWMs, the three TFs display similar specificity. Since three TFs have different PWMs, we also investigated the binding profiles at the twenty one loci and we found that, in certain cases, the use of the PWMs from the JASPAR database leads to differences in the predicted profile; see Figures \ref{fig:profileAllPositivesJasparGT}, \ref{fig:profileAllPositivesJasparHB} and \ref{fig:profileAllPositivesJasparKR} in the \sm. For example, for Kruppel, we observed a slightly better estimation of the binding profile at \emph{D}, \emph{H}, \emph{Kr}, \emph{cad}, \emph{ftz}, \emph{gt}, \emph{hkb}, \emph{os} and \emph{slp} loci and a slightly worse estimate of the profile at \emph{cnc}, \emph{croc}, \emph{kni}, \emph{opa}, \emph{prd}, \emph{run} and \emph{tll} loci compared to the case when the PWMs from BDTNP \cite{kaplan_2011} was used; compare Figure \ref{fig:profileAllPositivesKR} to Figure \ref{fig:profileAllPositivesJasparKR}  in the \sm.

\subsection{Genome-wide analysis of TF binding}

\begin{figure}
\centering
\includegraphics[width=0.7\textwidth]{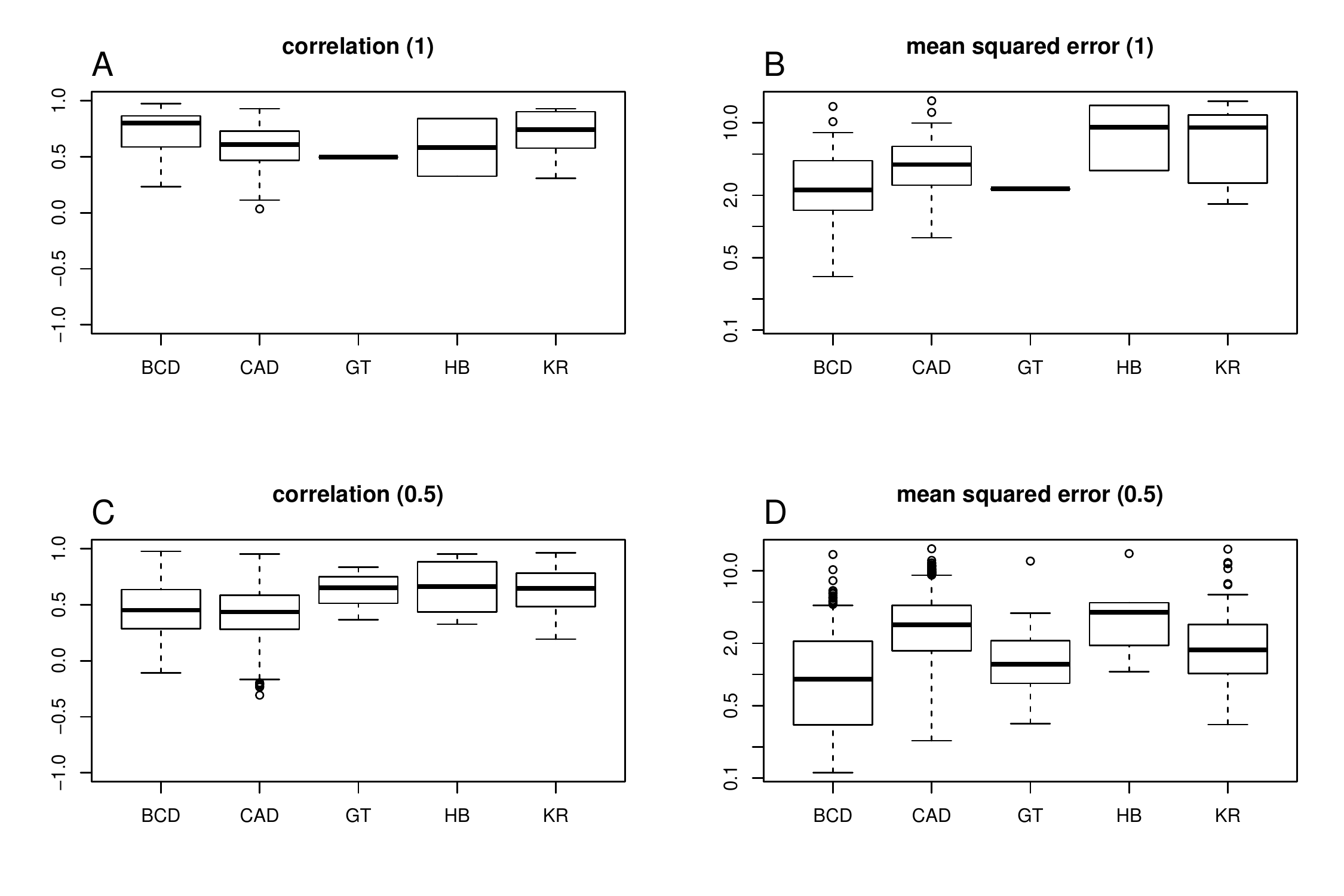}
\caption{\emph{Genome-wide quality of the fit}. The boxplots represent the $(A,C)$ correlation and $(B,D)$ mean squared error between the ChIP-seq data sets and the analytically estimated profiles. We partitioned the genome in $20\ Kbp$ regions and we kept only the regions that had at least one DNA accessible site ($4599$ regions). Next for each ChIP-seq data set we selected the regions where the mean ChIP-seq signal is higher than a proportion of the background (see Table \ref{tab:ChIPseqProfileStatistics} in the \sm). In $(A,B)$, we selected the regions with a mean ChIP-seq signal higher than the background ($>B$). In $(C,D)$, we selected the regions with a mean ChIP-seq signal higher than half the background ($>0.5\cdot B$). The numbers of DNA regions that display a mean ChIP-seq signal higher than the thresholds are listed in Table \ref{tab:GenomeWideNoOfRegions} in the \sm. In all subgraphs we used the set of parameters from Table \ref{tab:paramsModelAccRegions}. \label{fig:genomeWideQuality}}
\end{figure}

One advantage of our analytical model is that it can be used to predict the binding profiles genome-wide and, thus, we extended the analysis from the original twenty one loci to the entire genome. We partitioned the genome in $20\ Kbp$ regions, from which we removed regions that did not have any accessible site. For each ChIP-seq profile, we then selected the regions that display a ChIP-seq signal higher than the genome-wide background. We found that the quality of our model's predictions vary widely; see Figure \ref{fig:genomeWideQuality} \onel\ and \twol. In particular, there are regions where the correlation between our model predictions and the ChIP-seq profile is high, but at the same time regions where this correlation is low.

Kaplan \emph{et al.} \cite{kaplan_2011} found that, at loci with low binding (low ChIP-seq signal), the correlation between the statistical thermodynamics model and the ChIP-seq profile was low.  To test whether this is valid genome-wide, we also analysed regions where the mean signal is higher than half of the genome-wide background (leading in an increase in the number of investigated loci). Our results confirm that there is a decrease in the mean correlation when including regions with lower ChIP-seq signal; see Figure \ref{fig:genomeWideQuality} \threel. We also perform a Kolmogorov-Smirnov test that showed that in the case of Bicoid and Caudal this difference is statistically significant; see Figure \ref{fig:GenomeWideKSPvalue} in the \sm. This also means that, at least for regions with strong binding, the model predictions are highly correlated with the ChIP-seq profile as previously found \cite{kaplan_2011}; see Figure \ref{fig:genomeWideQuality}. Nevertheless, for regions with low binding, in addition to the reduction in the correlation we also observed a decrease in the mean squared error, which is statistically significant in the case of Bicoid, Caudal and Kruppel; see Figure \ref{fig:GenomeWideKSPvalue} in the \sm. Note that for Giant and Hunchback the difference is not statistically significant due to the small number of loci included in the analysis; see Table \ref{tab:GenomeWideNoOfRegions} in the \sm. This indicates that our model is able to correctly capture the low signal in those regions, but there is little or no correlation to the actual ChIP-seq signal. One explanation for this result is that, in those regions, there is little or no binding and what the ChIP-seq method recovers might be considered technical noise.

\subsection{Nucleus specific binding predictions}
\begin{figure}
\centering
\includegraphics[width=0.7\textwidth]{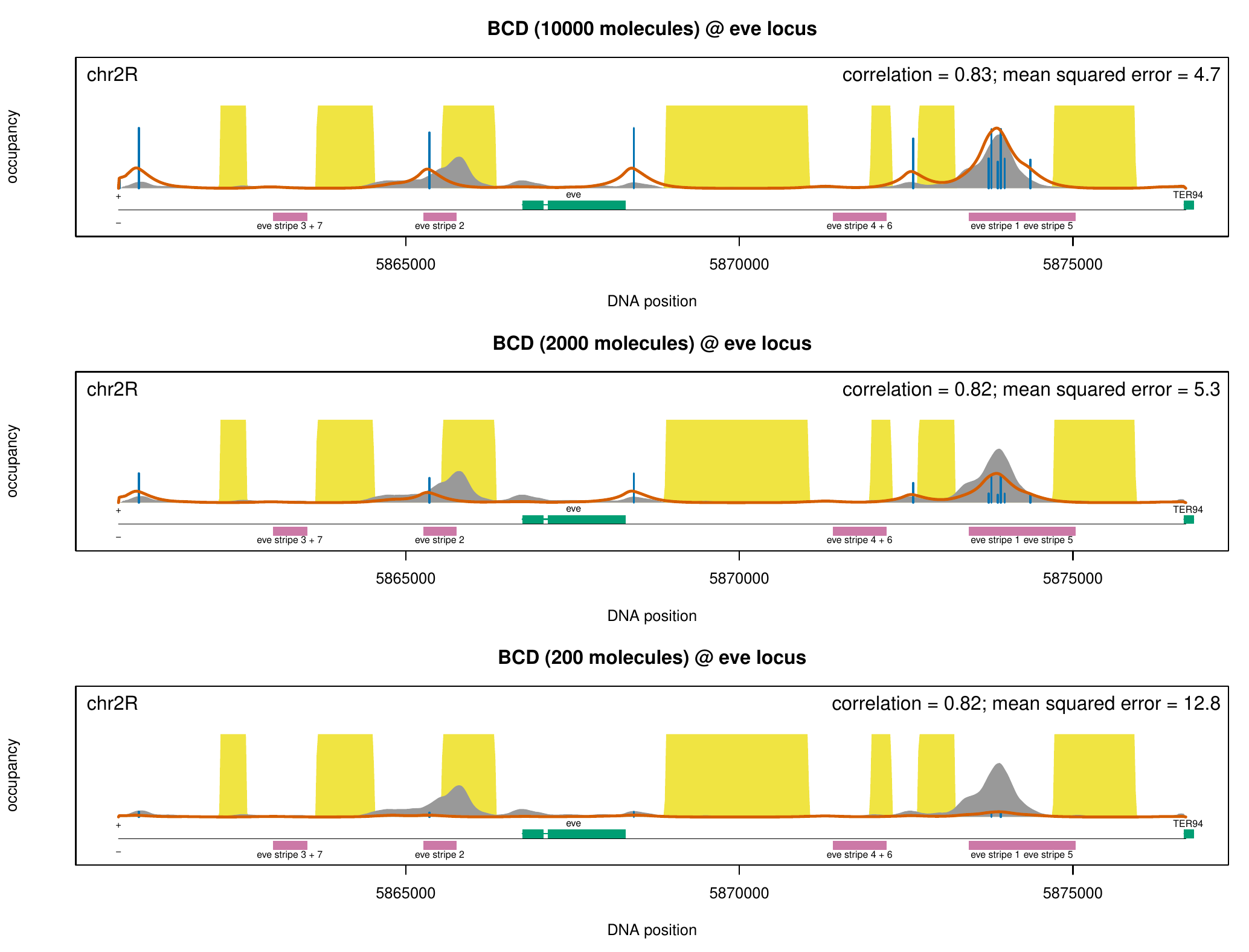}
\caption{\emph{Bicoid binding profile at eve locus for various abundances}. The grey shading represents the embryo-wide ChIP-seq profile of Bicoid, the red line represents the prediction of the analytical model, the yellow shading represents the inaccessible DNA and the vertical blue lines represent the percentage of occupancy of the site (we only displayed sites with an occupancy higher than $5\%$). We  consider three cases \one $N_{\textrm{BCD}}=10000\ molecules$ (anterior pole), \two $N_{\textrm{BCD}}=2000\ molecules$ (stripe 2 region) and \three $N_{\textrm{BCD}}=200\ molecules$ (posterior pole). We also assume a factor $\lambda_{\textrm{BCD}}=1.25$. The magenta rectangles mark the enhancers for the stripe formation; from left to right these are: \one \emph{eve stripe 3+7}, \two  \emph{eve stripe 2}, \three  \emph{eve stripe 4+6}, \four  \emph{eve stripe 1} and \five  \emph{eve stripe 5}.  \label{fig:profileBCDeveAbundanes}}
\end{figure}

Using our model we can investigate the binding profiles at various locations for which ChIP-seq data is not available. While the ChIP-seq profiles were generated for the entire embryo (and, thus, we are assuming average TF abundance over the pool of cells in an embryo), there is still no indication how these profiles look at specific locations along the anterior-posterior axis of the animal. This is important because, along the embryonic axis, the TF abundance can vary significantly, see \cite{stanojevic_1991,gregor_2007,abuarish_2010,grimm_2010,drocco_2011,kaplan_2011,little_2011,drocco_2012}. First, we generated the Bicoid binding profile in nuclei that are positioned at $40\%$ of egg-length along the A-P axis from the anterior pole (stripe 2) by assuming that there are $2000$ molecules of Bicoid in this region. In particular, we approximated at $40\%$ of egg-length along the A-P axis the number of specifically bound molecules is similar to the one computed for the embryo-wide ChIP-seq. Figure \ref{fig:profileBCDeveAbundanes} shows how this binding profile looks and confirms that Bicoid binds to the \emph{eve stripe 2} enhancer (chr2R:5865267-5865750), as opposed to the case of the posterior pole (with much lower concentration), where Bicoid does not bind to this enhancer. Note that we approximated that, at the posterior pole, the amount of Bicoid is ten times lower than at the \emph{eve stripe 2} enhancer \cite{gregor_2007}. 

Next, we generated the Bicoid binding profiles at the anterior region of the embryo assuming that, in this region, the number of specifically bound Bicoid molecules is approximately five times higher than at the \emph{eve stripe 2} enhancer (\cite{gregor_2007} approximated that at the anterior pole there are approximately four times more Bicoid molecules than at the \emph{eve stripe 2} enhancer). Figure \ref{fig:profileBCDeveAbundanes} shows that, at the anterior pole, there is significantly more binding of Bicoid to the \emph{eve stripe 2} enhancer, which raises the question of why there is no \emph{eve} expression in that region. Initially, it was assumed that Bicoid acts only as an activator for the \emph{eve stripe 2} enhancer \cite{stanojevic_1991,small_1991}. However, a recent study \cite{ilsley_2013} proposed that Bicoid has a dual role as both activator and repressor and this is controlled by its abundance, i.e., for low and medium abundances Bicoid activates \emph{eve}, while for high abundance it will repress it. Our results support (without providing a mechanistic explanation) that Bicoid cannot be an activator for \emph{eve} at high abundances, because this would mean that \emph{eve} should be expressed at the anterior regions of the embryo, which contradicts the experimental observations \cite{stanojevic_1991,small_1991} (assuming that expression equals occupancy).

\section{Discussion}
Gene regulation plays a significant role in cellular response to developmental, physiological or environmental signals. To better understand these processes, there is a need to move from genetic interaction models of gene regulation to more fine-grained models that include the regulatory sequence \cite{jaeger_2012}. In this manuscript, we proposed an analytical model that is able to compute genome-wide binding profiles of TFs, as opposed to more detailed computational models for the statistical thermodynamics framework that are limited to smaller DNA loci. Our model recapitulates the main driving forces in determining the genome-wide occupancy of a TF, namely: \one the PWM, \two DNA accessibility, \three the number of TF molecules that are bound to the DNA and \four the specificity of the TF (how well it discriminates between ``good'' and ``bad'' sequences) through the $\lambda$ factor \cite{berg_1987}. Frequently, we have data for the first two and we aim to determine the last two by fitting the profile predicted by the model to the profile generated experimentally (through ChIP-seq) \cite{roider_2007,he_2009,kaplan_2011,simicevic_2013,cheng_2013}.    

\subsection{Abundance of bound TF}

Previous studies quantified the accuracy of predictions by determining the set of parameters (usually the number of TF molecules that are bound to the genome) that maximises the correlation between the computationally and the experimentally generated profiles \cite{simicevic_2013,cheng_2013}. Here, we considered the ChIP-seq dataset for five TFs in the \dmel\ embryo during early development and computed the Pearson correlation coefficient between the analytical and the experimental profiles. In addition, we also computed the mean squared error between the computational and the ChIP-seq profiles. Our results confirm that DNA accessibility improves the predictions of the model \cite{kaplan_2011,simicevic_2013,cheng_2013} and show that the set of parameters that maximise the correlation also leads to high mean squared errors, while the set of parameters that minimises the mean squared errors leads to only a small decrease in the correlation (compared to the maximum). Furthermore, it seems that the correlation is less sensitive to changes in the TF abundance, but highly sensitive for changes in $\lambda$; e.g.  the orange regions in Figure \ref{fig:heatmapChIPseq_BCD_CAD_AccRegionsGroup070}($E,F$) are stretched horizontally. In contrast,  the mean squared error is highly sensitive to changes in the number of TF molecules, but less sensitive to changes in $\lambda$; e.g.  the blue regions in Figure \ref{fig:heatmapChIPseq_BCD_CAD_AccRegionsGroup070}($E,F$) are stretched vertically. Together, this suggests that correlation could be used to estimate the $\lambda$ factor and mean squared error to infer the amount of bound TF. Thus, our results indicate that when aiming to maximise the correlation, previous studies \cite{roider_2007,he_2009,kaplan_2011,simicevic_2013,cheng_2013} potentially overestimated the number of TF molecules that are bound to the DNA.

For example, some work suggests the of the number of Bicoid molecules in the early fly embryo to be around  $1.5\times 10^8\ molecules$ \cite{drocco_2011}. Bicoid displays a gradient along the anterior-posterior axis of the embryo with most of the TF located in the anterior pole. Assuming that the ChIP-seq signal mainly comes from the nuclei with high abundance, that  $\approx30\%$ of the nuclei display high Bicoid abundance and that there are $6000$ nuclei in the blastoderm embryo \cite{grimm_2010}, one can compute the average number of Bicoid molecules per nucleus to be $\approx 80000$. Some of the molecules will be localized to the nucleus while others will diffuse in the cytoplasm. Gregor \emph{et al.} \cite{gregor_2007_gradient}  estimated that only $40\%$ of Bicoid is nuclear, which means that the nuclear abundance of Bicoid is $\approx 30000\ molecules$. In a subsequent study, the same group as in \cite{drocco_2011} proposed a slightly lower abundance of Bicoid in the \dmel\ embryo, namely $4.5\times 10^7\ molecules$ \cite{drocco_2012}. Following the same logic, we computed that there are $10000\ molecules$ of Bicoid per nucleus and this indicates that the Bicoid nuclear abundance can be estimated to be between $10000-30000\ molecules$. In accordance with these estimates, Abu-Arish \emph{et al.} \cite{abuarish_2010} estimated the Bicoid nuclear abundance as $140\ nM$, which is equivalent to $12 000\ molecules$ per nucleus when using the estimate of nuclear volume from \cite{gregor_2007}.  There is a significant difference between the number of molecules that our model predicts ($\approx 1000-5000$ for all cases when DNA accessibility was included) and the number o molecules estimated in these experimental studies  ($10000-30000$). However, our model estimates the number of molecules that are actually bound to the DNA, while other previous studies of \cite{abuarish_2010,drocco_2011,drocco_2012} are based on the entire nuclear abundance of Bicoid.

Furthermore, TFs can bind specifically to high affinity site, but also non-specifically anywhere on the genome where they potentially perform one-dimensional sliding on the DNA \cite{riggs_1970a,berg_1981,vukojevic_2010,leith_2012,hammar_2012,chen_2014}. Nevertheless, experimental studies have shown that ChIP only recovers specific binding of the TF to the DNA \cite{carr_1999,toth_2000,rhee_2011,poorey_2013}. ChIP is a population average measurement, which means that what it reports is the proportion of cells in which a specific locus was bound. Due to their high affinity, specific sites will be occupied in the majority of the cells (nuclei). In contrast, individual non-specific sites will be occupied in a few cells, because there are many more similar low affinity sites in the genome \cite{mueller_2013}. Thus, ChIP data describes binding at the specific sites and this means that, when we estimate the number of bound TF molecules, in fact we estimate the number of specifically bound TF molecules.

In Table \ref{tab:FractionBoundMolecules} and Figure \ref{fig:FractionBoundMolecules} in the \sm, we summarised the results of a series of studies (note that this is not an exhaustive list) of the estimated percentage of specifically bound TFs \cite{phair_2004_method,mueller_2008,speil_2011,kloster_landsberg_2012,mazza_2012,gebhardt_2013,chen_2014,morisaki_2014}. These studies were performed in different mammalian cell lines (HeLa, 3134, H1299, MCF-7, U87, ES, NIH 3T3), using different techniques (Fluorescence Recovery After Photobleaching -FRAP, Fluorescence Correlation Spectroscopy - FCS, Single Molecule Tracking - SMT and Reflected Light-Sheet Microscope - RLSM) and in different conditions. The results indicate that the percentage of specifically bound TF ranges between $2.5\%$ and $99.7\%$ with a median of $\approx 20\%$. 

In the case of Bicoid, if only $20\%$ of the TF is specifically bound and there are between $10000-30000$ molecules in the nucleus, then the amount of specifically bound TF is between $2000-6000\ molecules$. These values are similar to the values that we calculate ($\approx 1000-5000\ molecules$) assuming different models for Bicoid (binary and continuous DNA accessibility data, including weak binding sites or using a different PWM). Furthermore, in \cite{zamparo_2009}, the authors proposed a lower limit for the nuclear abundance of the five TFs by analysing the  FlyEx database \cite{pisarev_2009}. Their values are much lower than what we estimate in the case of Bicoid. This can be explained as the authors in \cite{zamparo_2009} removed the highest $10\%$ measurements  when computing the averages. However, the nuclear abundances proposed in \cite{zamparo_2009} can be used to estimate the abundance in the nucleus for Caudal, Giant, Hunchback and Kruppel relative to Bicoid. We used this strategy to estimate the nuclear abundance of these four TFs (measured in number of molecules) and then we estimated the percentage of specifically bound TFs (based on the estimations of our analytical model and the nuclear abundance of TFs); see Table \ref{tab:TFabundanceEstimates} in the \sm. Our method supports an extensive body of literature that only a relatively small percentage ($30\%$ or less) of the molecules of TFs are bound specifically to the genome. 

Previous studies suggested that the TF abundance in eukaryotic systems can be high; e.g. \cite{biggin_2011} estimated that there are between $10^4$ and $3\times 10^5\ molecules$ per TF, while the same author later estimated that the median of TFs abundances in a mouse NIH 3T3 cell line is $7.1\times 10^4\ molecules$ \cite{li_2014_proteome}. A different group estimated that there are between  $250$ and $3\times 10^5\ molecules$ for each TF in mouse 3T3-L1 cells. Assuming that less than $30\%$ of these TF molecules are bound specifically to the genome, we estimate that the median number of TF molecules that are specifically bound is less than  $21,000 \ molecules$.

Since only specifically bound TFs seem to influence the transcription process \cite{morisaki_2014}, it is more important to know the exact amount of specifically bound TF, rather than the entire concentration in the nucleus \cite{foat_2006,roider_2007,zhao_2009,segal_2008}. Thus, the range of parameters found by our study will have a higher impact for further studies that model these biological systems, compared to other work that estimates the nuclear concentration of TFs. It worthwhile mentioning that the estimate for proportion of non-specifically bound TF is in the same range with the proportion of specifically bound TFs \cite{mazza_2012,mueller_2013}, which suggests that the amount of TF bound to the genome would be in similar ranges ($2000-40000\ molecules$). 

It is worthwhile noting that the accuracy of our method to estimate TF abundance is limited by the ChIP methodology to fully recover the quantitative aspects of TF binding. For example, the \emph{in silico} ChIP-seq profiles of lacI in \cite{zabet_2013_occupancy} seem to be similar for lacI abundances between $1$ and $1000\ molecules$, which suggests that our method will not be able to correctly estimate abundances lower than $1000\ molecules$. Thus, our method will perform best for cases where differences in TF abundance leads to strong differences in the ChIP profiles.

\subsection{The specificity of TFs}

Our model also predicted that TFs can display higher or lower specificity beyond the information content of the binding motif, through the coefficient that modulates the discrimination energy between strong and week binding sites ($\lambda$).  Our results show that the difference between the binding energy of strong and weak sites is high for Bicoid and Caudal and low for Giant, Hunchback and Kruppel. It is worthwhile noting that considering only the information content, a na\"{i}ve assumption would be that Hunchback and Kruppel have the highest specificity, but, when including the $\lambda$ scaling factor, these two TFs display the lowest specificity. 

In this context, one might ask if TF with low $\lambda$ cannot distinguish well between different DNA words, where does the high information content of their motif come from?  One hypothesis is that the methods used to determine TF specificity can potentially display technical biases. In fact, two different \emph{in vitro} methods, SELEX \cite{kaplan_2011} and bacterial one hybrid \cite{portales-casamar_2010}, lead to different PWM motifs for three of the TFs (Giant, Hunchback and Kruppel). When the motifs display higher information content (Giant in JASPAR and Hunchback and Kruppel in BDTNP), our method estimates a higher $\lambda$, which leads to lower specificities of the TFs. When the TFs display lower information content (Giant in BDTNP and Hunchback and Kruppel in JASPAR), our method estimates lower values for $\lambda$, which is consistent with the intuition that low information contents of the motifs will lead to low specificities. For the TFs that display similar PWM motifs in both sources (Bicoid and Caudal), we always estimate similar values for $\lambda$, which indicates that the specificity of the two TFs is given by the information content of the motifs. 

Nevertheless, Figures \ref{fig:profileAllPositivesHB} and \ref{fig:profileAllPositivesKR} in the \sm\ show that the ChIP-seq profiles of Hunchback and Kruppel display some sharp peaks, which suggest that these two TFs display higher specificity than predicted by our approach. This contradicts our findings and one explanation for the few narrow ChIP-seq peaks is that these two TFs bind cooperatively to the genome. In this scenario, in the few narrow peaks for Hunchback and Kruppel, these TFs co-localise with co-factor(s) and previous studies identified that this is the case for both TFs; e.g. \cite{cheng_2013}. This means that, by using our model, one could potentially underestimate the number of peaks in the binding profile.   

Finally, we obtained the highest correlation and lowest mean squared error between the ChIP-seq profile and our estimate for the TFs that display the highest specificity (Bicoid and Caudal). Thus, our model performs best in the case of TFs that can discriminate better between strong and weak binding sites. Note that we observed the same result also when investigating the binding profiles genome-wide; see Figure \ref{fig:genomeWideQuality}. This reduction in the accuracy of our model for regions with weak binding is not a direct consequence of our model being an analytical approximation of the full statistical thermodynamics model, because even exact solutions to the full model display reduced accuracy for regions where TFs do not bind strongly; e.g. \cite{kaplan_2011}. It is worthwhile noting that regions with weaker binding seem to also have a lower chance of driving expression \cite{fisher_2012,morisaki_2014} and might potentially be experimental artefacts; e.g. \cite{teytelman_2013}.  

\subsection{Additional factors that affect TF binding profiles}

Our analytical model can recapitulate observed genome-wide binding profiles  (e.g. for four of the TFs, the correlation is higher than $0.65$) especially at the loci with strong binding, but there are several loci, where our model under/overestimates the ChIP-seq profile; see Figure \ref{fig:bindingProfilesEve} and Figures  \ref{fig:profileAllPositivesBCD} - \ref{fig:profileAllPositivesKR} in the \sm. In this contribution, we systematically investigated potential causes for these differences. 

First, there is an inconsistency in the experimental data in the sense there are peaks in the ChIP-seq profile that are located in DNA inaccessible areas, e.g. there are peaks in the Bicoid ChIP-seq profile at \emph{run}, \emph{slp}, \emph{eve}, \emph{tll}, \emph{gt}, \emph{oc} loci that overlap with DNA that is marked as inaccessible; see Figure  \ref{fig:profileAllPositivesBCD} in the \sm. This indicates that either or both the DNA accessibility or the ChIP-seq data display some technical biases, e.g. \cite{teytelman_2013,meyer_2014}, and, in these cases, the analytical model assumes that the DNA accessibility data is accurate and predicts that there is no binding in DNA inaccessible areas. One solution is to use continuous data for DNA accessibility, where different areas display different levels of accessibility. When using continuous values for DNA accessibility data, we did not observe any improvements of our model's predictions. Nevertheless, we still observed ChIP-seq peaks for all five TFs that were overlapping with regions with reduce or no accessibility, thus, indicating the one or both data sets (ChIP-seq or DNase I) contain experimental biases; e.g. \cite{teytelman_2013,meyer_2014,sung_2014}. 

Alternatively, the underestimation of the peak height may be caused by PWM choice. Using motifs from the JASPAR database \cite{portales-casamar_2010}, we observed lower values for the correlation and higher values for the mean squared error compared to the case of using the PWMs from BDTNP \cite{kaplan_2011}. In addition, when we used different PWMs (from the JASPAR database) we found different estimates for the number of molecues that best explain the ChIP-seq data, but these values were within the same range ($2000-10000\ molecules$). This suggests that our estimates for the amount of bound TFs are not the exact values, but rather an estimate of the order of magnitude for the number of molecules that are bound specifically to the DNA. 

In this manuscript, we aim to deconvolute the contributions of different factors to the binding profiles of TFs. One of the most important factors that contribute to the binding profiles is the binding energy between the TF and the DNA words. Previous work \cite{berg_1987,stormo_2000,benos_2002} showed that the binding energy between a TF and the DNA is proportional to the PWM score and, thus, the binding energy can be approximated by a scaled PWM score ($E_{i}=w_i/\lambda$). In order to avoid introducing the effects of ``other factors'' in the binding energy estimation, one should consider that the PWM is representing only the binding frequency between the TF and the DNA words independent of other factors. Inferring the PWM motif from the ChIP-seq peaks would assume that DNA accessibility, TF cooperativity, crowding of molecules on the DNA, histone marks and others will affect the PWM. While by using a PWM derived from ChIP-seq data we might lead to better predictions of the binding profiles, we would not be able to distinguish between the real sources that drive the genomic occupancy and their relative contribution.  This is the rationale for testing PWMs derived from BDTNP \cite{kaplan_2011} and JASPAR\cite{portales-casamar_2010} and not investigating the case of the PWMs derived from ChIP-seq data.

Furthermore, we do not consider every aspect related to the binding of TFs to the genome, but binding energy (PWM scores and $\lambda$), TF abundance and DNA accessibility are sufficient  to explain most of the characteristics of the binding profiles for the TFs analysed in this study (Bicoid, Caudal, Giant, Hunchback and Kruppel). One aspect that our model does not include is cooperative binding to the DNA. Previous studies have shown that TF cooperativity can significantly impact the binding of TFs \cite{ackers_1982,buchler_2003,bintu_2005_model,bintu_2005b,chu_2009,sadka_2009} and that cooperativity can explain TF genomic binding \cite{cheng_2013}. However, it was found that the five TFs considered in this study display negligible or no cooperative interactions between them \cite{kaplan_2011,cheng_2013}, but these TFs seem to display cooperative interactions with other TFs \cite{cheng_2013,xu_2014,rhee_2014}. For example, the Bicoid binding profile seems to be significantly influenced by the maternally contributed factor Zelda, where the presence of Zelda increases the binding of Bicoid at the majority of loci and decreases it at a small set of loci \cite{xu_2014}. Modelling these binding profiles assuming cooperativity with other TFs could potentially improve our model predictions, but this requires further systematic investigation and  will be left to future research.

Our model does not implement competitive binding directly. Other models, e.g. \cite{kaplan_2011}, allow to model the competition between TFs explicitly, but the fact that we obtained similar correlation between the predicted profile and the ChIP-seq data indicates that there is negligible binding competition between the five TFs analysed in this study as also shown in \cite{segal_2008}.

Finally, we would like to point out that while our model was applied to a ChIP-seq dataset, it could also be used to investigate ChIP-chip and ChIP-exo \cite{rhee_2011} datasets as long as the appropriate length distribution of DNA fragments is included in the model (see \mm).

\section{Availability}
The R scripts used to perform the analysis can be downloaded from \url{https://github.com/nrzabet/ChIPseqProfile}

\section{Funding}
This work was supported by the Medical Research Council [G1002110]. BA is a Royal Society University Research Fellow.

\section{Acknowledgements}
We would like to thank Robert Foy, Xiaoyan Ma and Robert Stojnic for useful discussions and comments on the manuscript.

\appendix
    \begin{center}
      {\bf APPENDIX}
    \end{center}

\renewcommand{\thefigure}{A\arabic{figure}}
\renewcommand{\thetable}{A\arabic{table}}
\renewcommand{\thesection}{A\arabic{section}}
\renewcommand{\theequation}{A\arabic{equation}}
\setcounter{figure}{0}
\setcounter{equation}{0}
\setcounter{table}{0}
\setcounter{section}{0}

\section{Derivation of the analytical model} \label{sec:appendixAnalyticalModel}

Here, we derive an analytical solution for the statistical thermodynamics model, which makes it feasible to investigate genome-wide binding profiles of TFs, rather than being restricted to much shorter loci.  

\subsection{Binding energy between the TF and the DNA}
The amount of time a TF is bound to a specific site can be computed using the binding energy as \citep{gerland_2002}: 
\begin{equation}
\tau_{j} = \tau_{0}\exp\left(-E_{j}\right)= \tau_{0}\exp\left(\frac{1}{\lambda}w_{j}\right)\label{eq:tauPWM}
\end{equation}
where $j$ is the genomic coordinate of the site, $\tau_{0}$ is a scaling factor associated with each TF species (which takes into account an absolute strength between the TF and the DNA) and $E_{j}$ is the binding energy of site $j$. Note that the binding energy can be approximated by the PWM score $E_{j}=\frac{1}{\lambda}w_{j}$, where  $1/\lambda$ is a scaling factor which quantifies the penalty for differences from the preferred site \citep{berg_1987,roider_2007} and $w_{j}$ the PWM score at position $j$. 

We want to investigate the accuracy of our analytical model and, thus, we compare its results with the results from a detailed statistical thermodynamics model, as computed with GRiP \citep{zabet_2012_model,zabet_2012_grip,zabet_2012_subsystem,zabet_2012_review}; see section \ref{sec:GRiPequlibriumProfile}. As previously shown in \citep{zabet_2013_occupancy}, for highly abundant TFs, the estimation of the occupancy based on the PWM alone is not sufficient to accurately predict the profile found in simulations (and also by a simple statistical thermodynamics framework); see Figure \ref{fig:estimate070SDOlacI}. Note that in Figure \ref{fig:estimate070SDOlacI} we do not simulate the 1D random walk on the DNA, while in Figure 3 in \citep{zabet_2013_occupancy} the 1D random walk on the DNA is considered. 

\subsection{The number of bound molecules to the DNA}
We used our analytical solution for the statistical thermodynamics framework to include TF abundance in the model. Given that a TF has $N$ molecules bound to the DNA, the statistical weight that a site is unoccupied by a molecule can be written as \citep{ackers_1982,bintu_2005_model,bintu_2005b,chu_2009} 
\begin{equation}
Z\left(N\right) = \overbrace{\frac{\left(L\cdot n\right)!}{N!\left(L\cdot n-N\right)!}}^{\textrm{number of arrangements}} \times \underbrace{\exp{\left(-E^{ns}\right)}}_{\textrm{Boltzmann weight}} \label{eq:abundanceZ}
\end{equation}
where $E_x^{ns}$ is the binding free energy when the TF is non-specifically bound to the DNA (i.e. not to the target site), $L$ represents the total number of available sites (which can be approximated by the length of the DNA segment) and $n$ is the ploidy level (the number of copies of the genome, e.g.  for diploid genomes $n=2$). The total statistical weight is given by the sum of the statistical weight when the site is unoccupied and the statistical weight when the site is occupied. 
\begin{equation}
Z_{total}\left(N\right)  = \overbrace{Z\left(N\right)}^{\textrm{the site is empty}} +\overbrace{Z\left(N-1\right) \exp{\left(-E^{s}\right)}}^{\textrm{the site is occupied}} \label{eq:abundanceZoccupied}
\end{equation}
where $E^{s}$ binding free energy at the target site (where the TF is bound specific). Altogether the probability that the target site is occupied is given by the ratio between the statistical weight of the site being occupied over the total statistical weight.
\begin{equation}
P^{bound} = \frac{Z\left(N-1\right) \exp{\left(-E^{s}\right)}}{Z_{total}\left(N\right)} = \frac{1}{1+\frac{Z\left(N\right)} {Z\left(N-1\right) \exp{\left(-E^{s}\right)}}}
\end{equation}
This can be reduced to \citep{bintu_2005_model} :
\begin{equation}
P^{bound} \left(E_{j},N\right)= \frac{1}{1+\frac{L\cdot n-N+1}{N}\exp\left[\left(E^{s}-E^{ns}\right)\right]}= \frac{1}{1+\frac{\left(L\cdot n-N+1\right)\exp\left(-E^{ns}\right)}{N}\exp\left(E^{s}\right)}
\end{equation}

Given the size of the genome and the range of TF abundances reported in the literature, we can assume that the number of available sites is much larger than the number of bound molecules ($L\cdot n\gg N$) and, thus, $\left(L\cdot n-N+1\right)\approx L\cdot n$ \citep{bintu_2005_model}. In our model, we do not want to be constrained to only two binding energy levels (the binding energy for non-specific sites and for specific sites), but rather consider the full binding energy spectrum. If $E^{ns}$ represents the binding energy at other sites (not the current one), then $L\cdot n\cdot\exp\left(-E^{ns}\right)$ represents the average waiting time on the genome and can be approximated by $L\cdot n\cdot\left\langle\exp\left(-E_{i}\right)\right\rangle_{i}=L\cdot n\cdot\left\langle\exp\left(\frac{1}{\lambda}w_i\right)\right\rangle_i$ \citep{simicevic_2013}. This leads to the following probability of site $j$ being occupied by a TF molecule:
\begin{equation}
P^{bound}_j\left(\lambda, w, N\right)= \frac{1}{1+\frac{1}{N}L\cdot n\cdot\left\langle\exp\left(\frac{1}{\lambda}w_i\right)\right\rangle_i\exp\left(-\frac{1}{\lambda}w_j\right)} = \frac{N\exp\left(\frac{1}{\lambda}w_j\right)}{N\exp\left(\frac{1}{\lambda}w_j\right) + L\cdot n\cdot\left\langle\exp\left(\frac{1}{\lambda}w_i\right)\right\rangle_i}\label{eq:pboundAbundance}
\end{equation}

Figure \ref{fig:estimate070SDOlacIcn} shows that, by including the TF abundance in our model, the estimate of the occupancy obtained in the simulations improves significantly; compare Figure \ref{fig:estimate070SDOlacI} to  Figure \ref{fig:estimate070SDOlacIcn}. 

One approximation of our model is that we assume the mean value instead of the full distribution of PWM scores \citep{sheinman_2012}. However, we found that, by using this approximation (the mean value of the PWM scores) \citep{simicevic_2013}, the difference between the analytical and numerical results are negligible; see Figure \ref{fig:estimate070SDOlacIcn}. This approximation also holds in the case of TFs with lower specificity as we show in the \emph{Results} section of the main manuscript. 

\subsection{Including DNA accessibility data}
To apply our model to eukaryotic systems we need to include DNA accessibility as a parameter. The probability that a site $j$ is bound, in the case of $N$ bound molecules to the DNA and in presence of DNA accessibility data, can be rewritten as
\begin{equation}
P^{bound}_j\left(\lambda, w, N, a\right)=\frac{N\cdot a_j\cdot \exp\left(\frac{1}{\lambda}w_j\right)}{N\cdot a_j\cdot \exp\left(\frac{1}{\lambda}w_j\right) + L\cdot n\cdot\left\langle a_i \exp\left(\frac{1}{\lambda}w_i\right)\right\rangle_i}\label{eq:tauPWMabundanceAccA}
\end{equation}
where $a_j$ the probability that site $j$ on the genome is in accessible chromatin.

\subsection{Analysing ChIP-seq data with the analytical model}

To test the accuracy of our analytical model, we compare its estimated profile to ChIP-seq data and, thus, we need to convert the prediction of the analytical model to a profile of genomic occupancy. Given, $C_j$ as the experimentally determined ChIP-seq signal at position $j$ on the  genome, the equivalent occupancy based on the analytical estimate is \citep{simicevic_2013}  
\begin{equation}
A\left(j,\lambda, w, N\right)  = B + \left(M-B\right)\times P^{bound}_j \label{eq:PboundToChIPseq}
\end{equation}
where $B=\langle C\rangle$ is the background and $M=\max(C)$ is the maximum of the ChIP-seq signal.

\subsection{Modelling DNA accessibility data} \label{sec:ModellingDNAaccessibilityData}
Following the approach in \citep{kaplan_2011}, we modelled the probability of a site being accessible as
\begin{equation}
a_j = \frac{1}{1+\exp{\left(-\beta\cdot DD_j +\alpha \right)}}\label{eq:paccessible}
\end{equation}
where $DD_j$ is the DNase I read density and  $\alpha$ and $\beta$ are scaling parameters, which were estimated to be $\alpha=6.008$ and $\beta= 0.207$ in \citep{kaplan_2011}.  

\section{Computing the equilibrium binding profile with GRiP} \label{sec:GRiPequlibriumProfile}
To obtain the profile generated with a statistical thermodynamics approach, we performed stochastic simulations with our previously published computational tool GRiP \citep{zabet_2012_model,zabet_2012_grip,zabet_2012_subsystem} that simulates the facilitated diffusion of TFs \citep{hammar_2012}. In GRiP, all the molecules in the system are represented explicitly, along with steric hindrance effects. Three-dimensional diffusion is simulated using the Master Equation, an approximation which was previously proven to provide accurate results \citep{zon_2006}. The simulations were restricted to binding and unbinding to/from the DNA (we removed the one-dimensional random walk on the DNA from the simulations), which means that the results are equivalent to the ones predicted by the classical statistical thermodynamics framework. To compare the profiles generated by simulations ($S$) and the ones generated by the analytical approximation ($A$), we normalise all profiles to the highest values.

\begin{figure}[htp]
  \begin{center}
  		\includegraphics[width=0.7\textwidth]{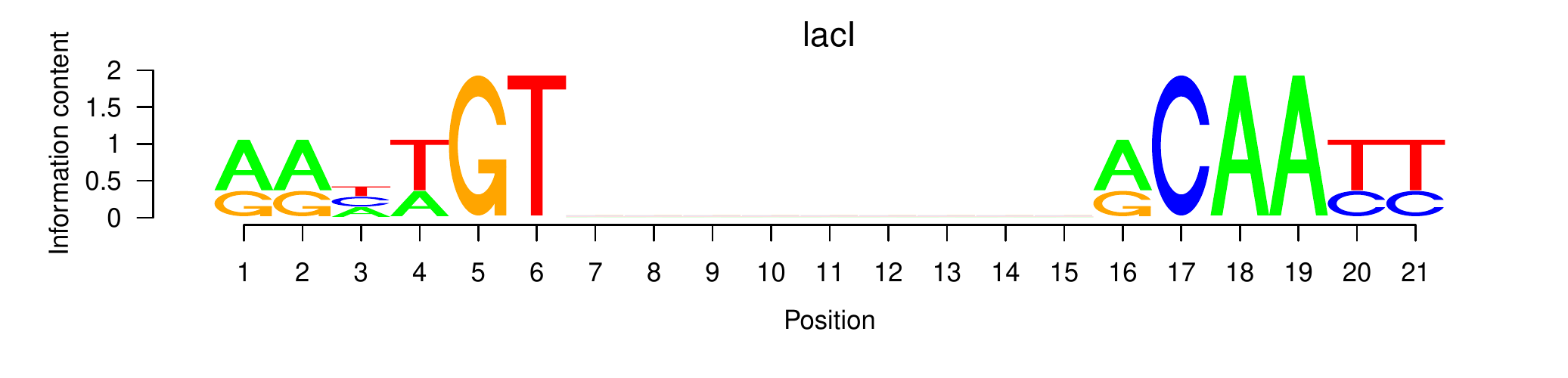}
  \end{center}
\caption{\emph{Sequence logo for lacI}. \citep{zabet_2012_subsystem}}\label{fig:lacIPWMmotif}
\end{figure}

\begin{figure}
\centering
\includegraphics[width=\textwidth]{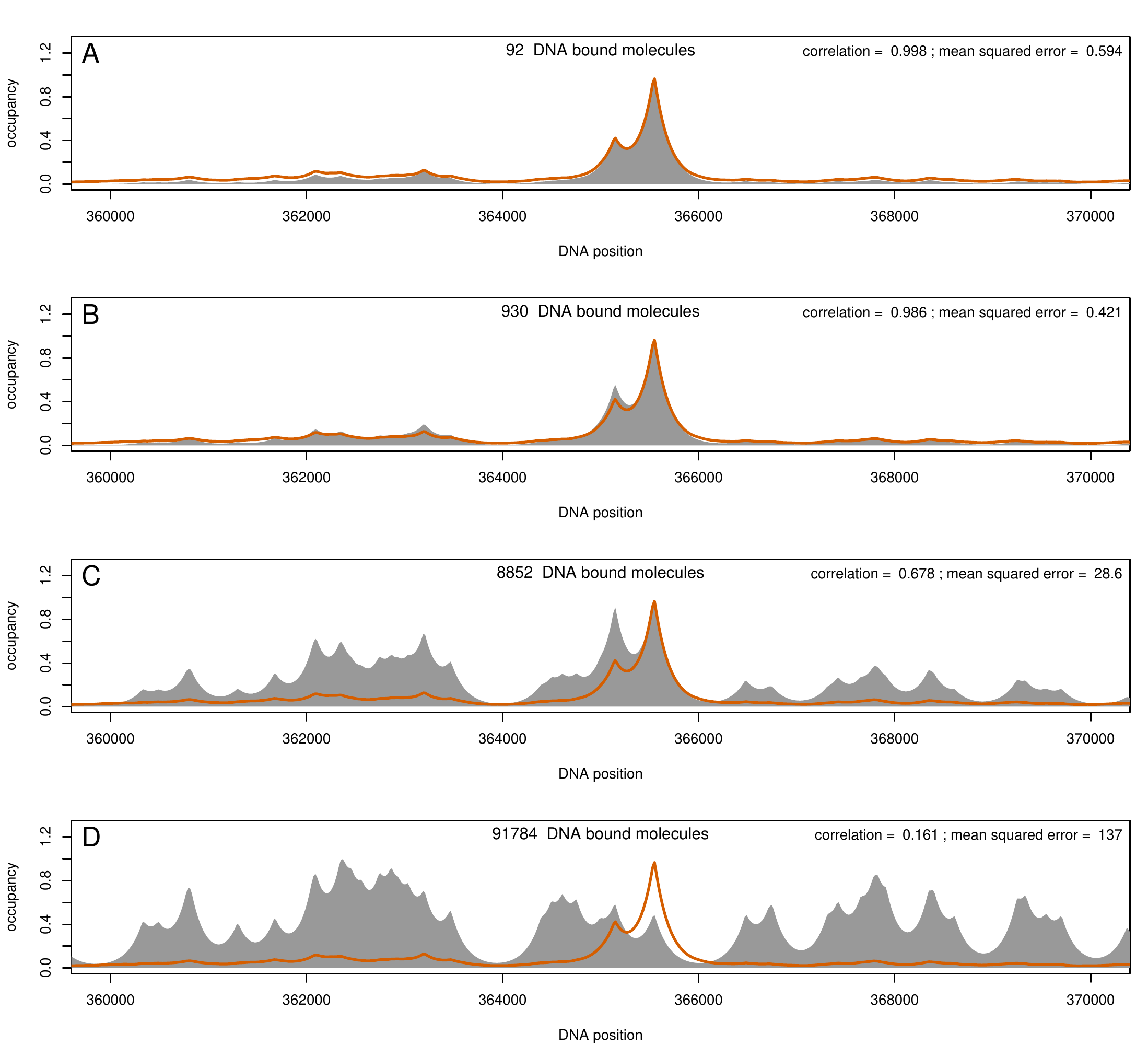}
\caption{\emph{Estimating the occupancy in the simulations (or statistical thermodynamic model) based on the PWM score alone}. We considered the case of the lac repressor and $20\ Kbp$ of DNA, which contain the $O_1$ site. In each chart, the filled region is the occupancy predicted base on the simulations ($S$), and the solid line is the occupancy estimated by the analytical solution; see equation \ref{eq:tauPWM}. We considered various lacI abundances: \onel\ $100$ lacI molecules (leading to $92$ lacI molecules being bound to the DNA), \twol\ $1000$ lacI molecules (i.e. $930$ lacI molecules), \threel\ $10000$ lacI molecules (i.e. $8852$ lacI molecules) and \fourl\ $100000$ lacI molecules (i.e. $91784$ lacI molecules). For each set of parameters, we consider $X=50$ independent simulations. We considered only the sites that have the binding energy of at least $70\%$ of the highest value (the strongest $81$ sites). We converted the single nucleotide resolution into expected occupancy profiles as proposed in \citep{kaplan_2011}, with a mean of $150\ bp$ and a standard deviation of $150\ bp$; see also \citep{zabet_2013_occupancy}. The inset in each panel indicates the Pearson coefficient of correlation and the mean squared error between the occupancy computed by the analytical model and the occupancy computed from the simulations, which is equivalent to the occupancy generated by the full statistical thermodynamics model. The x-axis represents the genomic location. \label{fig:estimate070SDOlacI}}
\end{figure}

\begin{figure}
\centering
\includegraphics[width=\textwidth]{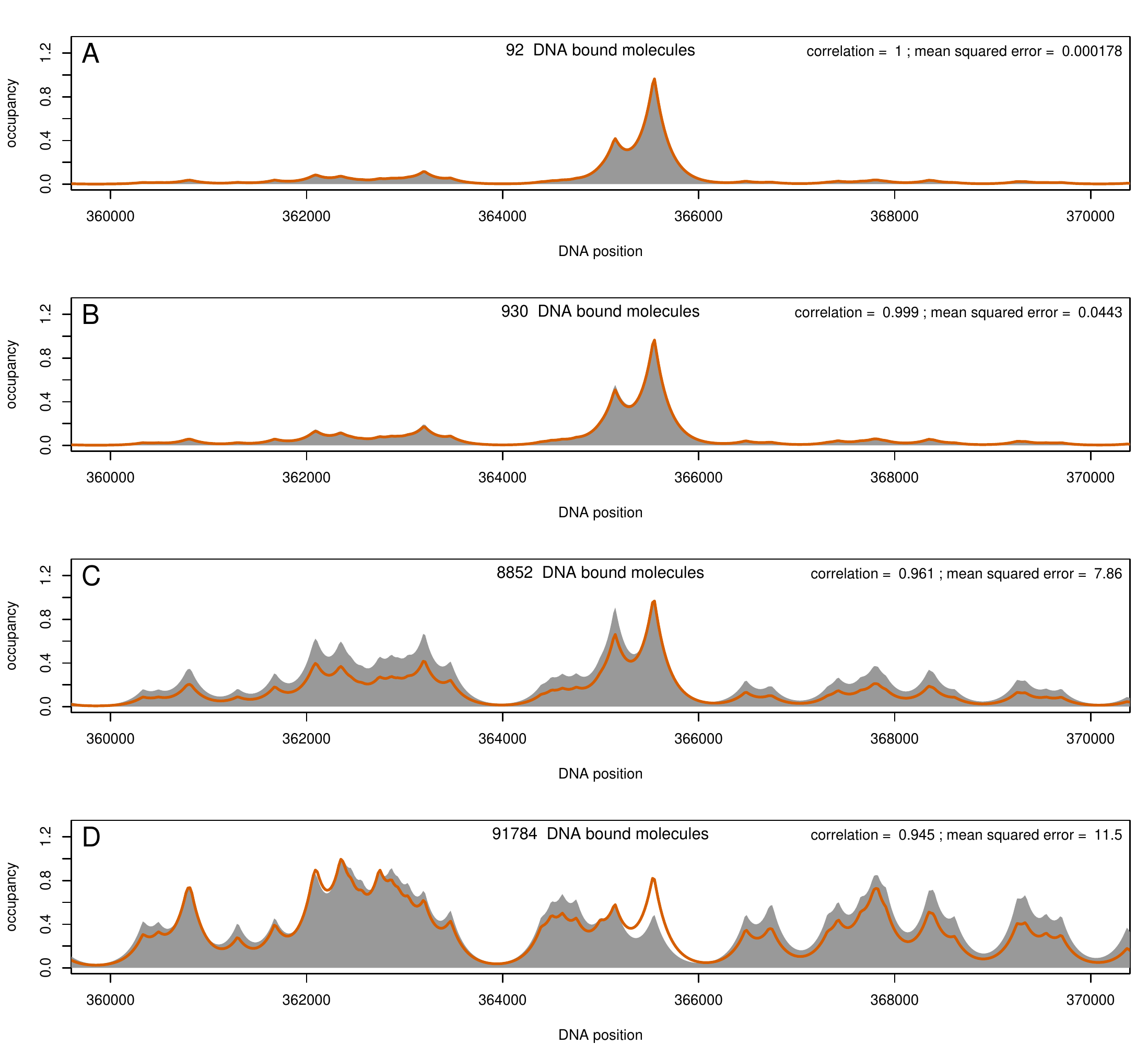}
\caption{\emph{Estimating the occupancy in the simulations (or statistical thermodynamic model) based on the PWM and TF abundance}. This is the same as Figure \ref{fig:estimate070SDOlacI}, except that equation \eqref{eq:pboundAbundance} was used to compute the occupancy, solid line. \label{fig:estimate070SDOlacIcn}}
\end{figure}

When comparing the analytical model to the results computed by GRiP, we considered the case of lac repressor (lacI), a well studied bacterial TF and $20\ Kbp$ of \ecoli\ DNA ( $355000..375000$ locus in the \emph{E.coli} K-12 genome). For lacI, we considered a motif constructed based on the three high-affinity sites of lacI in \citep{zabet_2012_subsystem}; see Figure \ref{fig:lacIPWMmotif}. The default parameters for the simulations with GRiP are listed in Table \ref{tab:modelTFparams}. Furthermore, we considered four cases with respect to lacI abundance and the set of parameters selected for each case ensured that each site is visited on average approximately $2000$ times; see Table \ref{tab:modelTFparamsCases}. We performed $50$ independent simulations for each set of parameters and computed the occupancy as time average in each simulation. Overall, this resulted in each site being visited on average $10^{5}$ times. 

Finally, to generate the ChIP profiles we used the method described by \citep{kaplan_2011}, where a mean segment length of $150\ bp$, a standard deviation of $150\ bp$ and no smoothing was used; the R implementation of this method is described in \citep{zabet_2013_occupancy}. The difference between the simulations and our analytical formula was quantified by: \one the Pearson coefficient of correlation and \two the mean squared error. 

\begin{table}
\centering
\begin{tabular}{| l | r | l |}
\hline
\textbf{parameter} & \textbf{lacI}  &  \textbf{notation}\\ \hline
copy number &  see Table \ref{tab:modelTFparamsCases} &  $n_{x}$\\ \hline
motif sequence &  see Figure \ref{fig:lacIPWMmotif} & \\ \hline
energetic penalty for mismatch &  $1K_BT$ & $\varepsilon^{*}_{x}$\\ \hline
nucleotides covered on left &  $0\ bp$ & $\delta^{\textrm{left}}_x$\\ \hline
nucleotides covered on right &  $0\ bp$ & $\delta^{\textrm{right}}_x$\\ \hline
association rate to the DNA &  see Table \ref{tab:modelTFparamsCases}  & $k^{a}_{x}$\\ \hline
unbinding probability &  $1.0$ & $P^{\textrm{unbind}}_{x}$\\ \hline
probability to slide left &  $0$ & $P^{\textrm{left}}_{x}$\\ \hline
probability to slide right &  $0$ & $P^{\textrm{right}}_{x}$\\ \hline
probability to dissociate completely when unbinding &  $1.0$ & $P^{\textrm{jump}}_{x}$\\ \hline
time bound at the target site &  $1.18E-6\ s$ & $\tau^{0}_{x}$\\ \hline
the size of a step to left &  $1\ bp$ & \\ \hline
the size of a step to right &  $1\ bp$ & \\ \hline
variance of repositioning distance after a hop &  $1\ bp$ & $\sigma^{2}_{\textrm{hop}}$\\ \hline
the distance over which a hop becomes a jump  &  $100\ bp$ & $d_{\textrm{jump}}$\\ 
\hline
\end{tabular}
\caption{\emph{TF species default parameters}}
\label{tab:modelTFparams}
\end{table}

\begin{table}
\centering
\begin{tabular}{| r | r | r | r |}
\hline
\textbf{lacI copy number} & \textbf{association rate to the DNA} & \textbf{simulation time} &  \textbf{bound molecules}\\ \hline
$100$ &  $2700$ &  $150\ s$ &  $92$\\ \hline
$1000$ &  $3100$ &  $15\ s$ &  $930$\\ \hline
$10000$ &  $4000$ &  $1.2\ s$ &  $8852$\\ \hline
$100000$ &  $20000$ &  $0.075\ s$ &  $91784$\\ \hline
\end{tabular}
\caption{\emph{The parameters for the six cases of lacI abundance}. The association rate was selected so that on average $90\%$ of the time the TF molecules are bound to the DNA.}
\label{tab:modelTFparamsCases}
\end{table}

\clearpage

\clearpage

\clearpage
\section{Data sets}

\begin{table}[h]
\centering
\begin{tabular}{| l | l | l | l |}
\hline
\textbf{Symbol} & \textbf{Gene name} & \textbf{Locus coordinates} &  \textbf{Length}\\ \hline
croc &  crocodile &  chr3L:21,461,001-21,477,000 &  16 Kb\\ \hline
cnc &  cap-n-collar &  chr3R:19,011,001-19,024,000 &  13 Kb\\ \hline
slp &  sloppy paired &  chr2L:3,820,001-3,840,000 &  20 Kb\\ \hline
kni &  knirps &  chr3L:20,683,260-20,695,259 &  12 Kb\\ \hline
hkb &  huckebein &  chr3R:169,001-181,000 &  12 Kb\\ \hline
D &  Dichaete &  chr3L:14,165,001-14,179,000 &  24 Kb\\ \hline
prd &  paired &  chr2L:120,77,501-12,095,500 &  18 Kb\\ \hline
H &  hairy &  chr3L:8,656,154-8,682,153 &  26 Kb\\ \hline
eve &  even skipped &  chr2R:5,860,693-5,876,692 &  16 Kb\\ \hline
cad &  caudal &  chr2L:20,767,501-20,786,500 &  19 Kb\\ \hline
oc &  ocelliless &  chrX:8,518,001-8,550,000 &  32 Kb\\ \hline
opa &  odd paired &  chr3R:670,001-696,000 &  26 Kb\\ \hline
ftz &  fushi tarazu &  chr3R:2,682,501-2,696,500 &  14 Kb\\ \hline
gt &  giant &  chrX:2,317,878-2,330,877 &  13 Kb\\ \hline
hb &  hunchback &  chr3R:4,513,501-4,531,500 &  18 Kb\\ \hline
Kr &  Kruppel &  chr2R:21,103,924-21,118,923 &  15 Kb\\ \hline
odd &  odd skipped &  chr2L:3,603,001-3,613,000 &  10 Kb\\ \hline
run &  runt &  chrX:20,548,001-20,570,000 &  22 Kb\\ \hline
fkh &  forkhead &  chr3R:24,396,001-24,420,000 &  24 Kb\\ \hline
tll &  tailless &  chr3R:26,672,001-26,684,000 &  12 Kb\\ \hline
os &  outstretched &  chrX:18,193,001-18,208,000 &  15 Kb\\ \hline
\end{tabular}
\caption{\emph{ Genes and coordinates for analysed sets loci}. These are the same \emph{loci} as considered in \citep{kaplan_2011}.}
\label{tab:kaplan2011loci}
\end{table}

\clearpage
\section{Analysis of the ChIP-seq profiles}
\subsection{Heatmaps for GT, HB and KR}

\begin{figure}
\centering
\includegraphics[width=\textwidth]{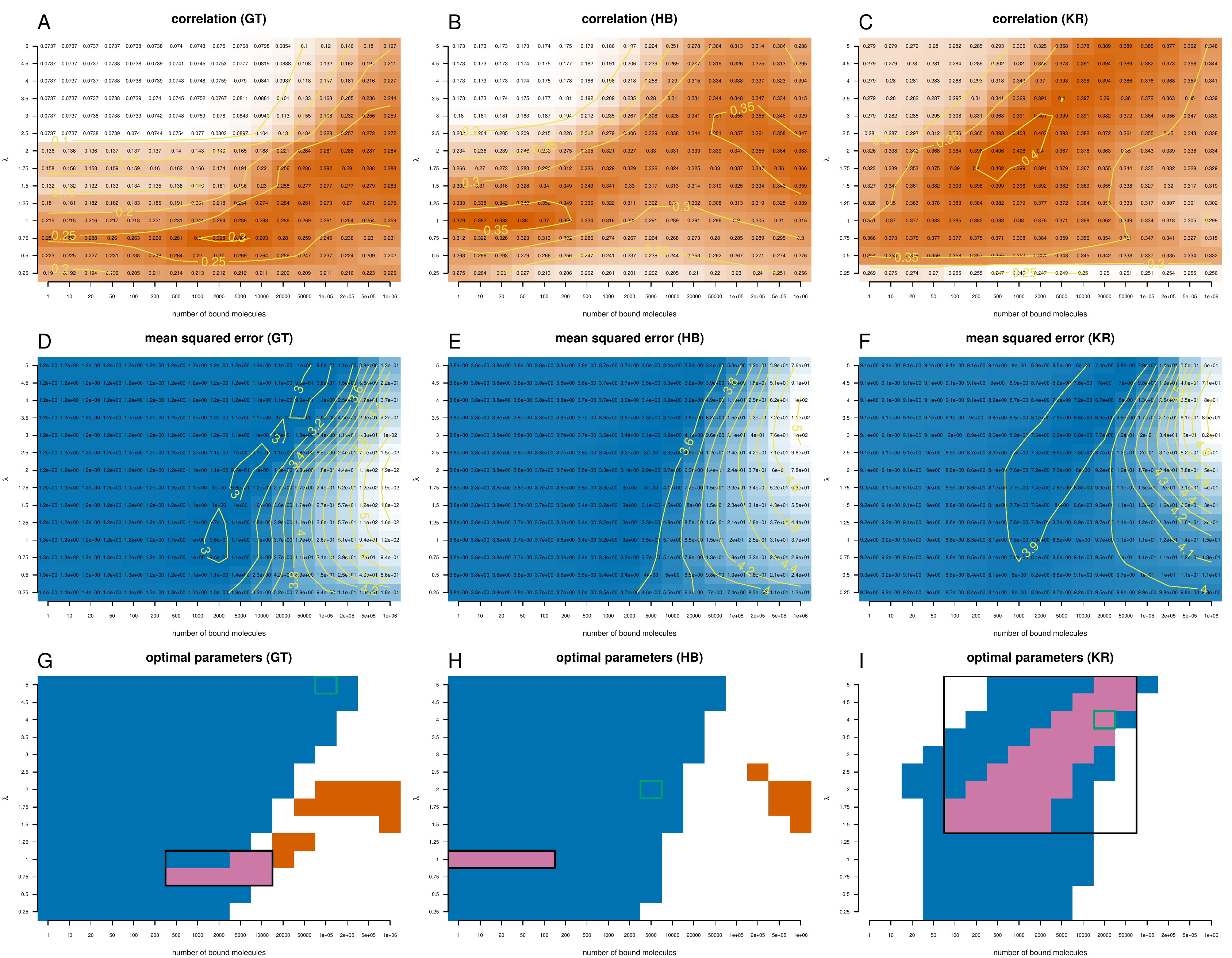}
\caption{
\emph{Quantifying the distances between Giant, Hunchback and Kruppel ChIP-seq profiles and the profiles derived with the analytical model}. We plotted heatmaps for the correlation ($A-C$) and mean squared error ($D-F$) between the analytical model and the ChIP-seq profile of Giant $(A,D)$, Hunchback $(B,E)$ and Kruppel $(C,F)$. We computed these values for different sets of parameters: $N\in[1,10^6]$ and $\lambda\in[0.25,5]$. We considered only the sites that have a PWM score higher than $70\%$ of the difference between the lowest and the highest score. ($A-C$) Orange colour indicates high correlation between the analytical model and the ChIP-seq profile, while white colour low correlation.  ($D-F$) Blue colour indicates low mean squared error between the analytical model and the ChIP-seq profile, while white colour high mean squared error. ($G-I$) We plotted the regions where the mean squared error is in the lower $12\%$ of the range of values (blue) and the correlation is the higher $12\%$ of the range of values (orange). With green rectangle we marked the optimal set of parameters in terms of mean squared error and with a black rectangle the intersection of the parameters for which the two regions intersect.  \label{fig:heatmapChIPseq_GT_HB_KR_NoaccGroup070}}
\end{figure}

\begin{figure}
\centering
\includegraphics[width=\textwidth]{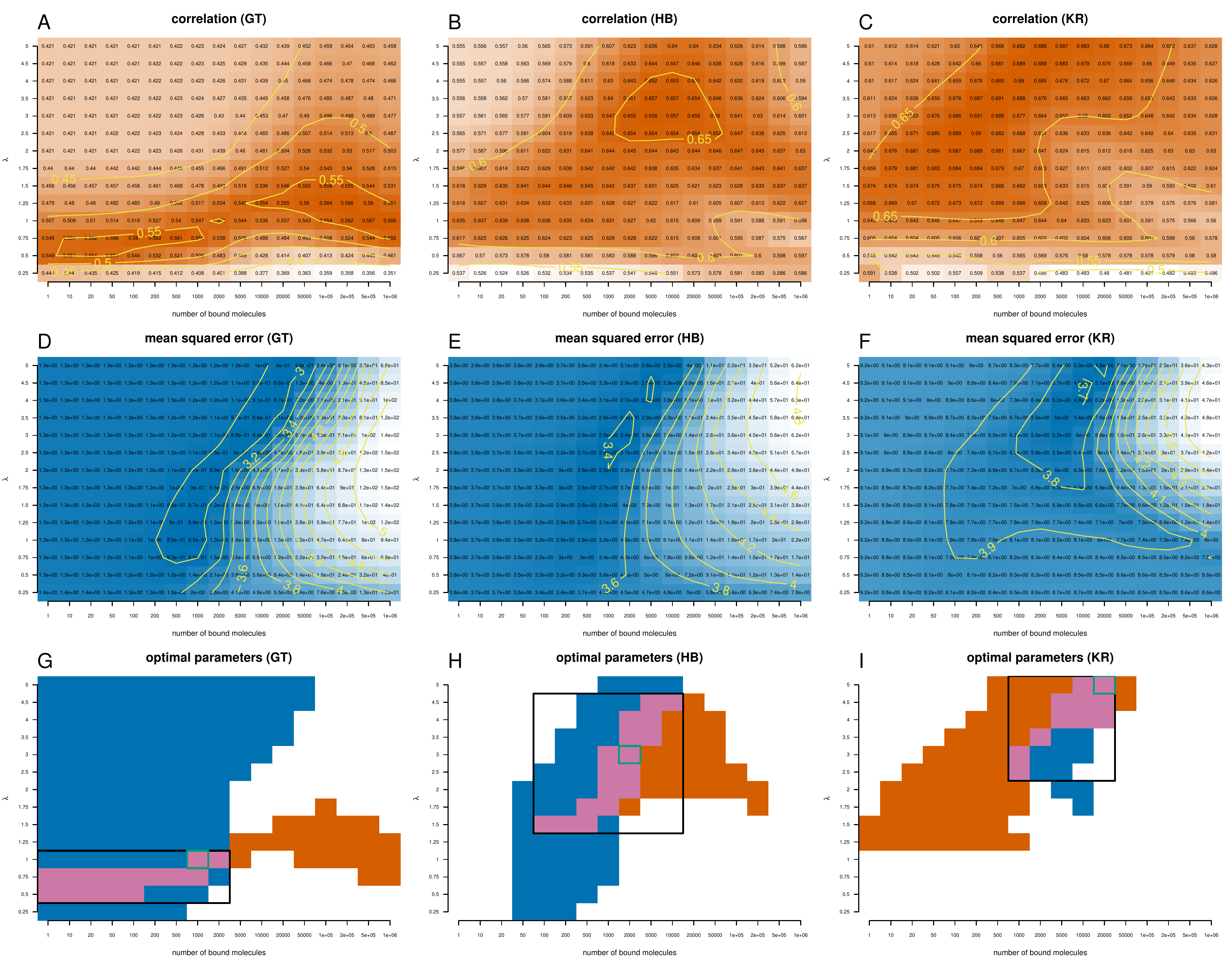}
\caption{\emph{Quantifying the distances between Giant, Hunchback and Kruppel ChIP-seq profiles and the profiles derived with the analytical model which includes DNA accessibility data}. This is the same as Figure \ref{fig:heatmapChIPseq_GT_HB_KR_NoaccGroup070}, except that we included binary DNA accessibility data in the analytical model. 
 \label{fig:heatmapChIPseq_GT_HB_KR_AccRegionsGroup070}}
\end{figure}

\clearpage
\subsection{Profiles for all 21 loci}

\begin{landscape}
\begin{figure}
\centering
\includegraphics[scale=0.8]{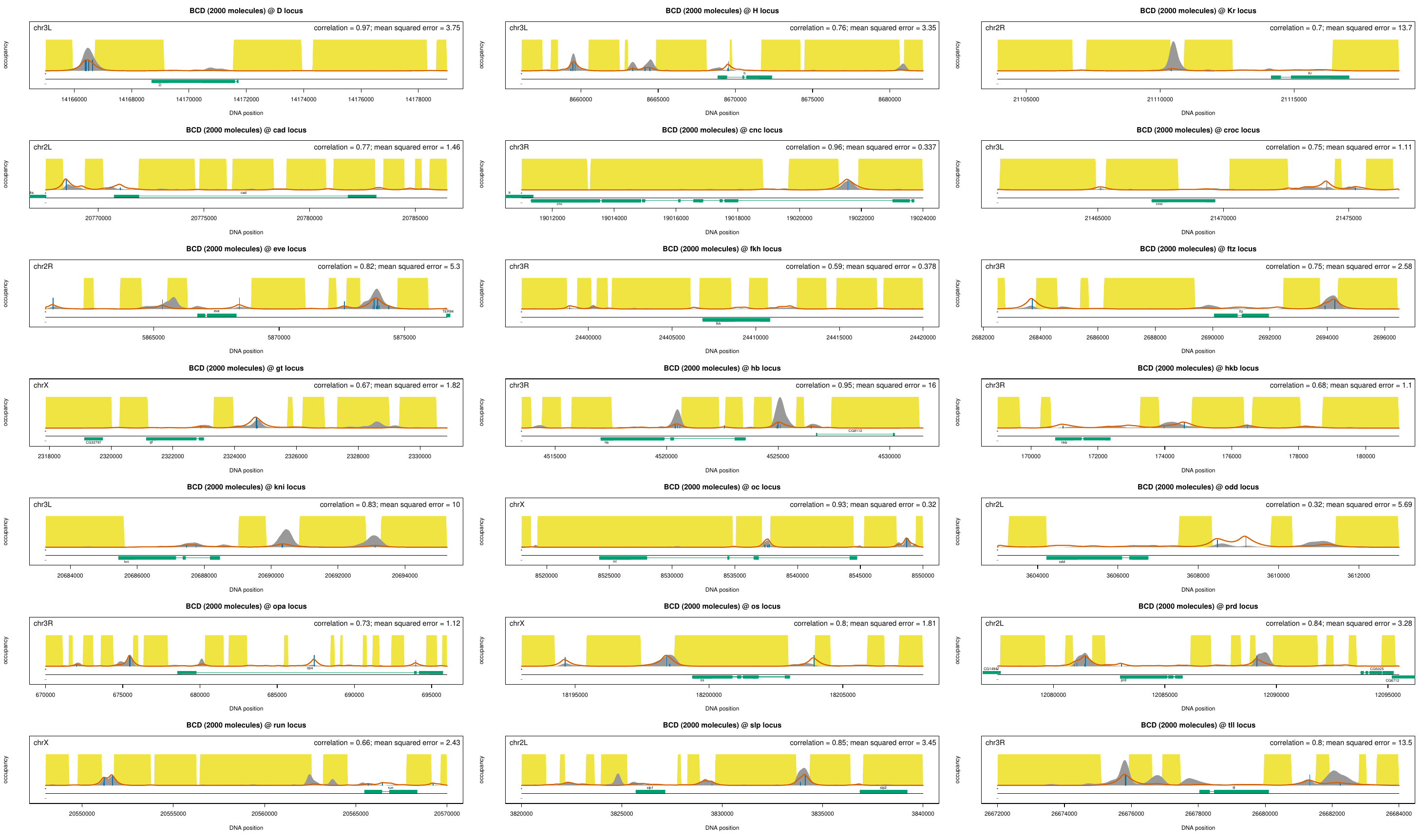}
\caption{\emph{Binding profiles for Bicoid at all 21 loci}.  The grey shading represents a ChIP-seq profile, the red line represents the prediction of the analytical model, the yellow shading represents the inaccessible DNA and the vertical blue lines represent the percentage of occupancy of the site (we only displayed sites with an occupancy higher than $5\%$). We considered the optimal set of parameters for Bicoid ($2000\ molecules$ and $\lambda=1.25$).}
 \label{fig:profileAllPositivesBCD}
\end{figure}
\end{landscape}

\begin{landscape}
\begin{figure}
\centering
\includegraphics[scale=0.8]{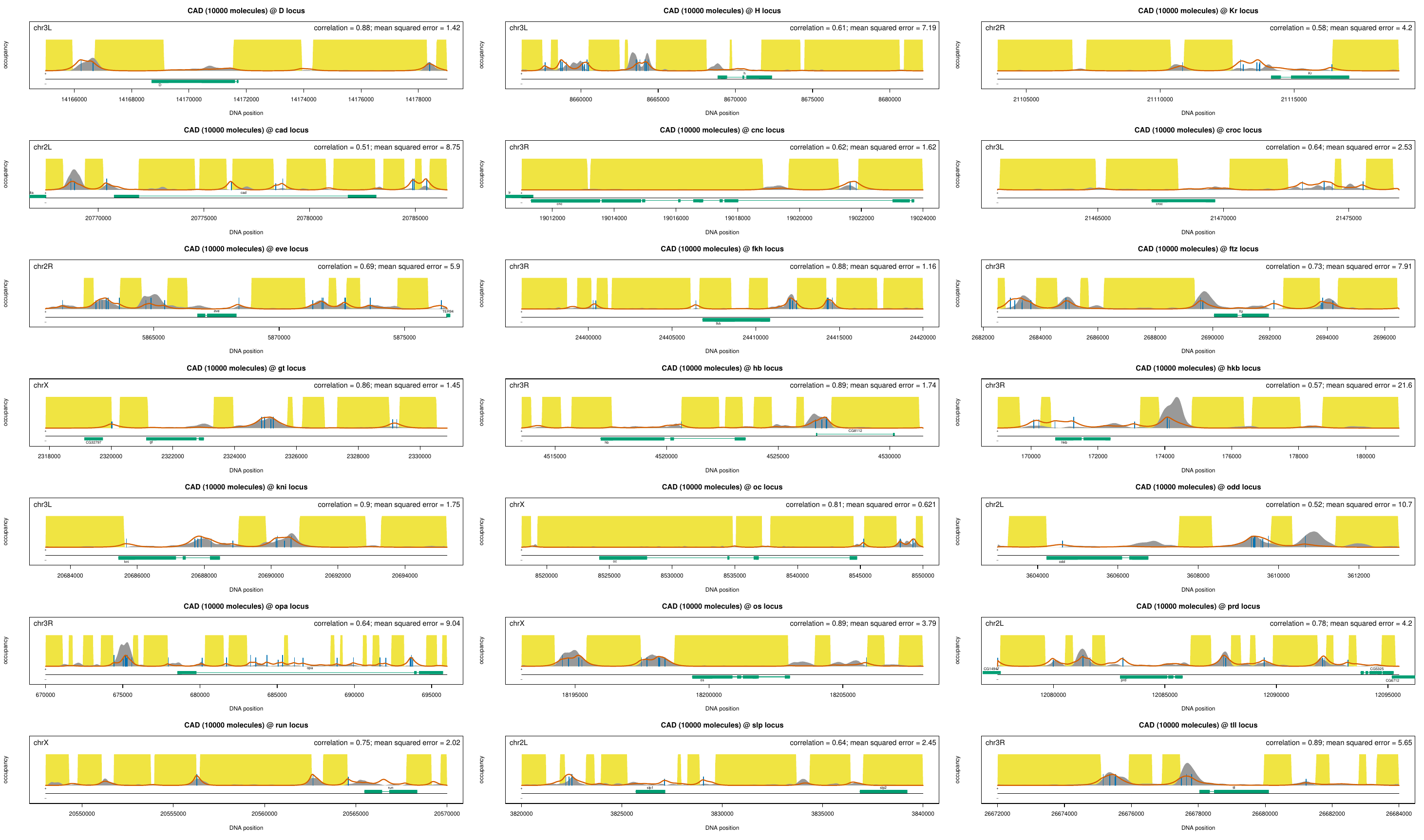}
\caption{\emph{Binding profiles for Caudal at all 21 loci}.  The grey shading represents a ChIP-seq profile, the red line represents the prediction of the analytical model, the yellow shading represents the inaccessible DNA and the vertical blue lines represent the percentage of occupancy of the site (we only displayed sites with an occupancy higher than $5\%$). We considered the optimal set of parameters for Caudal ($10000\ molecules$ and $\lambda=1.25$).}
 \label{fig:profileAllPositivesCAD}
\end{figure}
\end{landscape}

\begin{landscape}
\begin{figure}
\centering
\includegraphics[scale=0.8]{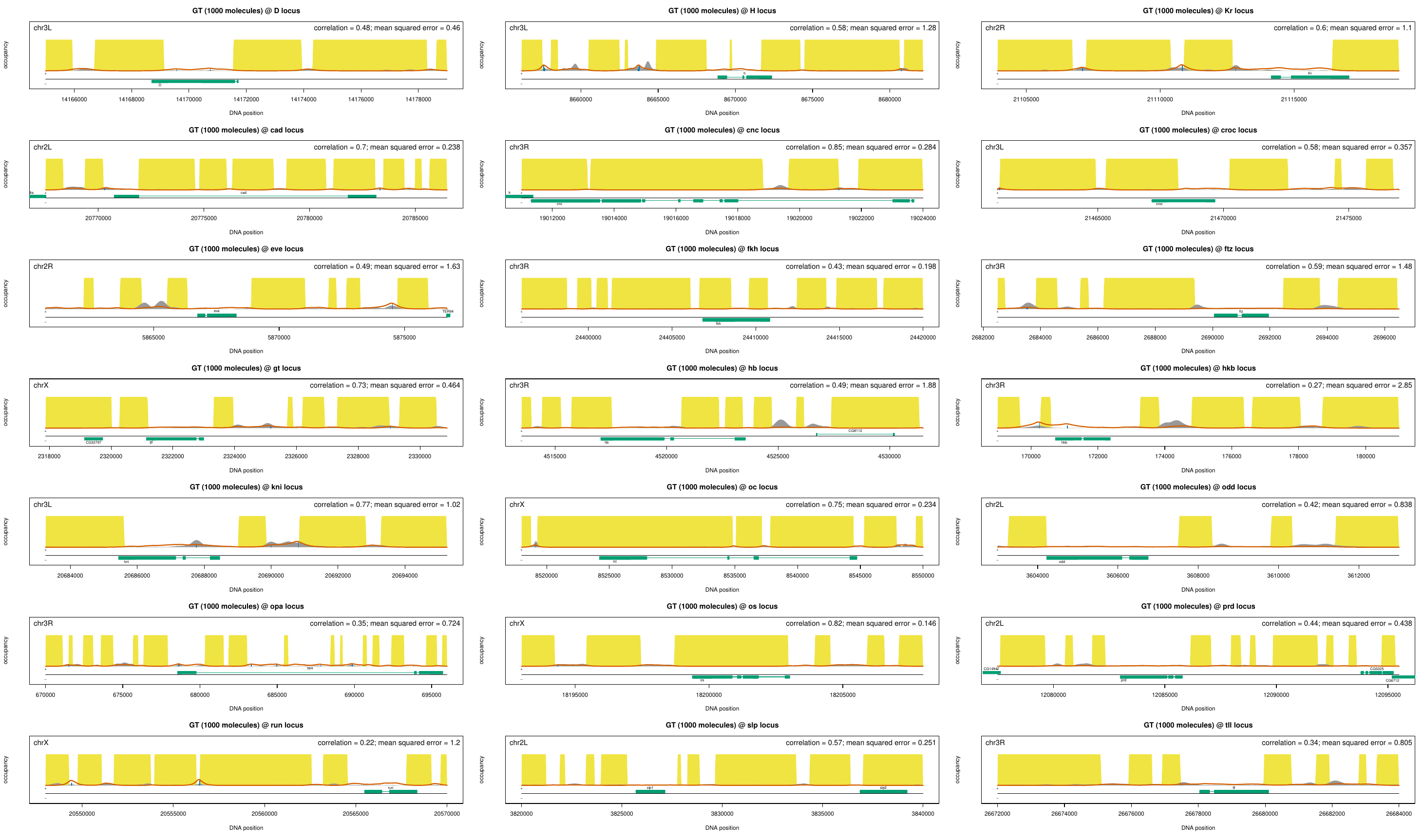}
\caption{\emph{Binding profiles for Giant at all 21 loci}.  The grey shading represents a ChIP-seq profile, the red line represents the prediction of the analytical model, the yellow shading represents the inaccessible DNA and the vertical blue lines represent the percentage of occupancy of the site (we only displayed sites with an occupancy higher than $5\%$). We considered the optimal set of parameters for Giant ($1000\ molecules$ and $\lambda=1.00$).}
 \label{fig:profileAllPositivesGT}
\end{figure}
\end{landscape}

\begin{landscape}
\begin{figure}
\centering
\includegraphics[scale=0.8]{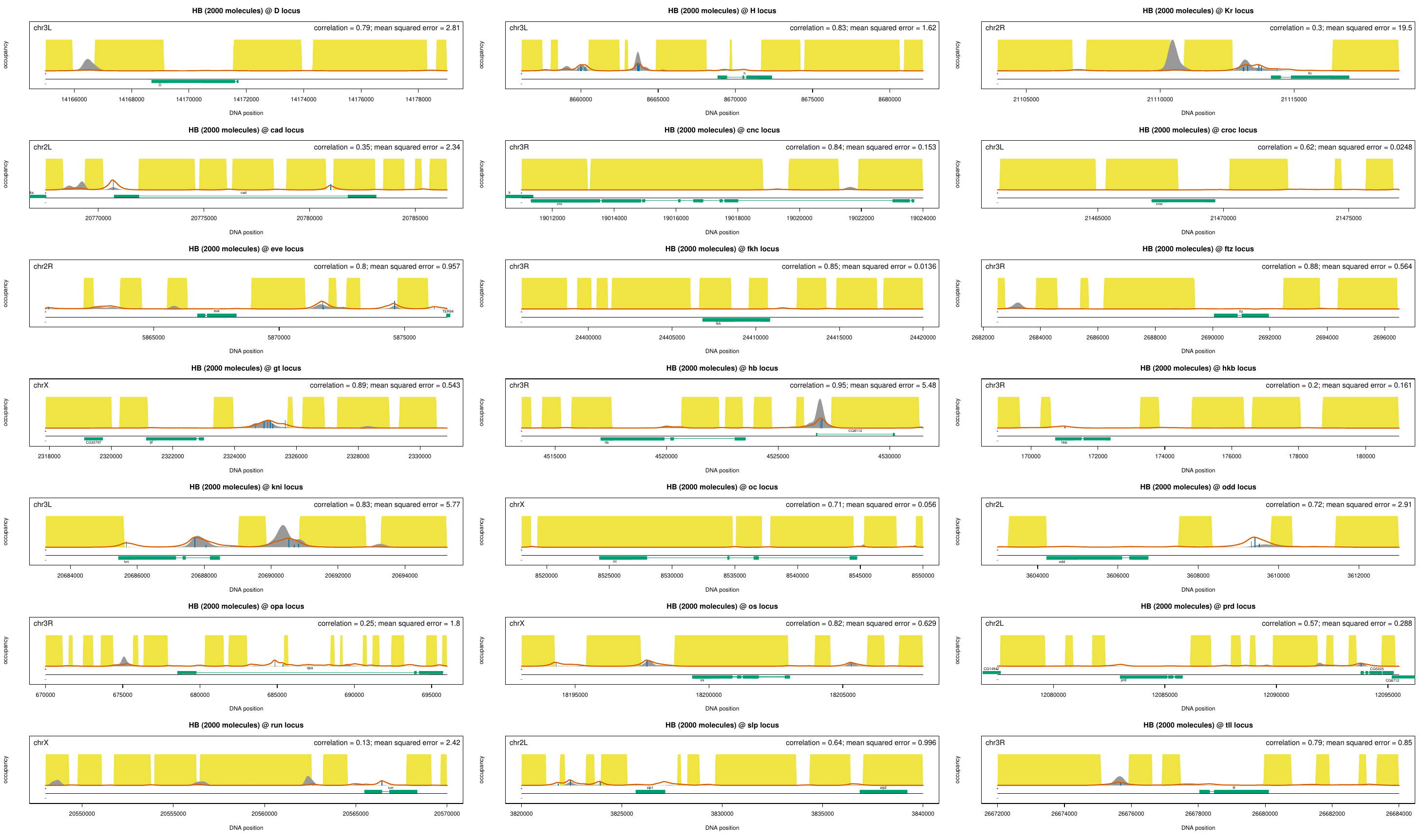}
\caption{\emph{Binding profiles for Hunchback at all 21 loci}.  The grey shading represents a ChIP-seq profile, the red line represents the prediction of the analytical model, the yellow shading represents the inaccessible DNA and the vertical blue lines represent the percentage of occupancy of the site (we only displayed sites with an occupancy higher than $5\%$). We considered the optimal set of parameters for Hunchback ($2000\ molecules$ and $\lambda=3.00$).}
 \label{fig:profileAllPositivesHB}
\end{figure}
\end{landscape}

\begin{landscape}
\begin{figure}
\centering
\includegraphics[scale=0.8]{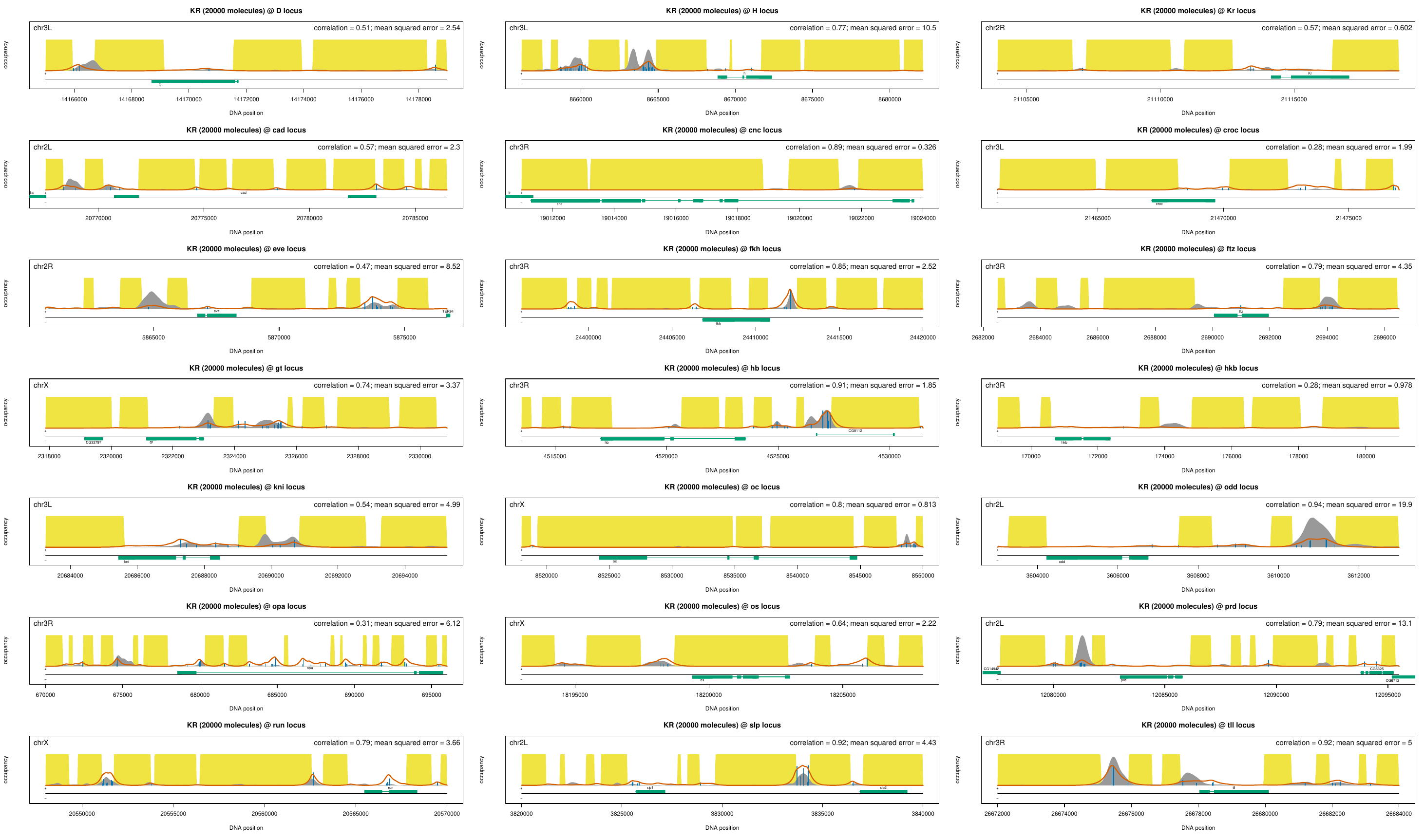}
\caption{\emph{Binding profiles for Kruppel at all 21 loci}.  The grey shading represents a ChIP-seq profile, the red line represents the prediction of the analytical model, the yellow shading represents the inaccessible DNA and the vertical blue lines represent the percentage of occupancy of the site (we only displayed sites with an occupancy higher than $5\%$). We considered the optimal set of parameters for Kruppel ($20000\ molecules$ and $\lambda=5.00$).}
 \label{fig:profileAllPositivesKR}
\end{figure}
\end{landscape}

\clearpage
\subsection{Including weak binding into the model}

\begin{figure}
\centering
\includegraphics[width=\textwidth]{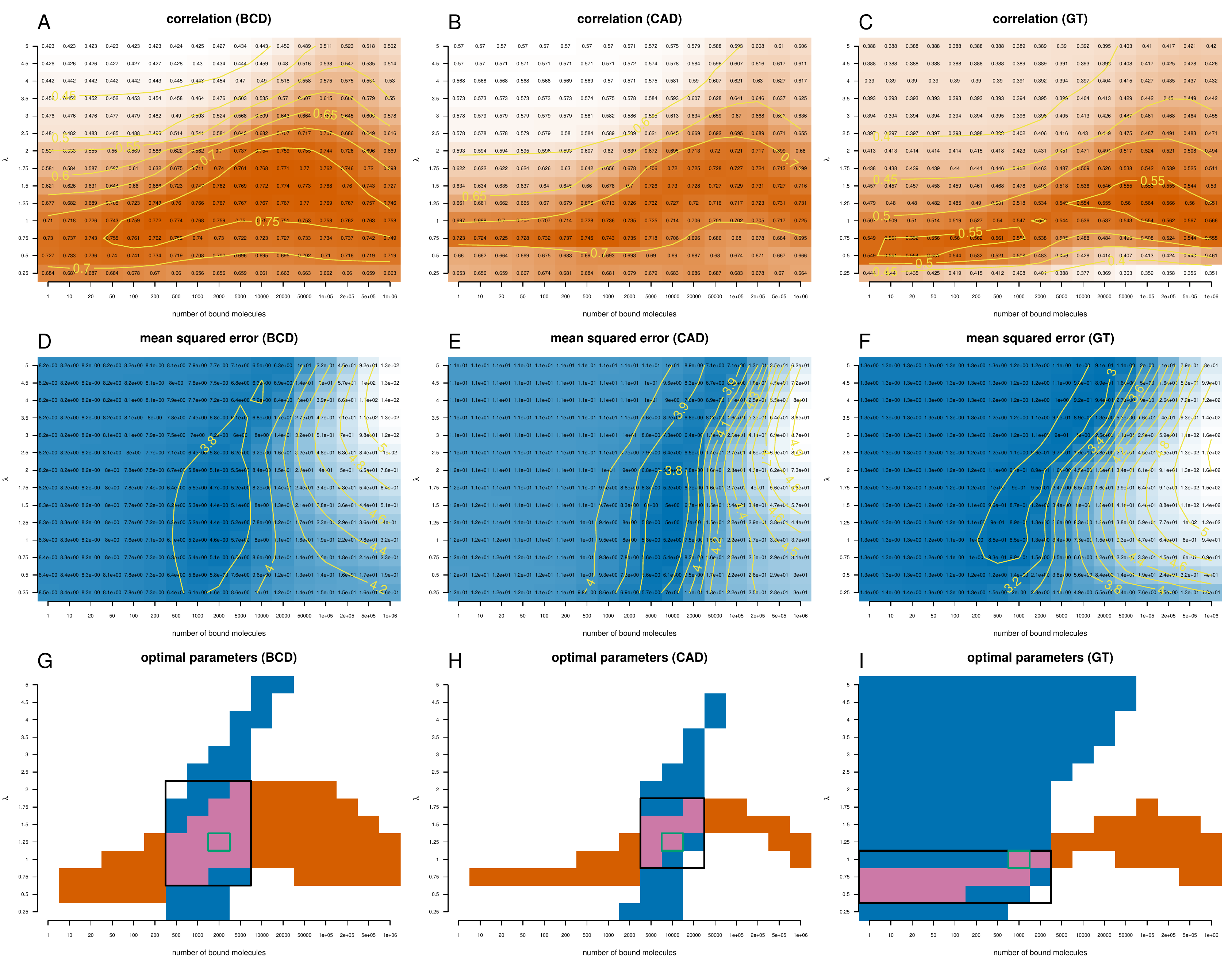}
\caption{\emph{The influence of weak binding on Bicoid, Caudal and Giant ChIP-seq profiles}. We plotted heatmaps for the correlation $(A-C)$ and mean squared error $(D-F)$  between the analytical model and the ChIP-seq profile of Bicoid $(A,D)$, Caudal $(B,E)$ and Giant $(C,F)$. The analytical model includes binary DNA accessibility data (the accessibility of any site can be either $0$ or $1$ depending on whether the site is accessible or not). We computed these values for different sets of parameters $N\in[1,10^6]$ and $\lambda\in[0.25,5]$. Colour code as above. Here, only considering sites that have a PWM score higher than $30\%$ of the difference between the lowest and the highest score. ($G-I$) We plotted the regions where the mean squared error is in the lower $12\%$ of the range of values (blue) and the correlation is the higher $12\%$ of the range of values (orange). With green rectangle we marked the optimal set of parameters in terms of mean squared error and with a black rectangle the intersection of the parameters for which the two regions intersect.  \label{fig:heatmapChIPseq_BCD_CAD_GT_AccRegionsGroup030}}
\end{figure}

\begin{figure}
\centering
\includegraphics[width=0.9\textwidth]{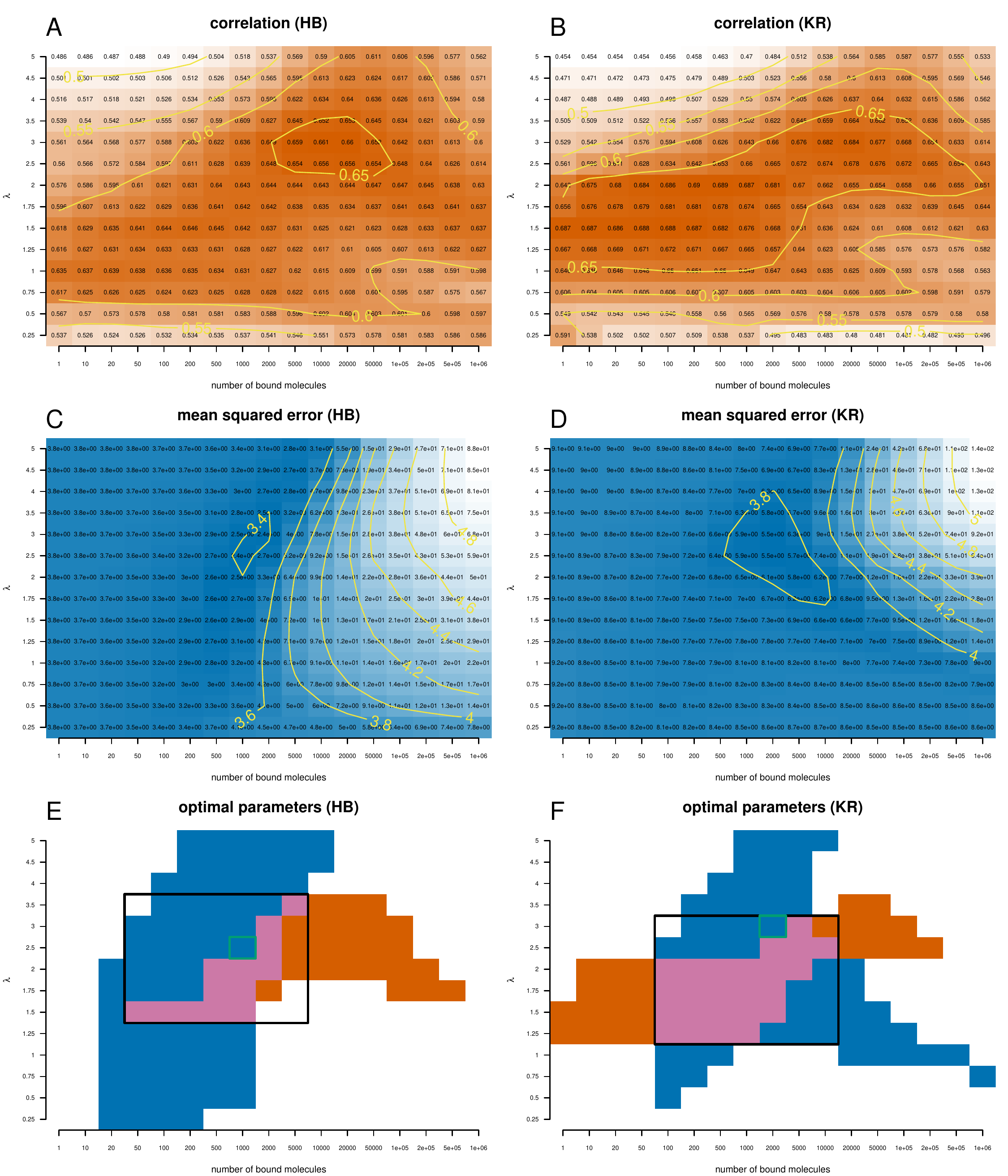}
\caption{\emph{The influence of weak binding on Hunchback and Kruppel ChIP-seq profiles}. We plotted heatmaps for the correlation \onel\ and \twol\ and mean squared error \threel\ and \fourl\ between the analytical model and the  ChIP-seq profile of Hunchback $(A,C)$ and Kruppel $(B,D)$. The analytical model includes binary DNA accessibility data (the accessibility of any site can be either $0$ or $1$ depending on whether the site is accessible or not). We computed these values for different sets of parameters: $N\in[1,10^6]$ and $\lambda\in[0.25,5]$. Colour code as above. PWM filtering as in Figure \ref{fig:heatmapChIPseq_BCD_CAD_GT_AccRegionsGroup030}. ($E,F$) We plotted the regions where the mean squared error is in the lower $12\%$ of the range of values (blue) and the correlation is the higher $12\%$ of the range of values (orange). With green rectangle we marked the optimal set of parameters in terms of mean squared error and with a black rectangle the intersection of the parameters for which the two regions intersect. \label{fig:heatmapChIPseq_HB_KR_AccRegionsGroup030}}
\end{figure}

\input{OptimalSetOfParametersTableAccessibilityRegionsWeakBinding.tex}

\clearpage
\subsection{Continuous DNA accessibility data}

\begin{figure}
\centering
\includegraphics[width=0.9\textwidth]{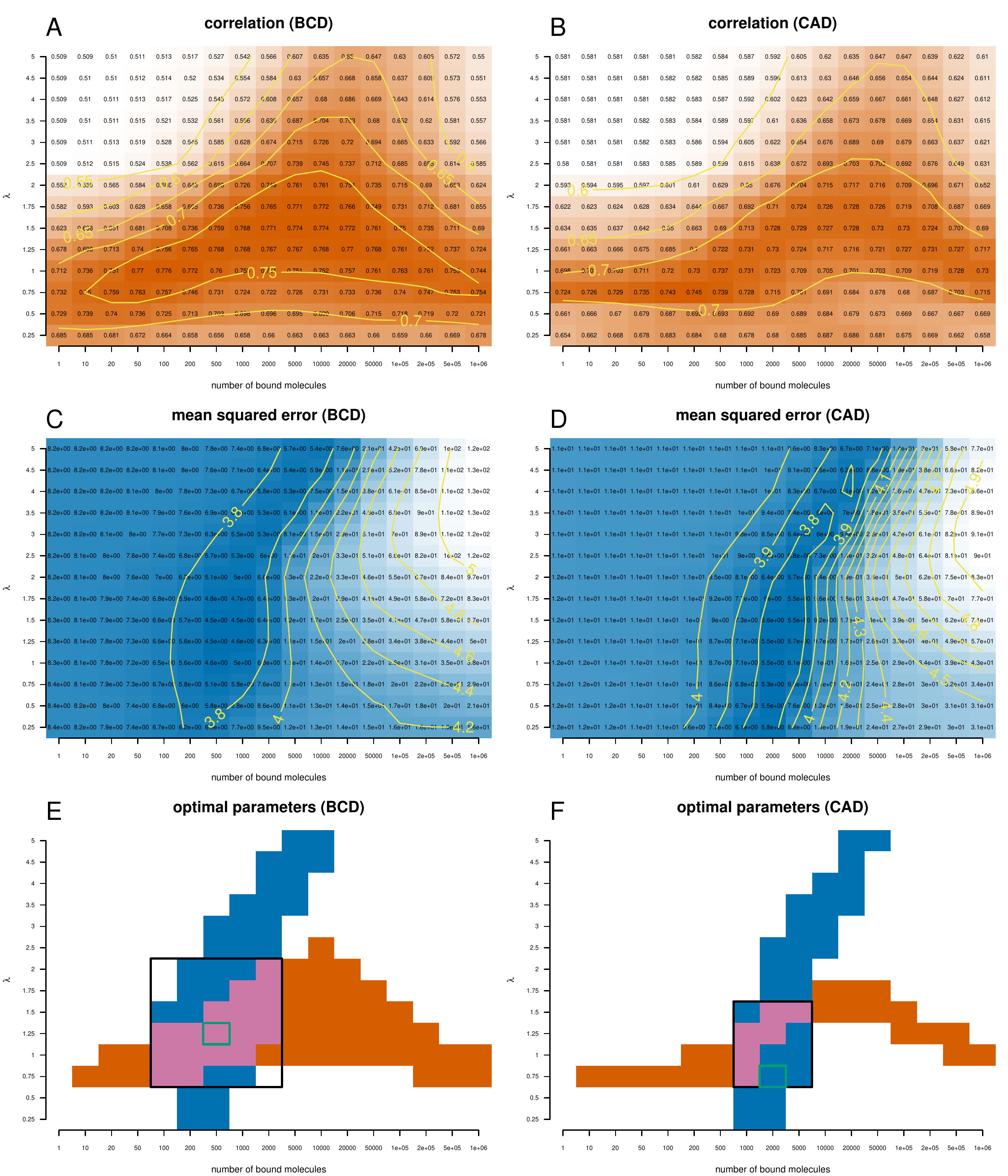}
\caption{\emph{Estimating the Bicoid and Caudal ChIP-seq profiles when assuming non-binary accessibility data}. We plotted heatmaps for the correlation \onel\ and \twol\ and mean squared error \threel\ and \fourl\ between the analytical model and the  ChIP-seq profile of Bicoid $(A,C)$ and Caudal $(B,D)$. The analytical model includes non-binary DNA accessibility data, by using equation \eqref{eq:paccessible} to compute the probability that a site is accessible. We computed these values for different sets of parameters: $N\in[1,10^6]$ and $\lambda\in[0.25,5]$. Colour code as above. We considered only the sites that have a PWM score higher than $70\%$ of the difference between the lowest and the highest score. ($E,F$) We plotted the regions where the mean squared error is in the lower $12\%$ of the range of values (blue) and the correlation is the higher $12\%$ of the range of values (orange). With green rectangle we marked the optimal set of parameters in terms of mean squared error and with a black rectangle the intersection of the parameters for which the two regions intersect. \label{fig:heatmapChIPseq_BCD_CAD_AccProbabilityGroup070}}
\end{figure}

\begin{figure}
\centering
\includegraphics[width=\textwidth]{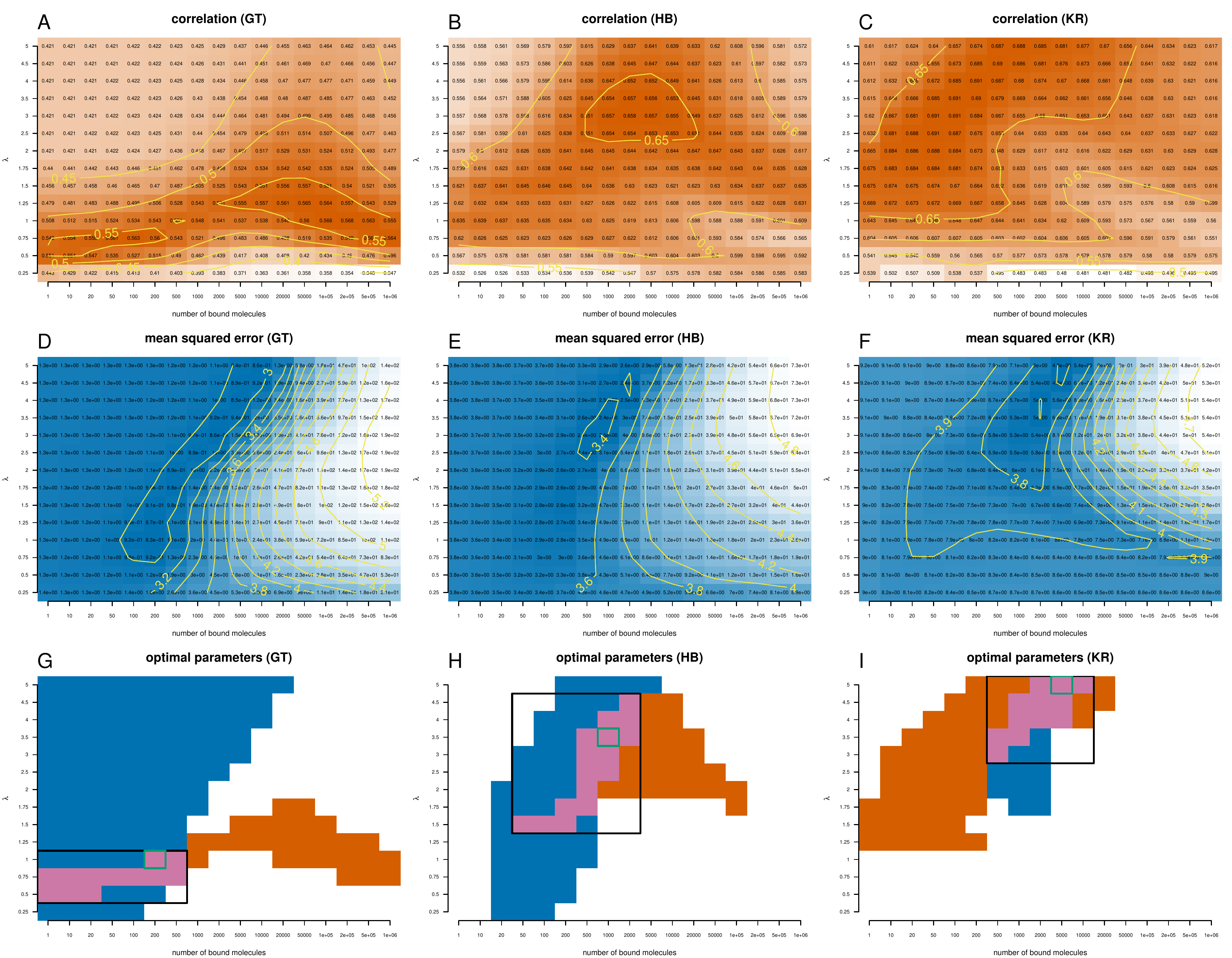}
\caption{\emph{Estimating the Giant, Hunchback and Kruppel  caudal ChIP-seq profiles when assuming non-binary accessibility data}. We plotted heatmaps for the correlation $(A-C)$ and mean squared error $(D-F)$ between the analytical model and the ChIP-seq profile of Giant $(A,D)$, Hunchback $(B,E)$ and Kruppel $(C,F)$. The analytical model includes non-binary DNA accessibility data, by using equation \eqref{eq:paccessible} to compute the probability that a site is accessible. We computed these values for different sets of parameters: $N\in[1,10^6]$ and $\lambda\in[0.25,5]$. Colour code as above. PWM filtering as before. ($G-I$) We plotted the regions where the mean squared error is in the lower $12\%$ of the range of values (blue) and the correlation is the higher $12\%$ of the range of values (orange). With green rectangle we marked the optimal set of parameters in terms of mean squared error and with a black rectangle the intersection of the parameters for which the two regions intersect. \label{fig:heatmapChIPseq_GT_HB_KR_AccProbabilityGroup070}}
\end{figure}

\input{OptimalSetOfParametersTableAccessibilityProbability.tex}

\clearpage
\subsection{Using the JASPAR PWMs}

\begin{figure}[htp]
  \begin{center}
  		\includegraphics[width=\textwidth]{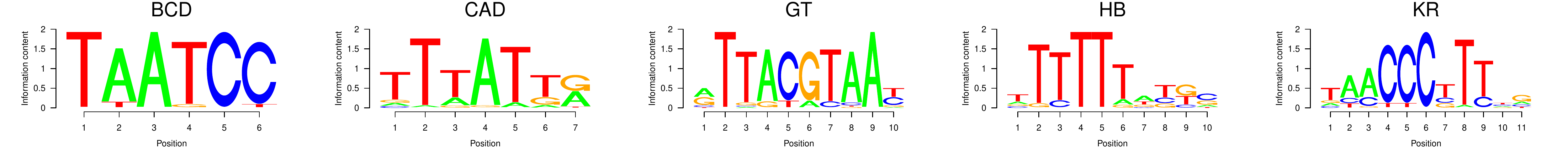}
  \end{center}
\caption{\emph{PWMs for the five TFs from the JASPAR database}. The graph plots the sequence logos for the following TFs: \one Bicoid, \two Caudal, \three Giant, \four Hunchback and  \five Kruppel \citep{portales-casamar_2010}. When computing the PWMs we used a pseudo count of $1$. The information content for the five motifs is: \one $I_{BCD}=11.0$, \two $I_{CAD}=8.9$, \three $I_{GT}=14.0$, \four $I_{HB}=10.4$ and  \five $I_{KR}=11.7$.}\label{fig:PWMmotifsJaspar}
\end{figure}

\begin{figure}
\centering
\includegraphics[width=0.9\textwidth]{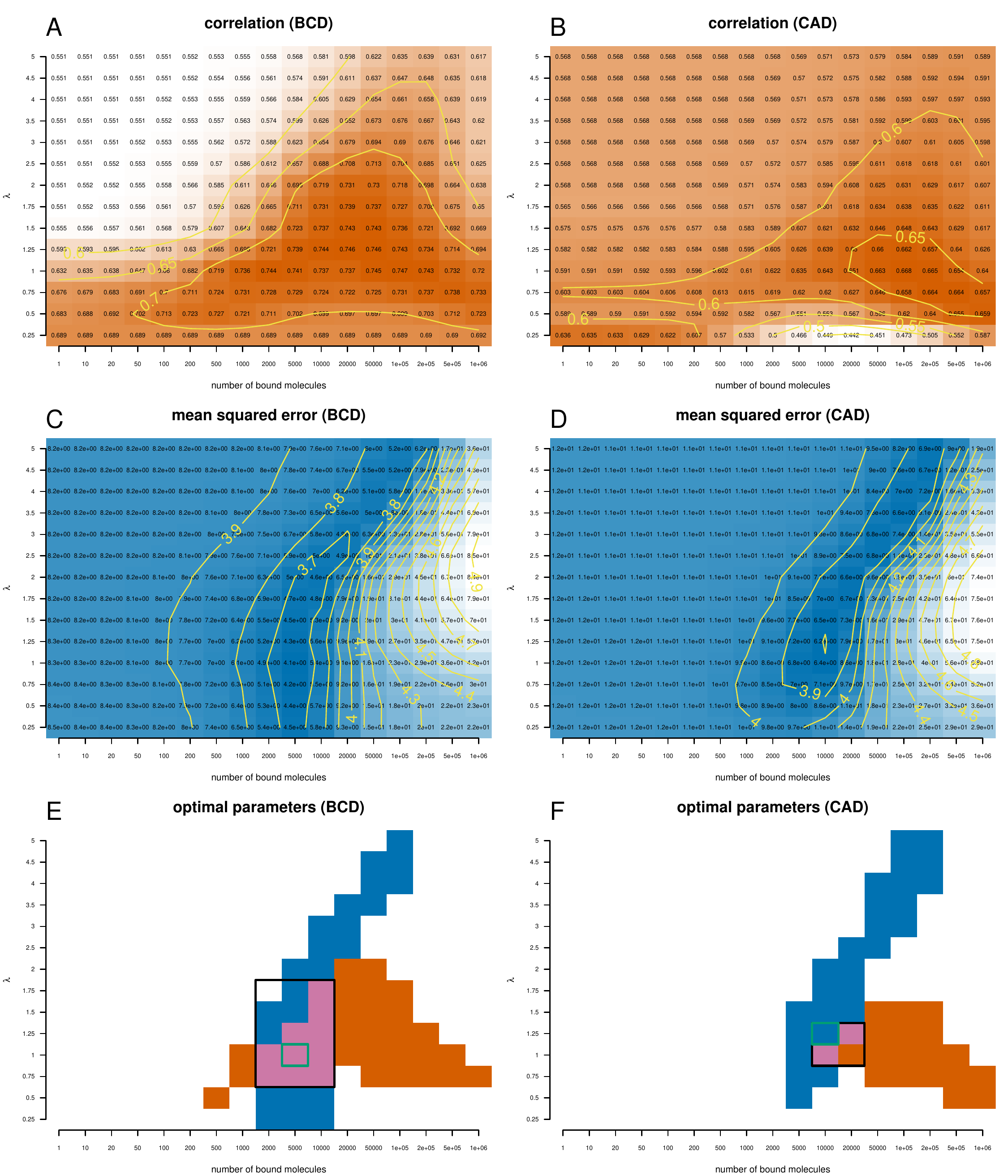}
\caption{\emph{Estimating the Bicoid and Caudal ChIP-seq profiles  when using the binding motif from JASPAR database}; see Figure \ref{fig:PWMmotifsJaspar}. We plotted heatmaps for the correlation \onel\ and \twol\ and mean squared error \threel\ and \fourl\  between the analytical model and the  ChIP-seq profile of Bicoid $(A,C)$ and Caudal $(B,D)$. The analytical model includes binary DNA accessibility data (the accessibility of any site can be either $0$ or $1$ depending on whether the site is accessible or not). We computed these values for different sets of parameters: $N\in[1,10^6]$ and $\lambda\in[0.25,5]$. Colour code as above. We considered only the sites that have a PWM score higher than $70\%$ of the difference between the lowest and the highest score. ($E,F$) We plotted the regions where the mean squared error is in the lower $12\%$ of the range of values (blue) and the correlation is the higher $12\%$ of the range of values (orange). With green rectangle we marked the optimal set of parameters in terms of mean squared error and with a black rectangle the intersection of the parameters for which the two regions intersect. \label{fig:heatmapChIPseq_BCD_CAD_AccRegionsJasparGroup070}}
\end{figure}

\begin{figure}
\centering
\includegraphics[width=\textwidth]{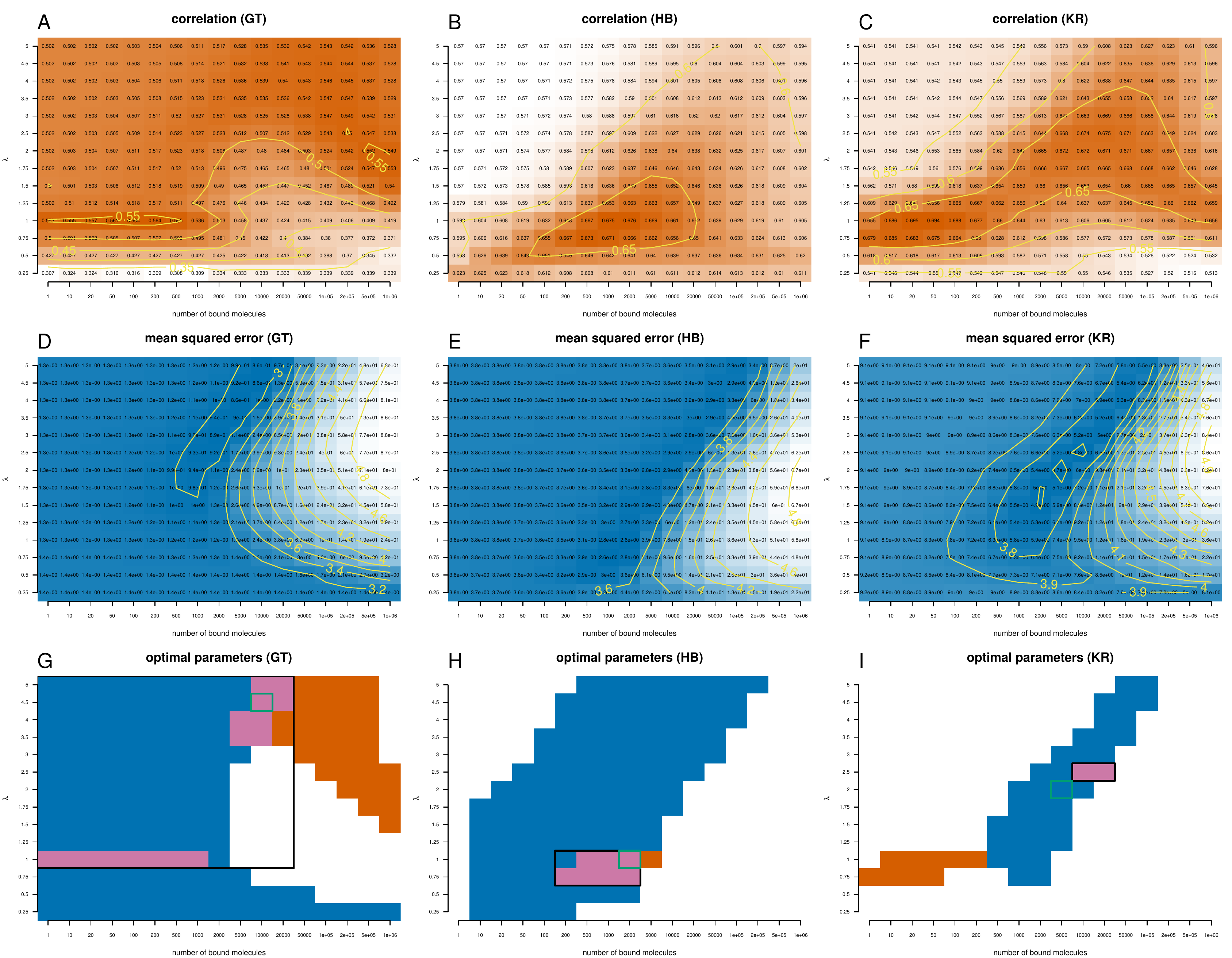}
\caption{\emph{Estimating the Giant, Hunchback, Kruppel and Caudal ChIP-seq profiles when using the binding motif from JASPAR database}; see Figure \ref{fig:PWMmotifsJaspar}. We plotted heatmaps for the correlation $(A-C)$ and mean squared error $(D-F)$ between the analytical model and the ChIP-seq profile of Giant $(A,D)$, Hunchback $(B,E)$ and Kruppel $(C,F)$. The analytical model includes binary DNA accessibility data (the accessibility of any site can be either $0$ or $1$ depending on whether the site is accessible or not). We computed these values for different sets of parameters: $N\in[1,10^6]$ and $\lambda\in[0.25,5]$. Colour code as above. PWM filtering as above. ($G-I$) We plotted the regions where the mean squared error is in the lower $12\%$ of the range of values (blue) and the correlation is the higher $12\%$ of the range of values (orange). With green rectangle we marked the optimal set of parameters in terms of mean squared error and with a black rectangle the intersection of the parameters for which the two regions intersect.
 \label{fig:heatmapChIPseq_GT_HB_KR_AccRegionsJasparGroup070}}
\end{figure}

\input{OptimalSetOfParametersTableAccessibilityRegionsJASPAR.tex}

\begin{landscape}
\begin{figure}
\centering
\includegraphics[scale=0.8]{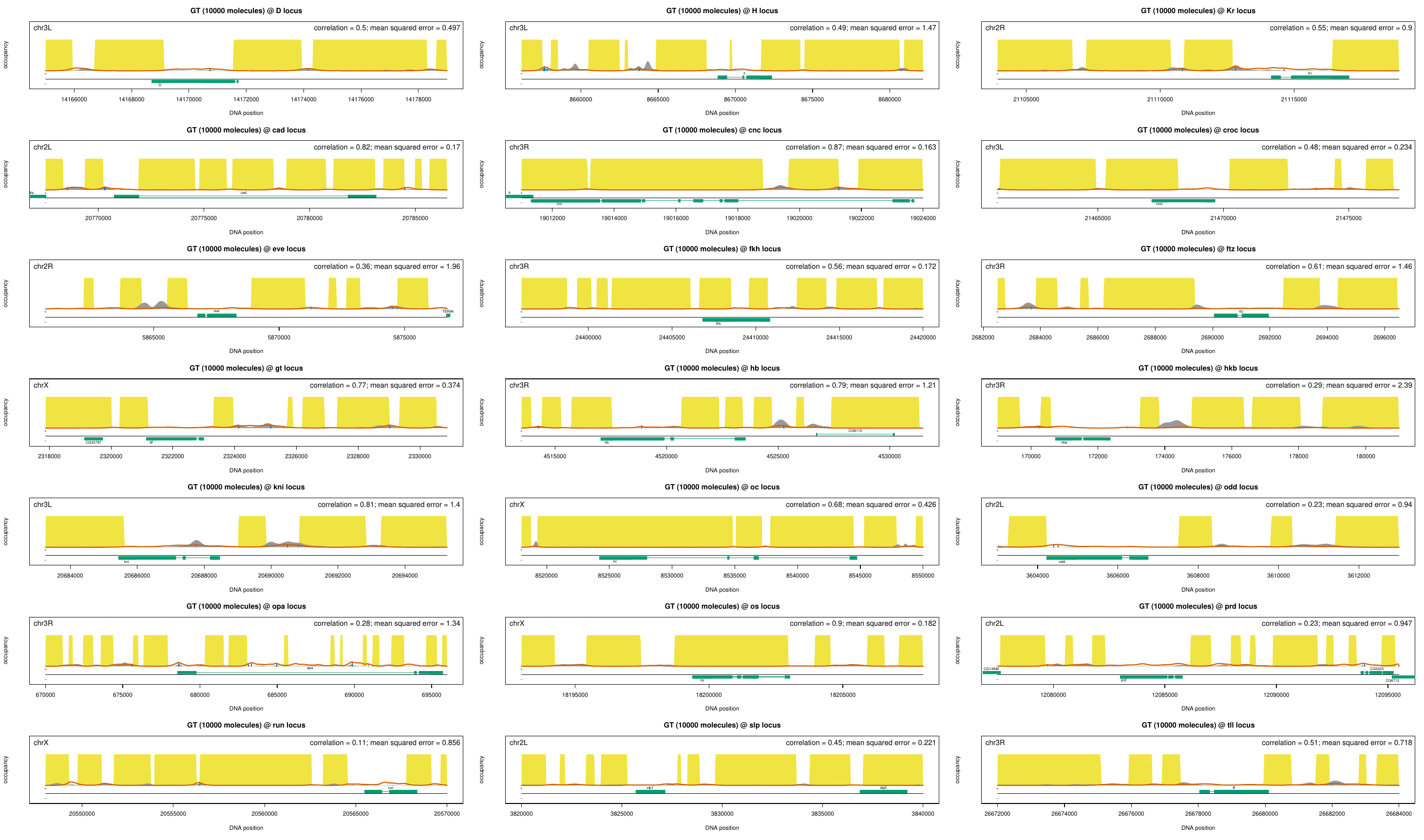}
\caption{\emph{Binding profiles for Giant at all 21 loci using the JASPAR database PWM}.  The grey shading represents a ChIP-seq profile, the red line represents the prediction of the analytical model, the yellow shading represents the inaccessible DNA and the vertical blue lines represent the percentage of occupancy of the site (we only displayed sites with an occupancy higher than $5\%$). We considered the optimal set of parameters for hunchback ($10000\ molecules$ and $\lambda=4.50$).}
 \label{fig:profileAllPositivesJasparGT}
\end{figure}
\end{landscape}

\begin{landscape}
\begin{figure}
\centering
\includegraphics[scale=0.8]{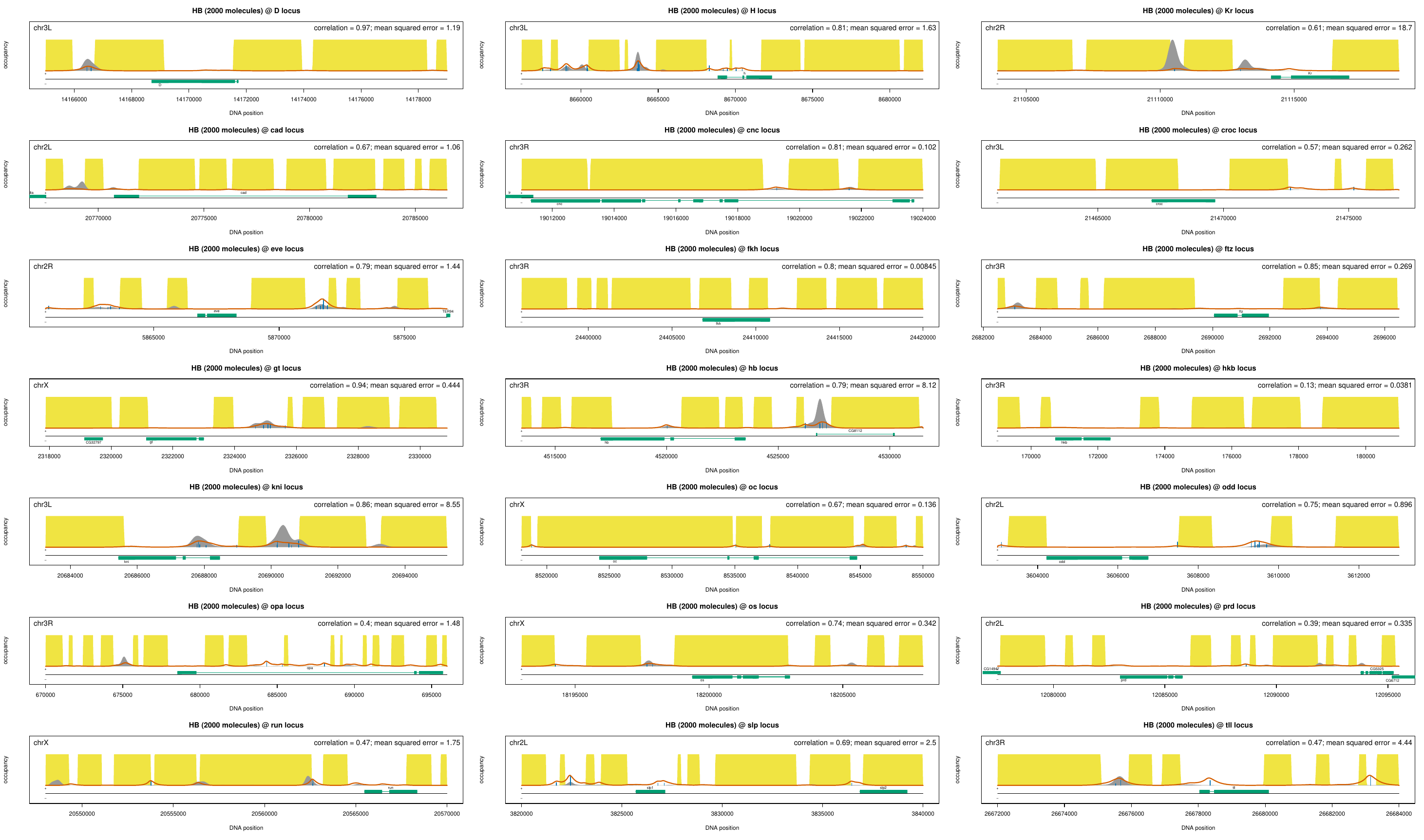}
\caption{\emph{Binding profiles for Hunchback at all 21 loci using the JASPAR database PWM}.  The grey shading represents a ChIP-seq profile, the red line represents the prediction of the analytical model, the yellow shading represents the inaccessible DNA and the vertical blue lines represent the percentage of occupancy of the site (we only displayed sites with an occupancy higher than $5\%$). We considered the optimal set of parameters for hunchback ($2000\ molecules$ and $\lambda=1.00$).}
 \label{fig:profileAllPositivesJasparHB}
\end{figure}
\end{landscape}

\begin{landscape}
\begin{figure}
\centering
\includegraphics[scale=0.8]{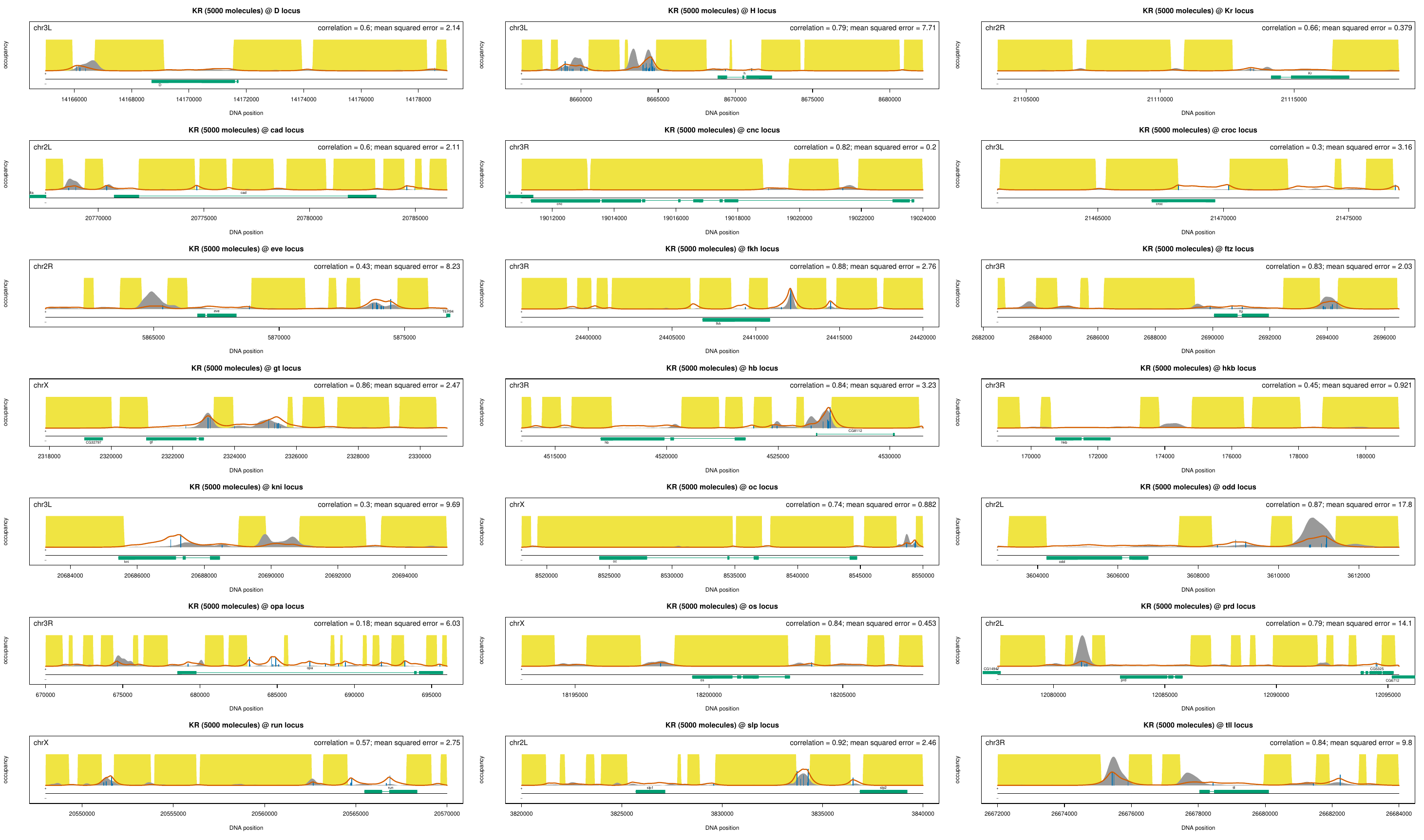}
\caption{\emph{Binding profiles for Kruppel at all 21 loci using the JASPAR database PWM}. The grey shading represents a ChIP-seq profile, the red line represents the prediction of the analytical model, the yellow shading represents the inaccessible DNA and the vertical blue lines represent the percentage of occupancy of the site (we only displayed sites with an occupancy higher than $5\%$). We considered the optimal set of parameters for Kruppel ($5000\ molecules$ and $\lambda=2.00$).}
 \label{fig:profileAllPositivesJasparKR}
\end{figure}
\end{landscape}

\clearpage
\subsection{Genome-wide analysis}
\input{ChIPseqProfileStatistics.tex}
\input{GenomeWideNoOfRegions.tex}

\begin{figure}
\centering
\includegraphics[width=0.6\textwidth]{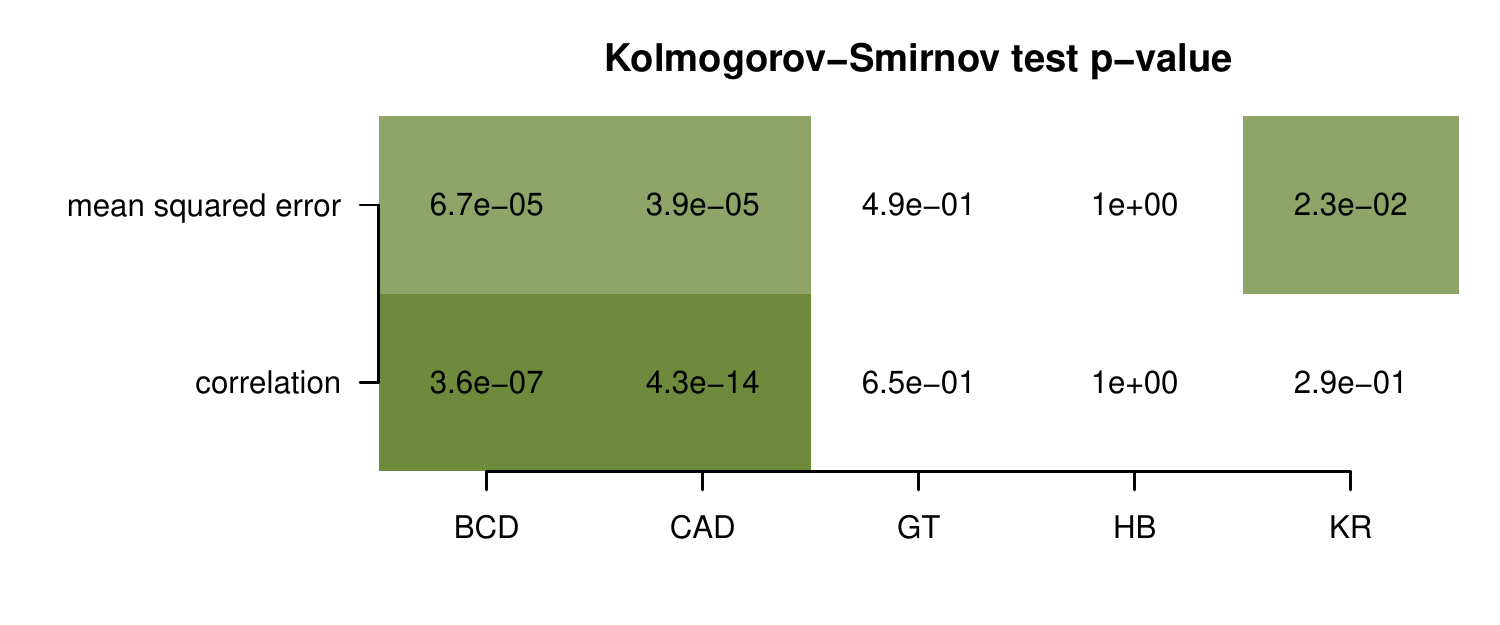}
\caption{\emph{The p-value of the Kolmogorov−Smirnov test between the two subsets of thresholds}. We performed a  Kolmogorov-Smirnov test between the case when we selected the regions with a mean ChIP-seq signal higher than the background ($>B$) and the case when we selected the regions with a mean ChIP-seq signal higher than half the background ($>0.5\cdot B$). We considered both the mean squared error and correlation coefficient for each of the five TFs.}
 \label{fig:GenomeWideKSPvalue}
\end{figure}

\begin{table}
\centering
\begin{tabular}{| l | l | r | l | l | l |}
\hline
TF & cell type & \% of bound TF &  experiment type & type of binding &  reference\\ \hline
C/EBP &  HeLa &  $35.5\%$ & FRAP & specific & \cite{phair_2004_method} \\ \hline
NF1 &  HeLa &  $58.1\%$ & FRAP & specific & \cite{phair_2004_method} \\ \hline
Jun &  HeLa &  $75.2\%$ & FRAP & specific & \cite{phair_2004_method} \\ \hline
Fos &  HeLa &  $54.2\%$ & FRAP & specific & \cite{phair_2004_method} \\ \hline
Myc &  HeLa &  $54.1\%$ & FRAP & specific & \cite{phair_2004_method} \\ \hline
Max &  HeLa &  $34.5\%$ & FRAP & specific & \cite{phair_2004_method} \\ \hline
Max &  3134 &  $16.0\%$ & FRAP & specific & \cite{mueller_2008} \\ \hline
Mad &  HeLa &  $21.2\%$ & FRAP & specific & \cite{phair_2004_method} \\ \hline
FBP &  HeLa &  $99.7\%$ & FRAP & specific & \cite{phair_2004_method} \\ \hline
XBP &  HeLa &  $18.6\%$ & FRAP & specific & \cite{phair_2004_method} \\ \hline
BRD4 &  HeLa &  $62.4\%$ & FRAP & specific & \cite{phair_2004_method} \\ \hline
p53 &  H1299 &  $7.0\%$ & FRAP/FCS/SMT & specific & \cite{mazza_2012} \\ \hline
p53 &  H1299 &  $2.5\%$ & SMT & specific & \cite{morisaki_2014} \\ \hline
GR &  MCF-7 &  $12.0\%$ & RLSM & specific & \cite{gebhardt_2013} \\ \hline
GR* &  MCF-7 &  $37.0\%$ & RLSM & specific & \cite{gebhardt_2013} \\ \hline
GR &  3134 &  $3.5\%$ & SMT & specific & \cite{morisaki_2014} \\ \hline
STAT1* &  HeLa &  $34.0\%$ & SMT & specific & \cite{speil_2011} \\ \hline
HSF1 &  U87 &  $30.0\%$ & FCS & specific & \cite{kloster_landsberg_2012} \\ \hline
HSF1* &  U87 &  $45.0\%$ & FCS & specific & \cite{kloster_landsberg_2012} \\ \hline
Sox2 &  ES &  $15.1\%$ & 2D SMT  & specific & \cite{chen_2014} \\ \hline
Sox2 (with Oct4*) &  NIH 3T3 &  $16.3\%$ & 2D SMT  & specific & \cite{chen_2014} \\ \hline
Sox2 (with Oct4) &  NIH 3T3 &  $16.9\%$ & 2D SMT  & specific & \cite{chen_2014} \\ \hline
Oct4 &  ES &  $14.4\%$ & 2D SMT  & specific & \cite{chen_2014} \\ \hline
Oct4  (with Sox2*) &  NIH 3T3 &  $18.4\%$ & 2D SMT  & specific & \cite{chen_2014} \\ \hline
Oct4  (with Sox2) &  NIH 3T3 &  $10.7\%$ & 2D SMT  & specific & \cite{chen_2014} \\ \hline
\end{tabular}
\caption{\emph{The fraction of bound molecules}. This is a (non exhaustive) list of the percentage of bound molecules of several TFs to the DNA as determined experimentally in previous works. The following methods were used:  Fluorescence Recovery After Photobleaching (FRAP), Fluorescence Correlation Spectroscopy (FCS), Single Molecule Tracking (SMT) and Reflected Light-Sheet Microscope (RLSM). The $*$ superscript indicates the TF in induced state. Note that if the same group published different measurements for the same TF, we only selected the latest value. In addition, if the data was fitted to a model assuming two modes of binding, we report only the faction of bound molecules in the slower moving component, which estimates the fraction of specifically bound TF. The distribution of the percentage of bound TFs is plotted in Figure \ref{fig:FractionBoundMolecules}.}
\label{tab:FractionBoundMolecules}
\end{table}

\begin{figure}
\centering
\includegraphics[width=0.6\textwidth]{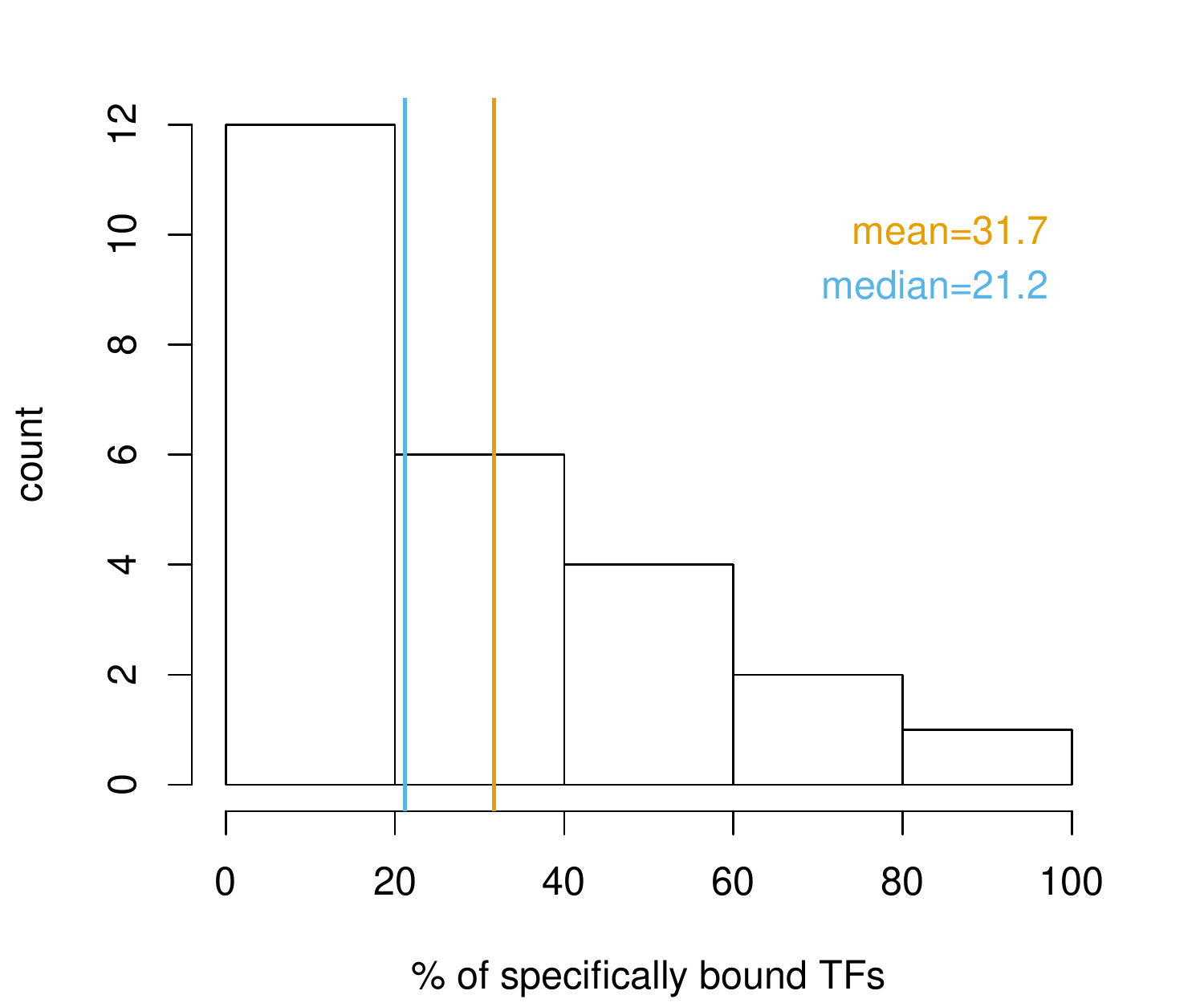}
\caption{\emph{The distribution of the percentage of specifically bound molecules for different TFs}. We plotted the data from Table \ref{tab:FractionBoundMolecules}.}
 \label{fig:FractionBoundMolecules}
\end{figure}

\input{TFabundanceEstimates.tex}

\clearpage
\bibliographystyle{unsrtnat}
\bibliography{mathematical_estimate.bib}

\end{document}

%% file: OptimalSetOfParametersTableNoAccessibility.tex
\begin{table}[ht]
\centering
\begin{tabular}{|l|r|r|r|r|}
  \hline
 & N & $\lambda$ & $MSE$ & $\rho$ \\ 
  \hline
BCD & 10000 & 1.25 & 5.29 (5.29) & 0.62 (0.62) \\ 
   \hline
CAD & 20000 & 0.25 & 8.82 (8.82) & 0.29 (0.30) \\ 
   \hline
GT & 1e+05 & 5.00 & 0.96 (0.96) & 0.12 (0.31) \\ 
   \hline
HB & 5000 & 2.00 & 2.93 (2.93) & 0.33 (0.38) \\ 
   \hline
KR & 20000 & 4.00 & 6.70 (6.70) & 0.39 (0.41) \\ 
   \hline
\end{tabular}
\caption{\emph{Set of parameters that minimises the difference between the ChIP-seq profile and the analytical model}. We also listed the values for the mean squared error ($MSE$) and correlation ($\rho$). The values in the parentheses represent the minimum mean squared error and the maximum correlation. We considered only the sites that have a PWM score higher than $70\%$ of the distance between the lowest and the highest score.} 
\label{tab:paramsModelNoAcc}
\end{table}

%% file: OptimalSetOfParametersTableAccessibilityRegions.tex
\begin{table}[ht]
\centering
\begin{tabular}{|l|r|r|r|r|}
  \hline
 & N & $\lambda$ & $MSE$ & $\rho$ \\ 
  \hline
BCD & 2000 & 1.25 & 4.40 (4.40) & 0.77 (0.77) \\ 
   \hline
CAD & 10000 & 1.25 & 5.03 (5.03) & 0.73 (0.75) \\ 
   \hline
GT & 1000 & 1.00 & 0.85 (0.85) & 0.55 (0.57) \\ 
   \hline
HB & 2000 & 3.00 & 2.38 (2.38) & 0.66 (0.66) \\ 
   \hline
KR & 20000 & 5.00 & 4.77 (4.77) & 0.68 (0.69) \\ 
   \hline
\end{tabular}
\caption{\emph{Set of parameters that minimises the difference between the ChIP-seq profile and the analytical model which includes DNA accessibility}. The accessibility of any site can be either $0$ or $1$ depending on whether the site is accessible or not. We also listed the values for the mean squared error ($MSE$) and correlation ($\rho$). The values in the parentheses represent the minimum mean squared error and the maximum correlation. We considered only the sites that have a PWM score higher than $70\%$ of the distance between the lowest and the highest score.} 
\label{tab:paramsModelAccRegions}
\end{table}

%% file: OptimalSetOfParametersTableAccessibilityRegionsWeakBinding.tex
\begin{table}[ht]
\centering
\begin{tabular}{|l|r|r|r|r|}
  \hline
 & N & $\lambda$ & $MSE$ & $\rho$ \\ 
  \hline
BCD & 2000 & 1.25 & 4.40 (4.40) & 0.77 (0.77) \\ 
   \hline
CAD & 10000 & 1.25 & 5.03 (5.03) & 0.73 (0.75) \\ 
   \hline
GT & 1000 & 1.00 & 0.85 (0.85) & 0.55 (0.57) \\ 
   \hline
HB & 1000 & 2.50 & 2.40 (2.40) & 0.64 (0.66) \\ 
   \hline
KR & 2000 & 3.00 & 5.46 (5.46) & 0.66 (0.69) \\ 
   \hline
\end{tabular}
\caption{\emph{Set of parameters that minimises the difference between the ChIP-seq profile when including weak binding}. The analytical model includes binary DNA accessibility data (the accessibility of any site can be either $0$ or $1$ depending on whether the site is accessible or not). We also listed the values for the mean squared error ($MSE$) and correlation ($\rho$). The values in the parentheses represent the minimum mean squared error and the maximum correlation. We considered only the sites that have a PWM score higher than $30\%$ of the distance between the lowest and the highest score.} 
\label{tab:paramsModelAccRegions030}
\end{table}

%% file: OptimalSetOfParametersTableAccessibilityProbability.tex
\begin{table}[ht]
\centering
\begin{tabular}{|l|r|r|r|r|}
  \hline
 & N & $\lambda$ & $MSE$ & $\rho$ \\ 
  \hline
BCD & 500 & 1.25 & 4.47 (4.47) & 0.77 (0.78) \\ 
   \hline
CAD & 2000 & 0.75 & 5.35 (5.35) & 0.72 (0.75) \\ 
   \hline
GT & 200 & 1.00 & 0.83 (0.83) & 0.54 (0.57) \\ 
   \hline
HB & 1000 & 3.50 & 2.38 (2.38) & 0.65 (0.66) \\ 
   \hline
KR & 5000 & 5.00 & 4.81 (4.81) & 0.68 (0.69) \\ 
   \hline
\end{tabular}
\caption{\emph{Set of parameters that minimises the difference between the ChIP-seq profile in the case of continous accessibility data}. The analytical model includes continous DNA accessibility data by using equation \eqref{eq:paccessible} to compute the probability that a site is accessible. We also listed the values for the mean squared error ($MSE$) and correlation ($\rho$). The values in the parentheses represent the minimum mean squared error and the maximum correlation. We considered only the sites that have a PWM score higher than $70\%$ of the distance between the lowest and the highest score.} 
\label{tab:paramsModelAccProbability070}
\end{table}

%% file: OptimalSetOfParametersTableAccessibilityRegionsJASPAR.tex
\begin{table}[ht]
\centering
\begin{tabular}{|l|r|r|r|r|}
  \hline
 & N & $\lambda$ & $MSE$ & $\rho$ \\ 
  \hline
BCD & 5000 & 1.00 & 4.06 (4.06) & 0.74 (0.75) \\ 
   \hline
CAD & 10000 & 1.25 & 6.23 (6.23) & 0.64 (0.67) \\ 
   \hline
GT & 10000 & 4.50 & 0.86 (0.86) & 0.54 (0.56) \\ 
   \hline
HB & 2000 & 1.00 & 2.56 (2.56) & 0.68 (0.68) \\ 
   \hline
KR & 5000 & 2.00 & 4.73 (4.73) & 0.67 (0.69) \\ 
   \hline
\end{tabular}
\caption{\emph{Set of parameters that minimises the difference between the ChIP-seq profiles when using the binding motif from JASPAR database} \citep{portales-casamar_2010}; see Figure \ref{fig:PWMmotifsJaspar}. Our model assumes binary DNA accessibility data (the accessibility of any site can be either $0$ or $1$ depending on whether the site is accessible or not). We also listed the values for the mean squared error ($MSE$) and correlation ($\rho$). The values in the parentheses represent the minimum mean squared error and the maximum correlation. We considered only the sites that have a PWM score higher than $70\%$ of the distance between the lowest and the highest score.} 
\label{tab:paramsModelAccRegionsJaspar070}
\end{table}

%% file: ChIPseqProfileStatistics.tex
\begin{table}[ht]
\centering
\begin{tabular}{|l|r|r|r|r|r|}
  \hline
 & BCD & CAD & GT & HB & KR \\ 
  \hline
$B$ & 0.03 & 0.03 & 0.04 & 0.04 & 0.04 \\ 
   \hline
$M$ & 1.85 & 1.63 & 4.97 & 16.23 & 5.23 \\ 
   \hline
\end{tabular}
\caption{\emph{Mean and maximum of the ChIP-seq signal for the five TFs}.} 
\label{tab:ChIPseqProfileStatistics}
\end{table}

%% file: GenomeWideNoOfRegions.tex
\begin{table}[ht]
\centering
\begin{tabular}{|r|r|r|r|r|r|}
  \hline
 & BCD & CAD & GT & HB & KR \\ 
  \hline
1.0 & 35 & 167 & 1 & 2 & 8 \\ 
   \hline
0.5 & 812 & 1984 & 32 & 6 & 83 \\ 
   \hline
\end{tabular}
\caption{\emph{The number of DNA segments with a ChIP-seq signal higher than the threshold}. We considered two thresholds $K=1.0$ and $K=0.5$. The mean ChIP-seq signal of the $20\ Kbp$ segment needs to be higher than $K\times B$.} 
\label{tab:GenomeWideNoOfRegions}
\end{table}

%% file: TFabundanceEstimates.tex
\begin{table}[ht]
\centering
\begin{tabular}{|l|r|r|r|r|}
  \hline
 & min abundance & max abundance & min specifically bound & max specifically bound \\ 
  \hline
BCD & 10000 &  30000 & 7\% & 20\% \\ 
   \hline
CAD & 12000 &  37000 & 27\% & 81\% \\ 
   \hline
GT & 23000 &  70000 & 1\% & 4\% \\ 
   \hline
HB & 11000 &  34000 & 6\% & 18\% \\ 
   \hline
KR & 37000 & 110000 & 18\% & 54\% \\ 
   \hline
\end{tabular}
\caption{\emph{The TF abundance in the nucleus and the percentage of specifically bound TF}. In the first two columns, we list the number of molecules that are in the nucleus for the five TFs. For our estimates for Bicoid nuclear abundance, see the Discussion section in the main text. In \cite{zamparo_2009}, the authors reanalysed the FlyEx data \cite{pisarev_2009} and proposed a lower limit for the nuclear abundance of the five TFs, but the proposed values are underestimates of the real values. For the last four TFs (Caudal, Giant, Hunchback and Kruppel), we considered the nuclear abundances of the four TFs relative to Bicoid nuclear abundance, as estimated in \cite{zamparo_2009} using the Poisson method, and then we multiplied these relative abundances with our estimates for the Bicoid nuclear abundance. In the last two columns we list the percentage of specifically bound TFs, based on the estimations of our method (Table 2 in the main text) and the values in the first two columns.} 
\label{tab:TFabundanceEstimates}
\end{table}

%% file: mathematical_estimate.bbl
\begin{thebibliography}{81}
\providecommand{\natexlab}[1]{#1}
\providecommand{\url}[1]{\texttt{#1}}
\expandafter\ifx\csname urlstyle\endcsname\relax
  \providecommand{\doi}[1]{doi: #1}\else
  \providecommand{\doi}{doi: \begingroup \urlstyle{rm}\Url}\fi

\bibitem[Park(2009)]{park_2009}
Peter~J. Park.
\newblock Chip-seq: advantages and challenges of a maturing technology.
\newblock \emph{Nature Reviews Genetics}, 10\penalty0 (10):\penalty0 669--680,
  2009.
\newblock \doi{10.1038/nrg2641}.

\bibitem[Wasserman and Sandelin(2004)]{wasserman_2004}
Wyeth~W. Wasserman and Albin Sandelin.
\newblock Applied bioinformatics for the identification of regulatory elements.
\newblock \emph{Nature Reviews Genetics}, 5:\penalty0 276--287, 2004.
\newblock \doi{10.1038/nrg1315}.

\bibitem[Hermsen et~al.(2006)Hermsen, Tans, and ten Wolde]{hermsen_2006}
Rutger Hermsen, Sander Tans, and Pieter~Rein ten Wolde.
\newblock Transcriptional regulation by competing transcription factor modules.
\newblock \emph{PLoS Comput Biol}, 2\penalty0 (12):\penalty0 1552--1560, 2006.
\newblock \doi{10.1371/journal.pcbi.0020164}.

\bibitem[Hoffman and Birney(2010)]{hoffman_2010}
Michael~M. Hoffman and Ewan Birney.
\newblock An effective model for natural selection in promoters.
\newblock \emph{Genome Resarch}, 20\penalty0 (5):\penalty0 685--692, 2010.
\newblock \doi{10.1101/gr.096719.109}.

\bibitem[Sheinman and Kafri(2012)]{sheinman_2012}
M~Sheinman and Y~Kafri.
\newblock How does the {DNA} sequence affect the {Hill} curve of
  transcriptional response?
\newblock \emph{Physical Biology}, 9\penalty0 (5):\penalty0 056006, 2012.
\newblock \doi{10.1088/1478-3975/9/5/056006}.

\bibitem[Djordjevic et~al.(2003)Djordjevic, Sengupta, and
  Shraiman]{djordjevic_2003}
Marko Djordjevic, Anirvan~M. Sengupta, and Boris~I. Shraiman.
\newblock A biophysical approach to transcription factor binding site
  discovery.
\newblock \emph{Genome Resarch}, 13\penalty0 (11):\penalty0 2381--2390, 2003.
\newblock \doi{10.1101/gr.1271603}.

\bibitem[Foat et~al.(2006)Foat, Morozov, and Bussemaker]{foat_2006}
Barrett~C. Foat, Alexandre~V. Morozov, and Harmen~J. Bussemaker.
\newblock Statistical mechanical modeling of genome-wide transcription factor
  occupancy data by {MatrixREDUCE}.
\newblock \emph{Bioinformatics}, 22\penalty0 (14):\penalty0 e141--e149, 2006.
\newblock \doi{10.1093/bioinformatics/btl223}.

\bibitem[Roider et~al.(2007)Roider, Kanhere, Manke, and Vingron]{roider_2007}
Helge~G. Roider, Aditi Kanhere, Thomas Manke, and Martin Vingron.
\newblock Predicting transcription factor affinities to {DNA} from a
  biophysical model.
\newblock \emph{Bioinformatics}, 23\penalty0 (2):\penalty0 134--141, 2007.
\newblock \doi{10.1093/bioinformatics/btl565}.

\bibitem[Zhao et~al.(2009)Zhao, Granas, and Stormo]{zhao_2009}
Yue Zhao, David Granas, and Gary~D. Stormo.
\newblock Inferring binding energies from selected binding sites.
\newblock \emph{PLoS Computational Biology}, 5\penalty0 (12):\penalty0
  e1000590, 2009.
\newblock \doi{10.1371/journal.pcbi.1000590}.

\bibitem[Simicevic et~al.(2013)Simicevic, Schmid, Gilardoni, Zoller, Raghav,
  Krier, Gubelmann, Lisacek, Naef, Moniatte, and Deplancke]{simicevic_2013}
Jovan Simicevic, Adrien~W Schmid, Paola~A Gilardoni, Benjamin Zoller, Sunil~K
  Raghav, Irina Krier, Carine Gubelmann, Frederique Lisacek, Felix Naef, Marc
  Moniatte, and Bart Deplancke.
\newblock Absolute quantification of transcription factors during cellular
  differentiation using multiplexed targeted proteomics.
\newblock \emph{Nature Methods}, 10\penalty0 (6):\penalty0 570--576, 2013.
\newblock \doi{10.1038/nmeth.2441}.

\bibitem[Zabet et~al.(2013)Zabet, Foy, and Adryan]{zabet_2013_occupancy}
Nicolae~Radu Zabet, Robert Foy, and Boris Adryan.
\newblock The influence of transcription factor competition on the relationship
  between occupancy and affinity.
\newblock \emph{PLoS ONE}, 8\penalty0 (9):\penalty0 e73714, 2013.
\newblock \doi{10.1371/journal.pone.0073714}.

\bibitem[Ilsley et~al.(2013)Ilsley, Fisher, Apweiler, DePace, and
  Luscombe]{ilsley_2013}
Garth~R Ilsley, Jasmin Fisher, Rolf Apweiler, Angela~H DePace, and Nicholas~M
  Luscombe.
\newblock Cellular resolution models for even skipped regulation in the entire
  \emph{Drosophila} embryo.
\newblock \emph{eLife}, 2, 2013.
\newblock \doi{10.7554/eLife.00522}.

\bibitem[Raveh-Sadka et~al.(2009)Raveh-Sadka, Levo, and Segal]{sadka_2009}
Tali Raveh-Sadka, Michal Levo, and Eran Segal.
\newblock Incorporating nucleosomes into thermodynamic models of transcription
  regulation.
\newblock \emph{Genome Research}, 19\penalty0 (8):\penalty0 1480--1496, 2009.
\newblock \doi{10.1101/gr.088260.108}.

\bibitem[He et~al.(2009)He, Chen, Hong, Fang, Sinha, Ng, and Zhong]{he_2009}
Xin He, Chieh-Chun Chen, Feng Hong, Fang Fang, Saurabh Sinha, Huck-Hui Ng, and
  Sheng Zhong.
\newblock A biophysical model for analysis of transcription factor interaction
  and binding site arrangement from genome-wide binding data.
\newblock \emph{PLoS ONE}, 4\penalty0 (12):\penalty0 e8155, 2009.
\newblock \doi{10.1371/journal.pone.0008155}.

\bibitem[Wasson and Hartemink(2009)]{wasson_2009}
Todd Wasson and Alexander~J. Hartemink.
\newblock An ensemble model of competitive multi-factor binding of the genome.
\newblock \emph{Genome Research}, 19:\penalty0 2101--2112, 2009.
\newblock \doi{10.1101/gr.093450.109}.

\bibitem[Kaplan et~al.(2011)Kaplan, Li, Sabo, Thomas, Stamatoyannopoulos,
  Biggin, and Eisen]{kaplan_2011}
Tommy Kaplan, Xiao-Yong Li, Peter~J. Sabo, Sean Thomas, John~A.
  Stamatoyannopoulos, Mark~D. Biggin, and Michael~B. Eisen.
\newblock Quantitative models of the mechanisms that control genome-wide
  patterns of transcription factor binding during early \emph{Drosophila}
  development.
\newblock \emph{PLoS Genetics}, 7\penalty0 (2):\penalty0 e1001290, 2011.
\newblock \doi{10.1371/journal.pgen.1001290}.

\bibitem[Cheng et~al.(2013)Cheng, Kazemian, Pham, Blatti, Celniker, Wolfe,
  Brodsky, and Sinha]{cheng_2013}
Qiong Cheng, Majid Kazemian, Hannah Pham, Charles Blatti, Susan~E. Celniker,
  Scot~A. Wolfe, Michael~H. Brodsky, and Saurabh Sinha.
\newblock Computational identification of diverse mechanisms underlying
  transcription factor-{DNA} occupancy.
\newblock \emph{PLoS Genetics}, 9\penalty0 (8):\penalty0 e1003571, 2013.
\newblock \doi{10.1371%2Fjournal.pgen.1003571}.

\bibitem[Ackers et~al.(1982)Ackers, Johnson, and Shea]{ackers_1982}
Gary~K. Ackers, Alexander~D. Johnson, and Madeline~A. Shea.
\newblock Quantitative model for gene regulation by lambda phage repressor.
\newblock \emph{PNAS}, 79:\penalty0 1129--1133, 1982.

\bibitem[Bintu et~al.(2005{\natexlab{a}})Bintu, Buchler, Garcia, Gerland, Hwa,
  Kondev, and Phillips]{bintu_2005_model}
Lacramioara Bintu, Nicolas~E. Buchler, Hernan~G. Garcia, Ulrich Gerland,
  Terence Hwa, Jane Kondev, and Rob Phillips.
\newblock Transcriptional regulation by the numbers: models.
\newblock \emph{Current Opinion in Genetics \& Development}, 15\penalty0
  (2):\penalty0 116--124, 2005{\natexlab{a}}.
\newblock \doi{10.1016/j.gde.2005.02.007}.

\bibitem[Chu et~al.(2009)Chu, Zabet, and Mitavskiy]{chu_2009}
Dominique Chu, Nicolae~Radu Zabet, and Boris Mitavskiy.
\newblock Models of transcription factor binding: Sensitivity of activation
  functions to model assumptions.
\newblock \emph{Journal of Theoretical Biology}, 257\penalty0 (3):\penalty0
  419--429, 2009.
\newblock \doi{10.1016/j.jtbi.2008.11.026}.

\bibitem[Berg and von Hippel(1987)]{berg_1987}
Otto~G. Berg and Peter~H. von Hippel.
\newblock Selection of {DNA} binding sites by regulatory proteins
  statistical-mechanical theory and application to operators and promoters.
\newblock \emph{Journal of Molecular Biology}, 193\penalty0 (4):\penalty0
  723--750, 1987.
\newblock \doi{10.1016/0022-2836(87)90354-8}.

\bibitem[Stormo and Zhao(2010)]{stormo_2010}
Gary~D. Stormo and Yue Zhao.
\newblock Determining the specificity of protein-{DNA} interactions.
\newblock \emph{Nature Reviews Genetics}, 11\penalty0 (11):\penalty0 751--760,
  2010.
\newblock \doi{10.1038/nrg2845}.

\bibitem[Bradley et~al.(2010)Bradley, Li, Trapnell, Davidson, Pachter, Chu,
  Tonkin, Biggin, and Eisen]{bradley_2010}
Robert~K. Bradley, Xiao-Yong Li, Cole Trapnell, Stuart Davidson, Lior Pachter,
  Hou~Cheng Chu, Leath~A. Tonkin, Mark~D. Biggin, and Michael~B. Eisen.
\newblock Binding site turnover produces pervasive quantitative changes in
  transcription factor binding between closely related \emph{Drosophila}
  species.
\newblock \emph{PLoS Biology}, 8:\penalty0 e1000343, 2010.
\newblock \doi{10.1371/journal.pbio.1000343}.

\bibitem[Li et~al.(2011)Li, Thomas, Sabo, Eisen, Stamatoyannopoulos, and
  Biggin]{li_2011}
Xiao-Yong Li, Sean Thomas, Peter Sabo, Michael Eisen, John Stamatoyannopoulos,
  and Mark Biggin.
\newblock The role of chromatin accessibility in directing the widespread,
  overlapping patterns of drosophila transcription factor binding.
\newblock \emph{Genome Biology}, 12\penalty0 (4):\penalty0 R34, 2011.
\newblock \doi{10.1186/gb-2011-12-4-r34}.

\bibitem[Riggs et~al.(1970)Riggs, Bourgeois, and Cohn]{riggs_1970a}
Arthur~D. Riggs, Suzanne Bourgeois, and Melvin Cohn.
\newblock The lac represser-operator interaction: {III.} kinetic studies.
\newblock \emph{Journal of Molecular Biology}, 53\penalty0 (3):\penalty0
  401--417, 1970.
\newblock \doi{10.1016/0022-2836(70)90074-4}.

\bibitem[Berg et~al.(1981)Berg, Winter, and von Hippel]{berg_1981}
Otto~G. Berg, Robert~B. Winter, and Peter~H. von Hippel.
\newblock Diffusion-driven mechanisms of protein translocation on nucleic
  acids. 1. models and theory.
\newblock \emph{Biochemistry}, 20\penalty0 (24):\penalty0 6929--6948, 1981.
\newblock \doi{10.1021/bi00527a028}.

\bibitem[Kabata et~al.(1993)Kabata, Kurosawa, I~Arai, Margarson, Glass, and
  Shimamoto]{kabata_1993}
H~Kabata, O~Kurosawa, M~Washizu I~Arai, SA~Margarson, RE~Glass, and
  N~Shimamoto.
\newblock Visualization of single molecules of rna polymerase sliding along
  dna.
\newblock \emph{Science}, 262\penalty0 (5139):\penalty0 1561--1563, 1993.
\newblock \doi{10.1126/science.8248804}.

\bibitem[Blainey et~al.(2006)Blainey, van Oijen, Banerjee, Verdine, and
  Xie]{blainey_2006}
Paul~C. Blainey, Antoine~M. van Oijen, Anirban Banerjee, Gregory~L. Verdine,
  and X.~Sunney Xie.
\newblock A base-excision dna-repair protein finds intrahelical lesion bases by
  fast sliding in contact with dna.
\newblock \emph{PNAS}, 103\penalty0 (15):\penalty0 5752--5757, 2006.
\newblock \doi{10.1073/pnas.0509723103}.

\bibitem[Elf et~al.(2007)Elf, Li, and Xie]{elf_2007}
Johan Elf, Gene-Wei Li, and X.~Sunney Xie.
\newblock Probing transcription factor dynamics at the single-molecule level in
  a living cell.
\newblock \emph{Science}, 316\penalty0 (5828):\penalty0 1191--1194, 2007.
\newblock \doi{10.1126/science.114196}.

\bibitem[Mirny et~al.(2009)Mirny, Slutsky, Wunderlich, Tafvizi, Leith, and
  Kosmrlj]{mirny_2009}
Leonid Mirny, Michael Slutsky, Zeba Wunderlich, Anahita Tafvizi, Jason Leith,
  and Andrej Kosmrlj.
\newblock How a protein searches for its site on {DNA}: the mechanism of
  facilitated diffusion.
\newblock \emph{Journal of Physics A: Mathematical and Theoretical},
  42\penalty0 (43):\penalty0 434013, 2009.
\newblock \doi{10.1088/1751-8113/42/43/434013}.

\bibitem[Hager et~al.(2009)Hager, McNally, and Misteli]{hager_2009}
Gordon~L. Hager, James~G. McNally, and Tom Misteli.
\newblock Transcription dynamics.
\newblock \emph{Molecular Cell}, 35\penalty0 (6):\penalty0 741--753, 2009.
\newblock \doi{10.1016/j.molcel.2009.09.005}.

\bibitem[Vukojevic et~al.(2010)Vukojevic, Papadopoulos, Terenius, Gehring, and
  Rigler]{vukojevic_2010}
Vladana Vukojevic, Dimitrios~K. Papadopoulos, Lars Terenius, Walter~J. Gehring,
  and Rudolf Rigler.
\newblock Quantitative study of synthetic hox transcription factor-dna
  interactions in live cells.
\newblock \emph{PNAS}, 107\penalty0 (9):\penalty0 4093--4098, 2010.
\newblock \doi{10.1073/pnas.0914612107}.

\bibitem[Hammar et~al.(2012)Hammar, Leroy, Mahmutovic, Marklund, Berg, and
  Elf]{hammar_2012}
Petter Hammar, Prune Leroy, Anel Mahmutovic, Erik~G. Marklund, Otto~G. Berg,
  and Johan Elf.
\newblock The lac repressor displays facilitated diffusion in living cells.
\newblock \emph{Science}, 336\penalty0 (6088):\penalty0 1595--1598, 2012.
\newblock \doi{10.1126/science.1221648}.

\bibitem[Zabet and Adryan(2012{\natexlab{a}})]{zabet_2012_model}
Nicolae~Radu Zabet and Boris Adryan.
\newblock A comprehensive computational model of facilitated diffusion in
  prokaryotes.
\newblock \emph{Bioinformatics}, 28\penalty0 (11):\penalty0 1517--1524,
  2012{\natexlab{a}}.
\newblock \doi{10.1093/bioinformatics/bts178}.

\bibitem[Zabet and Adryan(2012{\natexlab{b}})]{zabet_2012_review}
Nicolae~Radu Zabet and Boris Adryan.
\newblock Computational models for large-scale simulations of facilitated
  diffusion.
\newblock \emph{Molecular BioSystems}, 8\penalty0 (11):\penalty0 2815--2827,
  2012{\natexlab{b}}.
\newblock \doi{10.1039/C2MB25201E}.

\bibitem[Thomas et~al.(2011)Thomas, Li, Sabo, Sandstrom, Thurman, Canfield,
  Giste, Fisher, Hammonds, Celniker, Biggin, and
  Stamatoyannopoulos]{thomas_2011}
Sean Thomas, Xiao-Yong Li, Peter~J Sabo, Richard Sandstrom, Robert~E Thurman,
  Theresa~K Canfield, Erika Giste, William Fisher, Ann Hammonds, Susan~E
  Celniker, Mark~D Biggin, and John~A Stamatoyannopoulos.
\newblock Dynamic reprogramming of chromatin accessibility during
  \emph{Drosophila} embryo development.
\newblock \emph{Genome Biology}, 12\penalty0 (5):\penalty0 R43, 2011.
\newblock \doi{10.1186/gb-2011-12-5-r43}.

\bibitem[Adams et~al.(2000)Adams, Celniker, Holt, Evans, Gocayne, Amanatides,
  Scherer, Li, Hoskins, Galle, George, Lewis, Richards, Ashburner, Henderson,
  Sutton, Wortman, Yandell, Zhang, Chen, Brandon, Rogers, Blazej, Champe,
  Pfeiffer, Wan, Doyle, Baxter, Helt, Nelson, Gabor, Miklos, Abril, Agbayani,
  An, Andrews-Pfannkoch, Baldwin, Ballew, Basu, Baxendale, Bayraktaroglu,
  Beasley, Beeson, Benos, Berman, Bhandari, Bolshakov, Borkova, Botchan, Bouck,
  Brokstein, Brottier, Burtis, Busam, Butler, Cadieu, Center, Chandra, Cherry,
  Cawley, Dahlke, Davenport, Davies, Pablos, Delcher, Deng, Mays, Dew, Dietz,
  Dodson, Doup, Downes, Dugan-Rocha, Dunkov, Dunn, Durbin, Evangelista, Ferraz,
  Ferriera, Fleischmann, Fosler, Gabrielian, Garg, Gelbart, Glasser, Glodek,
  Gong, Gorrell, Gu, Guan, Harris, Harris, Harvey, Heiman, Hernandez, Houck,
  Hostin, Houston, Howland, Wei, Ibegwam, Jalali, Kalush, Karpen, Ke, Kennison,
  Ketchum, Kimmel, Kodira, Kraft, Kravitz, Kulp, Lai, Lasko, Lei, Levitsky, Li,
  Li, Liang, Lin, Liu, Mattei, McIntosh, McLeod, McPherson, Merkulov, Milshina,
  Mobarry, Morris, Moshrefi, Mount, Moy, Murphy, Murphy, Muzny, Nelson, Nelson,
  Nelson, Nixon, Nusskern, Pacleb, Palazzolo, Pittman, Pan, Pollard, Puri,
  Reese, Reinert, Remington, Saunders, Scheeler, Shen, Shue, Sidén-Kiamos,
  Simpson, Skupski, Smith, Spier, Spradling, Stapleton, Strong, Sun, Svirskas,
  Tector, Turner, Venter, Wang, Wang, Wang, Wassarman, Weinstock, Weissenbach,
  Williams, Woodage, Worley, Wu, Yang, Yao, Ye, Yeh, Zaveri, Zhan, Zhang, Zhao,
  Zheng, Zheng, Zhong, Zhong, Zhou, Zhu, Zhu, Smith, Gibbs, Myers, Rubin, and
  Venter]{adams_2000}
Mark~D. Adams, Susan~E. Celniker, Robert~A. Holt, Cheryl~A. Evans, Jeannine~D.
  Gocayne, Peter~G. Amanatides, Steven~E. Scherer, Peter~W. Li, Roger~A.
  Hoskins, Richard~F. Galle, Reed~A. George, Suzanna~E. Lewis, Stephen
  Richards, Michael Ashburner, Scott~N. Henderson, Granger~G. Sutton,
  Jennifer~R. Wortman, Mark~D. Yandell, Qing Zhang, Lin~X. Chen, Rhonda~C.
  Brandon, Yu-Hui~C. Rogers, Robert~G. Blazej, Mark Champe, Barret~D. Pfeiffer,
  Kenneth~H. Wan, Clare Doyle, Evan~G. Baxter, Gregg Helt, Catherine~R. Nelson,
  George~L. Gabor, Miklos, Josep~F. Abril, Anna Agbayani, Hui-Jin An, Cynthia
  Andrews-Pfannkoch, Danita Baldwin, Richard~M. Ballew, Anand Basu, James
  Baxendale, Leyla Bayraktaroglu, Ellen~M. Beasley, Karen~Y. Beeson, P.~V.
  Benos, Benjamin~P. Berman, Deepali Bhandari, Slava Bolshakov, Dana Borkova,
  Michael~R. Botchan, John Bouck, Peter Brokstein, Phillipe Brottier,
  Kenneth~C. Burtis, Dana~A. Busam, Heather Butler, Edouard Cadieu, Angela
  Center, Ishwar Chandra, J.~Michael Cherry, Simon Cawley, Carl Dahlke,
  Lionel~B. Davenport, Peter Davies, Beatriz~de Pablos, Arthur Delcher, Zuoming
  Deng, Anne~Deslattes Mays, Ian Dew, Suzanne~M. Dietz, Kristina Dodson,
  Lisa~E. Doup, Michael Downes, Shannon Dugan-Rocha, Boris~C. Dunkov, Patrick
  Dunn, Kenneth~J. Durbin, Carlos~C. Evangelista, Concepcion Ferraz, Steven
  Ferriera, Wolfgang Fleischmann, Carl Fosler, Andrei~E. Gabrielian, Neha~S.
  Garg, William~M. Gelbart, Ken Glasser, Anna Glodek, Fangcheng Gong, J.~Harley
  Gorrell, Zhiping Gu, Ping Guan, Michael Harris, Nomi~L. Harris, Damon Harvey,
  Thomas~J. Heiman, Judith~R. Hernandez, Jarrett Houck, Damon Hostin,
  Kathryn~A. Houston, Timothy~J. Howland, Ming-Hui Wei, Chinyere Ibegwam, Mena
  Jalali, Francis Kalush, Gary~H. Karpen, Zhaoxi Ke, James~A. Kennison,
  Karen~A. Ketchum, Bruce~E. Kimmel, Chinnappa~D. Kodira, Cheryl Kraft, Saul
  Kravitz, David Kulp, Zhongwu Lai, Paul Lasko, Yiding Lei, Alexander~A.
  Levitsky, Jiayin Li, Zhenya Li, Yong Liang, Xiaoying Lin, Xiangjun Liu,
  Bettina Mattei, Tina~C. McIntosh, Michael~P. McLeod, Duncan McPherson,
  Gennady Merkulov, Natalia~V. Milshina, Clark Mobarry, Joe Morris, Ali
  Moshrefi, Stephen~M. Mount, Mee Moy, Brian Murphy, Lee Murphy, Donna~M.
  Muzny, David~L. Nelson, David~R. Nelson, Keith~A. Nelson, Katherine Nixon,
  Deborah~R. Nusskern, Joanne~M. Pacleb, Michael Palazzolo, Gjange~S. Pittman,
  Sue Pan, John Pollard, Vinita Puri, Martin~G. Reese, Knut Reinert, Karin
  Remington, Robert D.~C. Saunders, Frederick Scheeler, Hua Shen,
  Bixiang~Christopher Shue, Inga Sidén-Kiamos, Michael Simpson, Marian~P.
  Skupski, Tom Smith, Eugene Spier, Allan~C. Spradling, Mark Stapleton, Renee
  Strong, Eric Sun, Robert Svirskas, Cyndee Tector, Russell Turner, Eli Venter,
  Aihui~H. Wang, Xin Wang, Zhen-Yuan Wang, David~A. Wassarman, George~M.
  Weinstock, Jean Weissenbach, Sherita~M. Williams, Trevor Woodage, Kim~C.
  Worley, David Wu, Song Yang, Q.~Alison Yao, Jane Ye, Ru-Fang Yeh, Jayshree~S.
  Zaveri, Ming Zhan, Guangren Zhang, Qi~Zhao, Liansheng Zheng, Xiangqun~H.
  Zheng, Fei~N. Zhong, Wenyan Zhong, Xiaojun Zhou, Shiaoping Zhu, Xiaohong Zhu,
  Hamilton~O. Smith, Richard~A. Gibbs, Eugene~W. Myers, Gerald~M. Rubin, and
  J.~Craig Venter.
\newblock The genome sequence of drosophila melanogaster.
\newblock \emph{Science}, 287\penalty0 (5461):\penalty0 2185--2195, 2000.
\newblock \doi{10.1126/science.287.5461.2185}.

\bibitem[Mueller et~al.(2013)Mueller, Stasevich, Mazza, and
  McNally]{mueller_2013}
Florian Mueller, Timothy~J. Stasevich, Davide Mazza, and James~G. McNally.
\newblock Quantifying transcription factor kinetics: At work or at play?
\newblock \emph{Crit Rev Biochem Mol Biol.}, 48\penalty0 (5):\penalty0
  492--514, 2013.
\newblock \doi{10.3109/10409238.2013.833891}.

\bibitem[Portales-Casamar et~al.(2010)Portales-Casamar, Thongjuea, Kwon,
  Arenillas, Zhao, Valen, Yusuf, Lenhard, Wasserman, and
  Sandelin]{portales-casamar_2010}
Elodie Portales-Casamar, Supat Thongjuea, Andrew~T. Kwon, David Arenillas,
  Xiaobei Zhao, Eivind Valen, Dimas Yusuf, Boris Lenhard, Wyeth~W. Wasserman,
  and Albin Sandelin.
\newblock {JASPAR} 2010: the greatly expanded open-access database of
  transcription factor binding profiles.
\newblock \emph{Nucleic Acids Research}, 38\penalty0 (suppl 1):\penalty0
  D105--D110, 2010.
\newblock \doi{10.1093/nar/gkp950}.

\bibitem[Stanojevic et~al.(1991)Stanojevic, Small, and Levine]{stanojevic_1991}
D~Stanojevic, S~Small, and M~Levine.
\newblock Regulation of a segmentation stripe by overlapping activators and
  repressors in the \emph{Drosophila} embryo.
\newblock \emph{Science}, 254\penalty0 (5036):\penalty0 1385--1387, 1991.
\newblock \doi{10.1126/science.1683715}.

\bibitem[Gregor et~al.(2007{\natexlab{a}})Gregor, Tank, Wieschaus, and
  Bialek]{gregor_2007}
Thomas Gregor, David~W. Tank, Eric~F. Wieschaus, and William Bialek.
\newblock Probing the limits to positional information.
\newblock \emph{Cell}, 130\penalty0 (1):\penalty0 153--164, 2007{\natexlab{a}}.
\newblock \doi{10.1016/j.cell.2007.05.025}.

\bibitem[Abu-Arish et~al.(2010)Abu-Arish, Porcher, Czerwonka, Dostatni, and
  Fradin]{abuarish_2010}
Asmahan Abu-Arish, Aude Porcher, Anna Czerwonka, Nathalie Dostatni, and Cecile
  Fradin.
\newblock High mobility of bicoid captured by fluorescence correlation
  spectroscopy: Implication for the rapid establishment of its gradient.
\newblock \emph{Biophysical Journal}, 99\penalty0 (4):\penalty0 L33 -- L35,
  2010.
\newblock \doi{10.1016/j.bpj.2010.05.031}.

\bibitem[Grimm and Wieschaus(2010)]{grimm_2010}
Oliver Grimm and Eric Wieschaus.
\newblock The bicoid gradient is shaped independently of nuclei.
\newblock \emph{Development}, 137\penalty0 (17):\penalty0 2857--2862, 2010.
\newblock \doi{10.1242/dev.052589}.

\bibitem[Drocco et~al.(2011)Drocco, Grimm, Tank, and Wieschaus]{drocco_2011}
Jeffrey~A. Drocco, Oliver Grimm, David~W. Tank, and Eric Wieschaus.
\newblock Measurement and perturbation of morphogen lifetime: Effects on
  gradient shape.
\newblock \emph{Biophysical Journal}, 101\penalty0 (8):\penalty0 1807--1815,
  2011.
\newblock \doi{10.1016/j.bpj.2011.07.025}.

\bibitem[Little et~al.(2011)Little, Tka\v{c}ik, Kneeland, Wieschaus, and
  Gregor]{little_2011}
Shawn~C. Little, Ga\v{s}per Tka\v{c}ik, Thomas~B. Kneeland, Eric~F. Wieschaus,
  and Thomas Gregor.
\newblock The formation of the bicoid morphogen gradient requires protein
  movement from anteriorly localized {mRNA}.
\newblock \emph{PLoS Biology}, 9\penalty0 (3):\penalty0 e1000596, 2011.
\newblock \doi{10.1371/journal.pbio.1000596}.

\bibitem[Drocco et~al.(2012)Drocco, Wieschaus, and Tank]{drocco_2012}
Jeffrey~A. Drocco, Eric Wieschaus, and David~W. Tank.
\newblock The synthesis-diffusion-degradation model explains bicoid gradient
  formation in unfertilized eggs.
\newblock \emph{Physical Biology}, 9\penalty0 (5):\penalty0 055004, 2012.
\newblock \doi{10.1088/1478-3975/9/5/055004}.

\bibitem[Small et~al.(1991)Small, Kraut, Hoey, Warrior, and Levine]{small_1991}
Stephen Small, Rachel Kraut, Timothy Hoey, Rahul Warrior, and Michael Levine.
\newblock Transcriptional regulation of a pair-rule stripe in
  \emph{Drosophila}.
\newblock \emph{Genes \& Development}, 5\penalty0 (5):\penalty0 827--839, 1991.
\newblock \doi{10.1101/gad.5.5.827}.

\bibitem[Jaeger et~al.(2012)Jaeger, Manu, and Reinitz]{jaeger_2012}
Johannes Jaeger, Manu, and John Reinitz.
\newblock \emph{Drosophila} blastoderm patterning.
\newblock \emph{Current Opinion in Genetics \& Development}, 22\penalty0
  (6):\penalty0 533--541, 2012.
\newblock \doi{10.1016/j.gde.2012.10.005}.

\bibitem[Gregor et~al.(2007{\natexlab{b}})Gregor, Wieschaus, McGregor, Bialek,
  and Tank]{gregor_2007_gradient}
Thomas Gregor, Eric~F. Wieschaus, Alistair~P. McGregor, William Bialek, and
  David~W. Tank.
\newblock Stability and nuclear dynamics of the bicoid morphogen gradient.
\newblock \emph{Cell}, 130\penalty0 (1):\penalty0 141--152, 2007{\natexlab{b}}.
\newblock \doi{10.1016/j.cell.2007.05.026}.

\bibitem[Leith et~al.(2012)Leith, Tafvizi, Huang, Uspal, Doyle, Fersht, Mirny,
  and van Oijen]{leith_2012}
Jason~S. Leith, Anahita Tafvizi, Fang Huang, William~E. Uspal, Patrick~S.
  Doyle, Alan~R. Fersht, Leonid~A. Mirny, and Antoine~M. van Oijen.
\newblock Sequence-dependent sliding kinetics of p53.
\newblock \emph{PNAS}, 109\penalty0 (41):\penalty0 16552--16557, 2012.
\newblock \doi{10.1073/pnas.1120452109}.

\bibitem[Chen et~al.(2014)Chen, Zhang, Li, Chen, Revyakin, Hajj, Legant, Dahan,
  Lionnet, Betzig, Tjian, and Liu]{chen_2014}
Jiji Chen, Zhengjian Zhang, Li~Li, Bi-Chang Chen, Andrey Revyakin, Bassam Hajj,
  Wesley Legant, Maxime Dahan, Timoth\'{e}e Lionnet, Eric Betzig, Robert Tjian,
  and Zhe Liu.
\newblock Single-molecule dynamics of enhanceosome assembly in embryonic stem
  cells.
\newblock \emph{Cell}, 156\penalty0 (6):\penalty0 1274--1285, 2014.
\newblock \doi{10.1016/j.cell.2014.01.062}.

\bibitem[Carr and Biggin(1999)]{carr_1999}
Alan Carr and Mark~D. Biggin.
\newblock A comparison of in vivo and in vitro {DNA}-binding specificities
  suggests a new model for homeoprotein {DNA} binding in \emph{Drosophila}
  embryos.
\newblock \emph{The EMBO Journal}, 18\penalty0 (6):\penalty0 1598--1608, 1999.
\newblock \doi{10.1093/emboj/18.6.1598}.

\bibitem[Toth and Biggin(2000)]{toth_2000}
Joseph Toth and Mark~D. Biggin.
\newblock The specificity of protein–{DNA} crosslinking by formaldehyde:
  \emph{in vitro} and in \emph{Drosophila} embryos.
\newblock \emph{Nucleic Acids Research}, 28\penalty0 (2):\penalty0 e4, 2000.
\newblock \doi{10.1093/nar/28.2.e4}.

\bibitem[Rhee and Pugh(2011)]{rhee_2011}
Ho~Sung Rhee and B.~Franklin Pugh.
\newblock Comprehensive genome-wide protein-dna interactions detected at
  single-nucleotide resolution.
\newblock \emph{Cell}, 147\penalty0 (6):\penalty0 1408--1419, 2011.
\newblock \doi{10.1016/j.cell.2011.11.013}.

\bibitem[Poorey et~al.(2013)Poorey, Viswanathan, Carver, Karpova, Cirimotich,
  McNally, Bekiranov, and Auble]{poorey_2013}
Kunal Poorey, Ramya Viswanathan, Melissa~N. Carver, Tatiana~S. Karpova,
  Shana~M. Cirimotich, James~G. McNally, Stefan Bekiranov, and David~T. Auble.
\newblock Measuring chromatin interaction dynamics on the second time scale at
  single-copy genes.
\newblock \emph{Science}, 342\penalty0 (6156):\penalty0 369--372, 2013.
\newblock \doi{10.1126/science.1242369}.

\bibitem[Phair et~al.(2004)Phair, Scaffidi, Elbi, Vecerová, Dey, Ozato, Brown,
  Hager, Bustin, and Misteli]{phair_2004_method}
Robert~D. Phair, Paola Scaffidi, Cem Elbi, Jaromíra Vecerová, Anup Dey, Keiko
  Ozato, David~T. Brown, Gordon Hager, Michael Bustin, and Tom Misteli.
\newblock Global nature of dynamic protein-chromatin interactions in vivo:
  Three-dimensional genome scanning and dynamic interaction networks of
  chromatin proteins.
\newblock \emph{Molecular and Cellular Biology}, 24\penalty0 (14):\penalty0
  6393--6402, 2004.
\newblock \doi{10.1128/MCB.24.14.6393-6402.2004}.

\bibitem[Mueller et~al.(2008)Mueller, Wach, and McNally]{mueller_2008}
Florian Mueller, Paul Wach, and James~G. McNally.
\newblock Evidence for a common mode of transcription factor interaction with
  chromatin as revealed by improved quantitative fluorescence recovery after
  photobleaching.
\newblock \emph{Biophysical Journal}, 94\penalty0 (8):\penalty0 3323--3339,
  2008.
\newblock \doi{10.1529/biophysj.107.123182}.

\bibitem[Speil et~al.(2011)Speil, Baumgart, Siebrasse, Veith, Vinkemeier, and
  Kubitscheck]{speil_2011}
Jasmin Speil, Eugen Baumgart, Jan-Peter Siebrasse, Roman Veith, Uwe Vinkemeier,
  and Ulrich Kubitscheck.
\newblock Activated {STAT1} transcription factors conduct distinct saltatory
  movements in the cell nucleus.
\newblock \emph{Biophysical Journal}, 101\penalty0 (11):\penalty0 2592--2600,
  2011.
\newblock \doi{10.1016/j.bpj.2011.10.006}.

\bibitem[Kloster-Landsberg et~al.(2011)Kloster-Landsberg, Herbomel, Wang,
  Derouard, Vourch, Usson, Souchier, and Delon]{kloster_landsberg_2012}
Meike Kloster-Landsberg, Gaetan Herbomel, Irène Wang, Jacques Derouard, Claire
  Vourch, Yves Usson, Catherine Souchier, and Antoine Delon.
\newblock Cellular response to heat shock studied by multiconfocal fluorescence
  correlation spectroscopy.
\newblock \emph{Biophysical Journal}, 103\penalty0 (6):\penalty0 1110--1119,
  2011.
\newblock \doi{10.1016/j.bpj.2012.07.041}.

\bibitem[Mazza et~al.(2012)Mazza, Abernathy, Golob, Morisaki, and
  McNally]{mazza_2012}
Davide Mazza, Alice Abernathy, Nicole Golob, Tatsuya Morisaki, and James~G.
  McNally.
\newblock A benchmark for chromatin binding measurements in live cells.
\newblock \emph{Nucleic Acids Research}, 40\penalty0 (15):\penalty0 e119, 2012.
\newblock \doi{10.1093/nar/gks701}.

\bibitem[Gebhardt et~al.(2013)Gebhardt, Suter, Roy, Zhao, Chapman, Basu,
  Maniatis, and Xie]{gebhardt_2013}
J~Christof~M Gebhardt, David~M Suter, Rahul Roy, Ziqing~W Zhao, Alec~R Chapman,
  Srinjan Basu, Tom Maniatis, and X~Sunney Xie.
\newblock Single-molecule imaging of transcription factor binding to {DNA} in
  live mammalian cells.
\newblock \emph{Nature Methods}, 10\penalty0 (5):\penalty0 421--426, 2013.
\newblock \doi{10.1038/nmeth.2411}.

\bibitem[Morisaki et~al.(2014)Morisaki, Muller, Golob, Mazza, and
  McNally]{morisaki_2014}
Tatsuya Morisaki, Waltraud~G. Muller, Nicole Golob, Davide Mazza, and James~G.
  McNally.
\newblock Single-molecule analysis of transcription factor binding at
  transcription sites in live cells.
\newblock \emph{Nature Communications}, 5:\penalty0 4456, 2014.
\newblock \doi{10.1038/ncomms5456}.

\bibitem[Zamparo and Perkins(2009)]{zamparo_2009}
Lee Zamparo and Theodore~J. Perkins.
\newblock Statistical lower bounds on protein copy number from fluorescence
  expression images.
\newblock \emph{Bioinformatics}, 25\penalty0 (20):\penalty0 2670--2676, 2009.
\newblock \doi{10.1093/bioinformatics/btp415}.

\bibitem[Pisarev et~al.(2009)Pisarev, Poustelnikova, Samsonova, and
  Reinitz]{pisarev_2009}
Andrei Pisarev, Ekaterina Poustelnikova, Maria Samsonova, and John Reinitz.
\newblock {FlyEx}, the quantitative atlas on segmentation gene expression at
  cellular resolution.
\newblock \emph{Nucleic Acids Research}, 37\penalty0 (suppl 1):\penalty0
  D560--D566, 2009.
\newblock \doi{10.1093/nar/gkn717}.

\bibitem[Biggin(2011)]{biggin_2011}
Mark~D. Biggin.
\newblock Animal transcription networks as highly connected, quantitative
  continua.
\newblock \emph{Developmental Cell}, 21\penalty0 (4):\penalty0 611--626, 2011.
\newblock \doi{10.1016/j.devcel.2011.09.008}.

\bibitem[Li et~al.(2014)Li, Bickel, and Biggin]{li_2014_proteome}
Jingyi~Jessica Li, Peter~J Bickel, and Mark~D Biggin.
\newblock System wide analyses have underestimated protein abundances and the
  importance of transcription in mammals.
\newblock \emph{PeerJ}, 2:\penalty0 e270, 2014.
\newblock \doi{10.7717/peerj.270}.

\bibitem[Segal et~al.(2008)Segal, Raveh-Sadka, Schroeder, Unnerstall, and
  Gaul]{segal_2008}
Eran Segal, Tali Raveh-Sadka, Mark Schroeder, Ulrich Unnerstall, and Ulrike
  Gaul.
\newblock Predicting expression patterns from regulatory sequence in
  \emph{Drosophila} segmentation.
\newblock \emph{Nature}, 451\penalty0 (7178):\penalty0 535 -- 540, 2008.
\newblock \doi{10.1038/nature06496}.

\bibitem[Fisher et~al.(2012)Fisher, Li, Hammonds, Brown, Pfeiffer, Weiszmann,
  MacArthur, Thomas, Stamatoyannopoulos, Eisen, Bickel, Biggin, and
  Celniker]{fisher_2012}
William~W. Fisher, Jingyi~Jessica Li, Ann~S. Hammonds, James~B. Brown,
  Barret~D. Pfeiffer, Richard Weiszmann, Stewart MacArthur, Sean Thomas,
  John~A. Stamatoyannopoulos, Michael~B. Eisen, Peter~J. Bickel, Mark~D.
  Biggin, and Susan~E. Celniker.
\newblock {DNA} regions bound at low occupancy by transcription factors do not
  drive patterned reporter gene expression in \emph{Drosophila}.
\newblock \emph{Proceedings of the National Academy of Sciences}, 109\penalty0
  (52):\penalty0 21330--21335, 2012.
\newblock \doi{10.1073/pnas.1209589110}.

\bibitem[Teytelman et~al.(2013)Teytelman, Thurtle, Rine, and van
  Oudenaarden]{teytelman_2013}
Leonid Teytelman, Deborah~M. Thurtle, Jasper Rine, and Alexander van
  Oudenaarden.
\newblock Highly expressed loci are vulnerable to misleading {ChIP}
  localization of multiple unrelated proteins.
\newblock \emph{PNAS}, 110\penalty0 (46):\penalty0 18602--18607, 2013.
\newblock \doi{10.1073/pnas.1316064110}.

\bibitem[Meyer and Liu(2014)]{meyer_2014}
Clifford~A. Meyer and X.~Shirley Liu.
\newblock Identifying and mitigating bias in next-generation sequencing methods
  for chromatin biology.
\newblock \emph{Nat Rev Genet}, 2014.
\newblock \doi{10.1038/nrg3788}.

\bibitem[Sung et~al.(2014)Sung, Guertin, Baek, and Hager]{sung_2014}
Myong-Hee Sung, Michael J. Guertin, Songjoon Baek, and Gordon L. Hager.
\newblock {DNase} footprint signatures are dictated by factor dynamics and
  {DNA} sequence.
\newblock \emph{Molecular Cell}, 2014.
\newblock \doi{10.1016/j.molcel.2014.08.016}.

\bibitem[Stormo(2000)]{stormo_2000}
Gary~D Stormo.
\newblock {DNA} binding sites: representation and discovery.
\newblock \emph{Bioinformatics}, 16\penalty0 (1):\penalty0 16--23, 2000.
\newblock \doi{10.1093/bioinformatics/16.1.16}.

\bibitem[Benos et~al.(2002)Benos, Lapedes, and Stormo]{benos_2002}
Panayiotis~V. Benos, Alan~S. Lapedes, and Gary~D. Stormo.
\newblock {Is there a code for protein?DNA recognition? Probab(ilistical)ly?}
\newblock \emph{BioEssays}, 24\penalty0 (5):\penalty0 466--475, 2002.
\newblock \doi{10.1002/bies.10073}.

\bibitem[Buchler et~al.(2003)Buchler, Gerland, and Hwa]{buchler_2003}
Nicolas~E. Buchler, Ulrich Gerland, and Terence Hwa.
\newblock On schemes of combinatorial transcription logic.
\newblock \emph{PNAS}, 100\penalty0 (9):\penalty0 5136--5141, 2003.
\newblock \doi{10.1073/pnas.0930314100}.

\bibitem[Bintu et~al.(2005{\natexlab{b}})Bintu, Buchler, Garcia, Gerland, Hwa,
  Kondev, and Phillips]{bintu_2005b}
Lacramioara Bintu, Nicolas~E. Buchler, Hernan~G. Garcia, Ulrich Gerland,
  Terence Hwa, Jane Kondev, and Rob Phillips.
\newblock Transcriptional regulation by the numbers: applications.
\newblock \emph{Current Opinion in Genetics \& Development}, 15\penalty0
  (2):\penalty0 125--135, 2005{\natexlab{b}}.
\newblock \doi{10.1016/j.gde.2005.02.006}.

\bibitem[Xu et~al.(2014)Xu, Chen, Ling, Yu, Struffi, and Small]{xu_2014}
Zhe Xu, Hongtao Chen, Jia Ling, Danyang Yu, Paolo Struffi, and Stephen Small.
\newblock Impacts of the ubiquitous factor {Zelda} on {Bicoid}-dependent {DNA}
  binding and transcription in \emph{Drosophila}.
\newblock \emph{Genes \& Development}, 28:\penalty0 608--621, 2014.
\newblock \doi{10.1101/gad.234534.113}.

\bibitem[Rhee et~al.(2014)Rhee, Cho, Zhai, Slattery, Ma, Mintseris, Wong,
  White, Celniker, Przytycka, Gygi, Obar, and Artavanis-Tsakonas]{rhee_2014}
David Y. Rhee, Dong-Yeon Cho, Bo~Zhai, Matthew Slattery, Lijia Ma, Julian
  Mintseris, Christina Y. Wong, Kevin P. White, Susan E. Celniker,
  Teresa M. Przytycka, Steven P. Gygi, Robert A. Obar, and Spyros
  Artavanis-Tsakonas.
\newblock Transcription factor networks in drosophila melanogaster.
\newblock \emph{Cell Reports}, 2014.
\newblock \doi{10.1016/j.celrep.2014.08.038}.

\bibitem[Gerland et~al.(2002)Gerland, Moroz, and Hwa]{gerland_2002}
Ulrich Gerland, J.~David Moroz, and Terence Hwa.
\newblock Physical constraints and functional characteristics of transcription
  factor-{DNA} interactions.
\newblock \emph{PNAS}, 99\penalty0 (19):\penalty0 12015--12020, 2002.
\newblock \doi{10.1073/pnas.192693599}.

\bibitem[Zabet and Adryan(2012{\natexlab{c}})]{zabet_2012_grip}
Nicolae~Radu Zabet and Boris Adryan.
\newblock {GRiP}: a computational tool to simulate transcription factor binding
  in prokaryotes.
\newblock \emph{Bioinformatics}, 28\penalty0 (9):\penalty0 1287--1289,
  2012{\natexlab{c}}.
\newblock \doi{10.1093/bioinformatics/bts132}.

\bibitem[Zabet(2012)]{zabet_2012_subsystem}
Nicolae~Radu Zabet.
\newblock System size reduction in stochastic simulations of the facilitated
  diffusion mechanism.
\newblock \emph{BMC Systems Biology}, 6\penalty0 (1):\penalty0 121, 2012.
\newblock \doi{10.1186/1752-0509-6-121}.

\bibitem[van Zon et~al.(2006)van Zon, Morelli, Tanase-Nicola, and ten
  Wolde]{zon_2006}
Jeroen~S. van Zon, Marco~J. Morelli, Sorin Tanase-Nicola, and Pieter~Rein ten
  Wolde.
\newblock Diffusion of transcription factors can drastically enhance the noise
  in gene expression.
\newblock \emph{Biophysical Journal}, 91\penalty0 (12):\penalty0 4350--4367,
  2006.
\newblock \doi{10.1529/biophysj.106.086157}.

\end{thebibliography}
